\documentclass[aps, prd, nofootinbib,showkeys]{revtex4}

\usepackage{amsmath}
\allowdisplaybreaks

\usepackage{makeidx}
\usepackage{amsfonts}
\usepackage[usenames,dvipsnames]{pstricks}
\usepackage{subfigure}
\usepackage{epsfig}
\usepackage{pst-grad} 
\usepackage{mathtools}
\usepackage[colorlinks,hyperindex]{hyperref}
\usepackage{array}
\usepackage{tikz}
\usetikzlibrary{decorations.markings,math,shapes.misc,patterns,arrows.meta,calc,patterns,angles,quotes}

\hypersetup
{	colorlinks,%
	citecolor=black,%
	linkcolor=black,%
	urlcolor=black,%
}

\usepackage{allpurpose}

\newcommand\nn{{\nonumber}}

\begin{document}

\title{The perturbative approach for the weak deflection angle}

\author{Junji Jia}
\email{junjijia@whu.edu.cn}
\affiliation{Center for Astrophysics \& MOE Key Laboratory of Artificial Micro- and Nano-structures, School of Physics and Technology, Wuhan University, Wuhan, 430072, China}

\date{\today}

\begin{abstract}
Both null and timelike rays experience trajectory bending in a gravitational field. In this work, we systematically develop a perturbative method to compute the deflection angle of rays with general velocity $v$ in arbitrary static and spherically symmetric spacetimes and in equatorial plane of arbitrary static and axisymmetric spacetimes. We show that the expansion in the large closest approach $x_0$ limit depends on the asymptotic behavior of the metric functions only, and the generated integrand is always integrable, resulting in a deflection angle in a series form of either $x_0$ or $b$, the impact parameter. Using this method, the deflection angles as series of both $x_0$ and $b$ are found in Schwarzschild, Reissner-Nordstr\"{o}m and Kerr-Newman spacetimes to 17-th, 15-th and 6-th orders respectively, for both lightrays and particles with general velocity. The effects of the impact parameter, velocity and other parameters of the spacatimes are briefly analyzed. Moreover, we show that for spacetimes whose metric functions are only asymptotically known, the deflection angle in the weak field limit can also be calculated. Furthermore, it is shown that the deflection angle in general static and spherically symmetric spacetime and equatorial plane of static and axisymmetric spacetime to the lowest non-trivial order, depends only on the impact parameter, velocity of the particle, and the effective ADM mass of the spacetime but not on other parameters such as charge or angular momentum. These deflection angles are used in an exact gravitational lensing equation and the corresponding apparent angles of the images of the source are also solved perturbatively.
\end{abstract}


\maketitle
\section{Introduction\label{secintro}}

One of the classical and important consequences of General Relativity (GR) is the deflection of lightlike geodesics in curved spacetimes.
One hundred years ago, Eddington's observation of the star position shift played a major role in helping GR won its acceptance by physicists and the public. Nowadays, the deflection of lightlike rays is usually observed by very long baseline radio interferometer \cite{Will:2014kxa} with an accuracy of sub-microarcsecond \cite{Fomalont:2003pd} . In 1991, the deflection of light due to Jupiter was observed \cite{treu} and that due to Saturn might be observed by Gaia \cite{Crosta:2005ch}.

The deflection of light leads to one important theoretical and observational tool in modern astronomy and cosmology: the gravitational lensing (GL). After the observation of the first GL in 1979 \cite{Walsh:1979nx}, many features, including luminous arcs \cite{lynds,soucail}, Einstein cross \cite{Huchra:1985zz}, Einstein rings \cite{hewitt}, GL of CMB \cite{Smith:2007rg,Das:2011ak,vanEngelen:2012va}, and supernovas \cite{Quimby:2013lfa,Nordin:2013cfa} and even composition features such as the SN Refsdal which combines the Einstein cross with the GL of supernova
\cite{Kelly:2014mwa}  have been observed. The GL were then used to study properties of the lens \cite{MiraldaEscude:1996ar}, coevolution of supermassive
black holes (BHs) and galaxies \cite{Peng:2006ew}, cosmological parameters such as large scale structure (for a review see \cite{Lewis:2006fu}), properties of the supernova \cite{Sharon:2014ija}, dark matter substructure \cite{Metcalf:2001ap, Metcalf:2001es}, and to discriminate alternative gravitational theories.
More recently, with the discovery of gravitational wave (GW) \cite{Abbott:2016blz,Abbott:2016nmj,Abbott:2017oio,TheLIGOScientific:2017qsa}, observation of its GL effects has also been proposed and put to use \cite{Collett:2016dey,Fan:2016swi}.

Although traditionally lightrays have been the main messenger  in the observation of GL, with the observation of extragalactic neutrinos from SN 1987A \cite{Hirata:1987hu, Bionta:1987qt} and blazer TXS 0506+056 \cite{IceCube:2018dnn,IceCube:2018cha} and the GWs, it is clear that these two kinds of messengers in principle can also go through the bending process and being observed in lensing scenarios. One of them, neutrinos, are long known to have non-zero small mass \cite{Tanabashi:2018oca}; while for GWs, its speed can also deviate from the speed of light in gravitational theories beyond GR \cite{Sakstein:2017xjx,Baker:2017hug}. Therefore in considering their lensing or GL of any other massive particles, in principle one should compute the deflection angle and time delay formulas using timelike geodesics. Currently, most theoretical works on the GL of these two messengers are still using corresponding values obtained using lightlike rays \cite{barrow1987lensing, Eiroa:2008ks, Fan:2016swi,Wei:2017emo,Yang:2018bdf}. Recently, some of us computed the deflection angles in Schwarzschild and Reissner-Nordstr\"{o}m (RN) spacetimes and time delay in Schwarzschild spacetime for general velocity \cite{Jia:2019hih}. It was shown that the difference of the apparent angles of lightrays and neutrinos can be correlated to the neutrino mass and mass hierarchy \cite{Pang:2018jpm,Jia:2015zon}. Moreover, the dependence of the deflection angle on messenger velocity can also cause small change of the BH shadow size in these spacetimes \cite{Pang:2018jpm,Wang:2019rvq}. These results show that in considering effects related to the trajectory bending of massive particles, timelike instead of lightlike geodesics should be used if a high accuracy is desired.

In GLs of both massless and massive particles, the GLs in the strong field limit is important for a few reasons, especially for their applications in strong field test of GR and alternative gravitational theories. In the strong field limit of arbitrary statics and spherically symmetric (SSS) spacetime, Bozza \cite{Bozza:2002zj} developed a general method to calculate the deflection angle for lightrays and showed that it always diverges logarithmically as $\ln(x-x_0)$ where $x_0$ is the closest approach of the ray. Many authors successfully calculated the deflection angles in the strong field limit for particular spacetimes using this method \cite{Eiroa:2003jf,Tsupko:2014wza,Bozza:2009yw,Jia:2015zon,Manna:2018jxb}.
However, from an observational point of view, all the light bending and GLs seen up to now are in the weak field limit, i.e., with small deflection angles. GL in the strong field limit will not be directly observable in near future experiments because of its high resolution  and magnification requirement. For example, to resolve the relativistic images produced by the Milky Way galaxy BH requires an angle resolution of $10^{-2}$ microarcsecond, which is 2 orders smaller than the resolution currently reachable \cite{Jia:2015zon}.

In this paper therefore, we still concentrate on the deflection angle and GL in the weak field limit, but for general velocity. We propose a general formalism for computing the deflection angle of particles with arbitrary velocity in SSS spacetimes and in the equatorial plane of stationary and axisymmetric (SAS) spacetimes. This procedure allows us to expand the deflection angle in the power of the reciprocal of the closest approach $x_0$ or impact parameter $b$, which is a large quantity in the weak field limit. Moreover, it will also be shown that this procedure can be used to calculate the deflection angle of metrics that are only known in the asymptotic region. This is particularly useful because in most spacetimes that have a complex matter distribution, the metric functions are usually not analytically solvable while their asymptotical behavior can usually be obtained using series method. We will also show that for all equatorial geodesics in any SAS and asymptotically flat spacetimes (note that SAS spacetimes cover SSS spacetimes), the deflection angle to the lowest non-trivial order always takes the form
\be
\alpha=\frac{2m}{b}\lb 1+\frac{1}{v^2}\rb\ee
where $m$ is the ADM mass of the spacetime, $b$ is the impact parameter and $v$ is the speed of the test particles at infinity. In other words, all other spacetime parameters such as effective charges or angular momentum will not influence the deflection angle at this order.  Again, we emphasis that all results in this paper apply not only to null rays but signals with general velocity. 

Previously, the weak deflection angle has been calculated mainly using two slightly different but yet connected approaches. The first and most traditional way is the direct integration method which tackles the integral for the deflection angle directly. The second, which is more recent and also very promising, is to utilize the Gauss-Bonnet theorem to find the deflection in a somewhat more indirect but elegant way 
\cite{Arakida:2017hrm,Ovgun:2018tua,Kumaran:2019qqp,Zhu:2019ura,Ovgun:2019qzc,Javed:2019kon}.
From this point of view, our work leans more towards the first category. 

We arrange the paper in the following way. In Sec. \ref{secdaexp}, we present the perturbative procedure for computing the deflection angle in SSS spacetimes to any desired order. The results are then used in Sec. \ref{seceg} to find the deflection angle to the minus 17th order of $b$ in Schwarzschild spacetime and minus 15th order in RN spacetime. We also give an example (the SU(2) Yang-Mills-Einstein solution) for computing the deflection angle in asymptotically known spacetimes. In Sec. \ref{secsas}, the deflection angle in the equatorial plane of Kerr and Kerr-Newman (KN) spacetimes are computed to the minus 6th order of $b$. We then in Sec. \ref{secgl} show how all these deflection angles can be used to find the apparent angle in GL perturbatively. Lastly, a few possible applications and directions of extension are discussed in Sec. \ref{secdiscussion}.  Throughout the paper we use the geometric unit $G=c=1$.

\section{Deflection angle in SSS spacetimes\label{secdaexp}}

For general SSS spacetime, the metric can always be written as
\be \dd s^2=-A(x)\dd t^2+B(x)\dd x^2+C(x)(\dd\theta^2+\sin^2\theta\dd \phi^2). \label{sss1} \ee
For this metric,  the geodesic equations after two first integrals take the form
\begin{subequations}\label{schgeo}
\begin{align}
\dot{t}=&\frac{E}{A(x)}, \label{eqdtdl}\\
\dot{\phi}=&\frac{L}{C(x)}, \label{eqdpdl}\\
\dot{x}^2=&\frac{1}{B(x)}\lb \kappa -\frac{E^2}{A(x)}+\frac{L^2}{C(x)}\rb, \label{eqdxdl}
\end{align}
\end{subequations}
where $\dot{~}$ denotes the derivative with respect to the proper time (or affine parameter) $\lambda$. Here we have already set $\theta(\lambda)=\pi/2$ without losing any generality and $\kappa=0,~1$ for lightlike and timelike particles respectively.
$E$ is the energy of the lightlike ray or that of unit mass of the test particles at infinity, which can be related to their velocity $v$ at infinity through
\be E=\frac{1}{\sqrt{1-v^2} }. \label{einv}\ee
$L$ is the angular momentum of unit mass of the test particle at infinity, satisfying
\be
|L|=|\mathbf{p}\times \mathbf{r}|=\frac{v}{\sqrt{1-v^2}}b=b\sqrt{E^2-1}, \label{linv} \ee
where $b$ is the impact parameter of the ray. Here and henceforth, we assign the impact parameter to be a positive constant, while the angular momentum $L$ in principle can change sign.

Using Eqs. \eqref{schgeo}-\eqref{linv}, it is easy to show that after some simple algebra, one can integrate $\dd\phi/\dd x$  to obtain the deflection angle of a ray both originating and propagating from infinite radius as \cite{Virbhadra:1998dy}
\be
\alpha(x_0)=I(x_0)-(2 n+1)\pi\ee
where the change of the angular coordinate, $I(x_0)$, is
\be
I(x_0)=2\int_{x_0}^\infty \frac{\sqrt{B(x)\lb E^2-\kappa A(x_0)\rb }}{\sqrt{C(x)}\sqrt{E^2\lsb \frac{C(x)}{A(x)}\frac{A(x_0)}{C(x_0)}-1\rsb +\kappa A(x_0)\lb 1-\frac{C(x)}{C(x_0)}\rb }}\dd x. \label{idef}
\ee
Here $n$ is an integer such that $-\pi\leq\alpha(x_0)<\pi$. In the weak field limit, we should always choose $n=0$.

Setting $\kappa=0$, Eq. \eqref{idef} reduces to the change of the angular coordinate of lightlike ray given in Ref. \cite{Virbhadra:1998dy,Bozza:2002zj}
\be
I(x_0)=2\int_{x_0}^\infty \frac{\sqrt{B(x)}}{\sqrt{C(x)}\sqrt{\frac{C(x)}{A(x)}\frac{A(x_0)}{C(x_0)}-1}}\dd x .
\ee
The closest distance of approach $x_0$ can be linked with $L$ by setting $\dot{x}=0$ in Eq. \eqref{eqdxdl}. Further using Eqs. \eqref{einv} and \eqref{linv}, one  has the following relation between the impact parameter $b$ and $x_0$
\bea
\frac{1}{b}&=&\frac{\sqrt{E^2-1}}{|L|}\label{blerel}\\
&=&\frac{\sqrt{E^2-1}}{\sqrt{E^2-\kappa A(x_0)}}\sqrt{\frac{A(x_0)	}{C(x_0)}}\equiv l\lb \frac{1}{x_0}\rb. \label{br0rel}
\eea

As we stated in Sec. \ref{secintro}, we are interested in the weak field limit of the deflection angle. To calculate $\alpha(x_0)$ in weak deflection limit, then we should let $x_0$ approach infinity. The main content of this paper is then to show that the change of the angular coordinate $I(x_0)$ and consequently the deflection angle $\alpha(x_0)$ can be expanded as a power series of $\frac{1}{x_0}$ and the integral can always be carried out. Moreover, we find a practical procedure to calculate the coefficient of each power, for both null and timelike rays.

To carry out the expansion and integration of Eq. \eqref{idef}, we first change variable from $x$ to $u=x/x_0$ so that $I(x_0)$ becomes
\bea
I(x_0)&=&2\int_0^1 \frac{x_0}{u^2} \frac{\sqrt{B(x_0/u)(E^2-\kappa A(x_0))}}{\sqrt{C(x_0/u)}\sqrt{E^2\lsb \frac{C(x_0/u)}{A(x_0/u)}\frac{A(x_0)}{C(x_0)}-1\rsb+\kappa A(x_0)\lb 1-\frac{C(x_0/u)}{C(x_0)}\rb }}\dd u\equiv \int_0^1 y(x_0,u)\dd u, \label{angtrans}
\eea
where $y(x_0,u)$ stands for the integrand.
This change of variable effectively transforms a partial dependence of $I(x_0)$ on $x_0$ through the lower limit of the integral to a full dependence on $x_0$ through the integrand $y(x_0,u)$. For this integrand, since in the weak field limit, $x_0$ is much larger than any characteristic length intrinsic to the spacetime, we can always do an asymptotic expansion in powers of $\displaystyle \frac{1}{x_0}$ to find
\bea
&&y(x_0,u)\nonumber\\
&=&\frac{1}{x_0^0} \lim_{z\to\infty} \lcb -\frac{2z A\lb \frac{z}{u}\rb  B\lb \frac{z}{u}\rb  \sqrt{C(z)g(z)}}
{u^2 C\lb \frac{z}{u}\rb  h(z,u)}\rcb \nonumber\\
&&+\frac{1}{x_0^1} \lim_{z\to\infty} \lcb  -\frac{2}{u^2}\lsb \frac{\sqrt{g(z)}}{h(z,u)}\lb  \frac{1}{\sqrt{C\lb \frac{z}{u}\rb }}\lb  \sqrt{C(z)}\lb  \sqrt{B\lb \frac{z}{u}\rb } \lb  \sqrt{A \lb \frac{z}{u}\rb }\lb -z^2 +\frac{\kappa z^3 A^\prime(z)}{2 g(z)}\rb  \right.\right.\right.\right.\right.\right.\nonumber\\
&&\left.\left.\left.\left.-\frac{z^3  A^\prime \lb \frac{z}{u}\rb }{2 A\lb \frac{z}{u}\rb }\rb  -\frac{z^3 \sqrt{ A\lb \frac{z}{u}\rb } B^\prime \lb \frac{z}{u}\rb }{2 B\lb \frac{z}{u}\rb }\rb  - \frac{z^3\sqrt{A\lb \frac{z}{u}\rb  B\lb \frac{z}{u}\rb }C^\prime(z)}{2\sqrt{C(z)}}\rb +
\frac{z^3\sqrt{A\lb \frac{z}{u}\rb B\lb \frac{z}{u}\rb C(z)}C^\prime \lb \frac{z}{u}\rb }{2 C\lb \frac{z}{u}\rb ^{3/2}}\rb
\nonumber\\
&&+\frac{z^3\sqrt{g(z)A\lb \frac{z}{u}\rb B\lb \frac{z}{u}\rb C(z)}}{2\sqrt{C\lb \frac{z}{u}\rb } h(z,u)^3} \lb  -A^\prime(z)\lb  \kappa A\lb \frac{z}{u}\rb  C(z) +C\lb \frac{z}{u}\rb  g\lb \frac{z}{u}\rb \rb
+E^2\lb C(z)A^\prime \lb \frac{z}{u}\rb +A\lb \frac{z}{u}\rb C^\prime(z)\rb \right.\nonumber\\
&&\left.\left.\left.+A(z)\lb \kappa \lb \lb C\lb \frac{z}{u}\rb -C(z)\rb  A^\prime \lb \frac{z}{u}\rb  -A\lb \frac{z}{u}\rb  C^\prime(z)\rb -C^\prime \lb \frac{z}{u}\rb g\lb \frac{z}{u}\rb \rb \rb \rsb\rcb \nonumber\\
&&+\frac{1}{x_0^2}\lim_{z\to\infty}y_2\lb g_{\mu\nu}(z),g_{\mu\nu}\lb \frac{z}{u}\rb ,g_{\mu\nu}(z)^\prime,g_{\mu\nu}\lb \frac{z}{u}\rb ^\prime,g_{\mu\nu}(z)^{\prime\prime},g_{\mu\nu}\lb \frac{z}{u}\rb ^{\prime\prime}\rb  +\mathcal{O}\lb \frac{1}{x_0^3}\rb  \label{expabcfull}\\
&\equiv&\sum_{n=0}\frac{y_n(x_0,u)}{x_0^n}\label{expabcsim}
\eea
where $^\prime$ denotes the derivative with respect to $z$, and
\bea
g(z)&=&E^2-\kappa A(z),\\
h(z,u)&=&\sqrt{A(z)C\lb \frac{z}{u}\rb  g\lb \frac{z}{u}\rb -A\lb \frac{z}{u}\rb  C(z) g(z)},
\eea
and $y_n(x_0,u)~(n\geq 2)$ in Eqs. \eqref{expabcfull} and \eqref{expabcsim} denote the coefficient of $\displaystyle \left(\frac{1}{x_0}\right)^n$.
Setting $\kappa=0$ in Eq. \eqref{expabcfull} yields for lightlike rays,
\bea
y(x_0,u)
&=&\frac{1}{x_0^0} \lim_{z\to\infty} \lcb -\frac{2 z\sqrt{B\lb \frac{z}{u}\rb  }}{u^2\sqrt{C\lb \frac{z}{u}\rb  } \sqrt{\frac{A(z)
   C\lb \frac{z}{u}\rb  }{A\lb \frac{z}{u}\rb  C(z)}-1}}\rcb \nonumber\\
   &&+\frac{1}{x_0^1} \lim_{z\to\infty} \lcb \frac{z^2}{2}     \lsb  z\lb \frac{B\lb \frac{z}{u}\rb  }{C\lb \frac{z}{u}\rb }\rb ^\prime \sqrt{\frac{C\lb \frac{z}{u}\rb }{B\lb \frac{z}{u}\rb }}-2\sqrt{\frac{B\lb \frac{z}{u}\rb  }{C\lb \frac{z}{u}\rb }}\rsb  \lb \frac{A(z) C\lb \frac{z}{u}\rb  }{A\lb \frac{z}{u}\rb  C(z)}-1\rb ^{-1/2} \right. \nonumber \\
   &&\left.+ \frac{z^3}{2}\sqrt{\frac{B\lb \frac{z}{u}\rb  }{C\lb \frac{z}{u}\rb }}  \lb  \frac{A(z)C\lb \frac{z}{u}\rb  } {A\lb \frac{z}{u}\rb  C(z)}\rb ^\prime   \lb \frac{A(z) C\lb \frac{z}{u}\rb  }{A\lb \frac{z}{u}\rb  C(z)}-1\rb ^{-3/2} \rcb \nonumber\\
&&+\frac{1}{x_0^2}\lim_{z\to\infty}y_2 \lsb g_{\mu\nu}(z),g_{\mu\nu}\lb \frac{z}{u}\rb , g_{\mu\nu}(z)^\prime,g_{\mu\nu}\lb \frac{z}{u}\rb ^\prime, g_{\mu\nu}(z)^{\prime\prime}, g_{\mu\nu}\lb \frac{z}{u}\rb ^{\prime\prime}\rsb  +\mathcal{O}\lb \frac{1}{x_0}\rb^3 \nonumber\\
&&  \label{expabcnull}
\eea

Integrand \eqref{expabcfull} and \eqref{expabcnull} have a few remarkable properties.
It is seen that the order $\displaystyle \lb \frac{1}{x_0}\rb^0$ term involves only the limits of the metric functions, and the order $\displaystyle \lb \frac{1}{x_0}\rb^1$  terms involve up to the first order derivative of the metric functions. The $\displaystyle \lb\frac{1}{x_0}\rb^2$  term is not shown explicitly here because of its length, but we can show that it only contains terms up to second derivative of the metric functions. In general, it is found that this expansion can always be carried out to arbitrary desired power of $\displaystyle \frac{1}{x_0}$  and the coefficient for the $\displaystyle \lb \frac{1}{x_0}\rb^n$  order involves up to the $n$-th derivative of the metric functions. Note that although the expressions might appear long and tedious as the order increases, the involved mathematics are only taking derivatives and finding limits, and therefore still algebraically simple. In this sense, this expansion is straightforward and tractable, especially when the metric functions are explicitly known. As will be seen in Sec. \ref{seceg}, for explicit spacetime metrics the coefficients for all orders of $\displaystyle \lb\frac{1}{x_0}\rb^n$  are concise. Therefore, length of the results of $I(x_0)$ obtained after integrating $u$ in Eqs. \eqref{expabcfull} and \eqref{expabcnull} will also be much shorter comparing to the length of the integrand.

The second property, which is simple but fundamental, is that the expansions use {\it only the asymptotic behavior} of the metric functions to the desired order. In other words, from the point of view of the deflection angle $\alpha(x_0)$, the rays with asymptotically large $x_0 $ do not experience how the spacetime is curved in the central region. Spacetime in the large $x_0$ region to various orders of $\displaystyle \frac{1}{x_0}$  is enough to determine the deflection angle to the corresponding order. Although intuitively this seems simple, it is this property that allows us to determine the deflection angle in spacetimes whose metric function are not fully but only asymptotically known. We shall see in Sec. \ref{secunknownmetric} how the deflection in colored BH spacetime can be calculated.

The third property of these expansions, is that they are always integrable if the metric functions $A(x),~B(x)$ and $C(x)/x^2$ are finite at large $x$ and have asymptotical expansions of the form
\be
A(x)=\sum_{n=0}\frac{a_n}{x^n},~ B(x)=\sum_{n=0}\frac{b_n}{x^n},~\frac{C(x)}{x^2}=\sum_{n=0}\frac{c_n}{x^n} ~(n=0,~1,~2,\cdots)\label{expabc}
\ee
where $a_n$, $b_n$ and $c_n$ are finite. These conditions are certainly satisfied by many familiar spacetimes such as Schwarzschild, RN, Hayward \cite{Hayward:2005gi}, Bardeen \cite{bardeen}, Gibbons-Maeda-Garfinkle-Horowitz-Strominger (GMGHS) \cite{Gibbons:1982ih,Gibbons:1987ps,Garfinkle:1990qj} and Janis-Newman-Winicour \cite{Janis:1968zz} metrics. With these conditions, clearly all the terms in various numerators in $y_n(x_0,u)$ in Eq. \eqref{expabcfull} can be expanded as series of their arguments, which become powers of $u$. The only non-trivial parts are the denominators in $y_n(x_0,u)$, i.e., the terms $1/h(z,u)^n$. In the limit $z\to\infty$, they can be simplified as
\be \lim_{z\to\infty}\frac{1}{h(z,u)^n}=\lsb \frac{1}{a_0(E-\kappa a_0)c_0z^2}\frac{u^2}{(1-u^2)}\rsb ^{n/2}. \ee
A further transformation of variable $u\to\cos\theta$ will transform the expansion \eqref{expabcsim} into the following form
\be
I(x_0)=\int_1^0\sum_{n=0}\frac{y_n(u)}{x_0^n}\dd u
\xrightarrow{u\to \cos\theta} \sum_{n=0}\frac{1}{x_0^n}\int_0^{\frac{\pi}{2}}\frac{\displaystyle \sum_{m=0}^ny_{n,m}\cos^m\theta}{(1+\cos\theta)^n}\dd \theta .
\label{iynmexp}\ee
Here $y_{n,m}$ are functions obtained after the transformation and they do not depend on the integration variable. Because of the integrability of functions of type $\displaystyle \frac{\cos^m\theta}{(1+\cos\theta)^n}~(m\leq n)$, the expansion \eqref{expabcfull} becomes always integrable under conditions Eq. \eqref{expabc}.

In short, with the integrand expanded as \eqref{expabcfull}, one can then do the integral of $u$ to get the change of the angular coordinate as a power series of $\displaystyle \frac{1}{x_0}$
\be
I(x_0)=\sum_{n=0}\frac{1}{x_0^n}\int_1^0y_n(u)\dd u \equiv \sum_{n=0}\frac{I_n}{x_0^n}, \label{iton}\ee
where $I_n$ is the $n$-th order coefficient in $I(x_0)$.
It is important to note that the integral here can always be carried out because of the integrability of Eq. \eqref{iynmexp}.

Because the impact parameter $b$ is much easier to be linked to observable quantities in GLs, it is also desirable to write the expansion \eqref{iton} in series of $\displaystyle \frac{1}{b}$. In that case, we can attempt to directly solve Eq. \eqref{br0rel} to obtain the inverse function
\be
\frac{1}{x_0}=l^{-1}\lb \frac{1}{b}\rb  \label{x0inb}
\ee
and then substituting into Eq. \eqref{expabcfull} and expanding around large $b$ to obtain the desired expansion
\be
I(b)=\sum_{n=0}\frac{I_n^\prime}{b^n}. \label{itonp}
\ee
Indeed, because what is needed is only the {\it expansion} of \eqref{x0inb} by not necessarily the inverse function $l^{-1}\lb \frac{1}{b}\rb$ itself, one can use the Lagrange inversion theorem to directly find this expansion without solving $l^{-1}\lb \frac{1}{b}\rb$ explicitly
\be
\frac{1}{x_0}=\sum_{n=0}\frac{1}{n!}
\lim_{w\to 0} \frac{\dd^{n-1}}{\dd w^{n-1}} \lcb \lsb \frac{w}{l(w)-l(0)}\rsb ^n\rcb
\frac{1}{b^n}.\label{lthm}\ee
This will be especially useful for those metric functions in Eq. \eqref{br0rel} that are hard to find the inverse function $l^{-1}\lb \frac{1}{b}\rb$ analytically.

Moreover, for ultrarelativistic massive particles, we can also further expand the integrand \eqref{expabcfull} around $v=1$ and integrate to find $I(x_0)$ or $I(b)$ as a series expansion of the velocity difference
\be
I=\sum_{n=0}\lsb \frac{1}{x_0^n}\sum_{m=0} I_{n,m} (1-v)^m \rsb =\sum_{n=0}\lsb \frac{1}{b^n} \sum_{m=0} I_{n,m}^\prime (1-v)^m \rsb. \ee
Apparently, the leading terms in the $m$ summation $\displaystyle \sum_{n=0} \frac{I_{n,0}}{x_0^n}$ or $\displaystyle \sum_{n=0} \frac{I_{n,0}^\prime}{b^n}$ will be just the change of the angular coordinate of lightlike rays.

\section{Application to particular SSS spacetimes\label{seceg}}

In this section, we apply the procedure in Sec. \ref{secdaexp} to some known SSS spacetimes, namely the Schwarzschild, RN and the colored BH spacetimes. We will find the deflection angles in the weak deflection limit for general particle velocity, i.e., for both lightlike and timelike particles. The state of art of the deflection angle for Schwarzschild metric is to the sixth order of $\displaystyle \frac{1}{x_0}$  and $\displaystyle \frac{1}{b}$ for lightlike rays \cite{Keeton:2005jd}, and for RN metric to the third order of $\displaystyle \frac{1}{x_0}$ and $\displaystyle \frac{1}{b}$ for lightlike rays \cite{Keeton:2005jd}, and to various orders for some other interesting metrics as well \cite{Wei:2015qca, Batic:2014loa, Ghaffarnejad:2014zva, Renzini:2017bqg, Chiba:2017nml,Cao:2018lrd}.
In the following, we present the change of the angular coordinate to much higher orders for signals with general velocity.
As we explained before, although the idea of this formalism is powerful and clear, the calculations, especially taking high order derivatives, simplifications and integration are tedious. Therefore in most cases we will avoid showing the intermediate steps.

\subsection{Deflection angle in the weak field limit in Schwarzschild spacetime}
For the Schwarzschild spacetime, the metric functions take the form
\be
A(x)=1-\frac{2m}{x},~B(x)=\frac{1}{A(x)},~C(x)=x^2. \label{schmetricf}\ee
After substituting into Eq. \eqref{expabcfull}, simplifying and integrating over $u$, we find $I(x_0)$ to the order $\displaystyle \frac{1}{x_0^{17}}$
\be I_\mathrm{S}(x_0,v)=\sum_{n=0}^{17}S_n\lb \frac{m}{x_0}\rb ^n+ \mathcal{O}\lb \frac{m}{x_0}\rb ^{18} ,\label{angschfull}\ee
with
\begin{subequations}\label{sndetails}
\begin{align}
S_0=&\pi,\\
S_1=&2\lb 1+\frac{1}{v^2}\rb,\\
S_2=&\frac{3\pi}{4}+(3\pi-2)\frac{1}{v^2}-2\frac{1}{v^4},\\
S_3=&\frac{10}{3}+\lb26-\frac{3\pi}{2}\rb\frac{1}{v^2}-3(2\pi-3)\frac{1}{v^4} +\frac{7}{3}\frac{1}{v^6},\\
S_4=&\frac{105\pi}{64}+\lb\frac{93\pi}{4}-18\rb\frac{1}{v^2}+\lb\frac{69\pi}{4}-86\rb\frac{1}{v^4} +(12\pi-23)\frac{1}{v^6}-3\frac{1}{v^8},
\end{align}
\end{subequations}
and the $S_5$ to $S_{17}$ terms are given in Eq. \eqref{sndetailssupp} because of their excessive length.

For lightlike particles, setting $v=1$ in Eq. \eqref{angschfull} yields the result
\be I_\mathrm{S,\gamma}(x_0)=\sum_{n=0}^{17}S_{n,\gamma}\lb \frac{m}{x_0}\rb ^n+ \mathcal{O}\lb \frac{m}{x_0}\rb ^{18} ,\label{angschlight}\ee
with
\begin{subequations}\label{sndetailsv1}
\begin{align}
S_{0,\gamma}=&\pi,\\
S_{1,\gamma}=&4,\\
S_{2,\gamma}=& \frac{15\pi}{4}-4,\\
S_{3,\gamma}=&\frac{122}{3}-\frac{15\pi}{2},\\
S_{4,\gamma}=& \frac{3465\pi}{64}-130,
\end{align}
\end{subequations}
with $S_{5,\gamma}$ to $S_{17,\gamma}$ given in Eq. \eqref{sndetailsv1supp}.

For ultrarelativistic particles, their change of the angular coordinates can be obtained by expanding \eqref{angschfull} around \eqref{angschlight} as series of $(1-v)$. To the order $(1-v)^1$, this takes the form
\bea
I_\mathrm{S}(x_0,v\to 1)&=&I_\mathrm{S,\gamma}(x_0)+(1-v)
\lsb
  4\lb\frac {m} {x_ 0} \rb^{1}
  +\lb 6 \pi -12\rb \lb\frac {m} {x_ 0} \rb^{2}
  +\lb 102-27 \pi\rb \lb\frac {m} {x_ 0} \rb^{3}\right.\nn\\
&&\left.
  +\lb\frac{375 \pi }{2}-542\rb \lb\frac {m} {x_ 0} \rb^{4}
+ \mathcal{O}\lb \frac{m}{x_0}\rb ^{5}\rsb +\mathcal{O}(1-v)^2.\label{angschvcinx0}\eea
Higher order terms in $\lb \frac{m}{x_0}\rb$ are given in \eqref{eq:isvto1full} and terms of order $(1-v)^n~(n\geq 2)$ can be easily obtained too.

In order to write the expansion \eqref{angschfull} in power series of $\frac{1}{b}$, we first substitute metric functions \eqref{schmetricf} and $\kappa=1$ into Eq. \eqref{br0rel}
\be
\frac{1}{b}=\frac{1}{x_0}\frac{\sqrt{E^2-1}}{\sqrt{E^2-\lb 1-\frac{2m}{x_0}\rb}}\sqrt{1-\frac{2m}{x_0}}.\ee
Using Eq. \eqref{lthm}, we can find $\frac{m}{x_0}$ in power series of $\frac{m}{b}$ as
\be
\frac{m}{x_0}=\sum_{n=1}^{17}C_{\mathrm{S},n}\lb \frac{m}{b}\rb^n +\mathcal{O}\lb \frac{m}{b}\rb^{18}  \label{eq:x0inbschgeneral}
\ee
where the coefficients are
\begin{subequations}\label{x0inbsch}
\begin{align}
C_{\mathrm{S},1}=&1,\\
C_{\mathrm{S},2}=&\frac{1}{v^2},\\
C_{\mathrm{S},3}=&\frac{2}{v^2}+\frac{1}{2v^4},\\
C_{\mathrm{S},4}=&4\lb\frac{1}{v^2}+\frac{1}{v^4}\rb,
\end{align}
\end{subequations}
and the $C_{\mathrm{S},5}$ to $C_{\mathrm{S},17}$ terms are given in Eq. \eqref{x0inbschsupp}.

Putting Eq. \eqref{eq:x0inbschgeneral} into Eq. \eqref{sndetails}, one finally obtain the change of the angular coordinate in the power series of $\displaystyle \frac{1}{b}$ for general velocity
\be
I_\mathrm{S}(b,v)=\sum_{n=0}^{17}S^\prime_n\lb \frac{m}{b}\rb^n+ \mathcal{O}\lb \frac{m}{b}\rb ^{18} ,\label{angschinbfull}\ee
with
\begin{subequations}\label{angschinb}
\begin{align}
S^\prime_0=&\pi ,\\
S^\prime_1=&2\lb 1  +\frac{1}{v^2} \rb  ,\\
S^\prime_2=&\frac{3\pi}{4}\lb 1+\frac{4}{v^2}\rb,\\
S^\prime_3=&2\lb \frac{5}{3}+\frac{15}{v^2}+\frac{5}{v^4}-\frac{1}{3 v^6}\rb,\\
S^\prime_4=&\frac{105\pi}{64}\lb 1 +\frac{16}{v^2} +\frac{16}{v^4}\rb ,
\end{align}
\end{subequations}
and $S^\prime_5$ to $S^\prime_{17}$ are given in Eq. \eqref{angschinbsupp}.
Comparing the powers of $\frac{1}{b}$ and $\frac{1}{v}$ in each summand of \eqref{angschinbfull}, we can see that
\be
\lim_{b\to\infty,v\to 0}\frac{S_{n+2}}{S_n}\lb\frac{m}{b}\rb^2 \to \frac{1}{v^{2+\delta}}\lb\frac{m}{b}\rb^2, \ee
where $\delta=0,~1$ for odd and even $n$ respectively.
Therefore in order for the entire expansion to converge, the velocity shall not be indefinitely small. Rather, the range of convergence for $v$ is $\displaystyle v\in (\mathcal{O}\lb \frac{m}{b}\rb,1]$. Note that when $b$ is large, this range covers velocities of all interested rays or particles that are (potentially) timelike. These include supernova neutrinos and GWs in some generalized theories of GR.

After substituting $v=1$ we obtain
\be
I_\mathrm{S,\gamma}(b,v)=\sum_{n=0}^{17}S^\prime_{n,\gamma}\lb \frac{m}{b}\rb^n+ \mathcal{O}\lb \frac{m}{b}\rb ^{18} ,\label{angschlightinbgeneral}\ee
where
\begin{subequations}\label{angschinblight}
\begin{align}
S^\prime_{0,\gamma}=&\pi ,\\
S^\prime_{1,\gamma}=&4 ,\\
S^\prime_{2,\gamma}=&\frac{15\pi}{4},\\
S^\prime_{3,\gamma}=&\frac{128}{3},\\
S^\prime_{4,\gamma}=&\frac{3465\pi}{64},
\end{align}
\end{subequations}
and $S^\prime_{5,\gamma}$ to $S^\prime_{17,\gamma}$ are given in Eq. \eqref{angschinbsupp}.

Again, for the ultrarelativistic particle, the expansion in powers of $(1-v)^1$ is
\be
I_\mathrm{S}(b,v\to1)=I_\mathrm{S,\gamma}(b)+(1-v)\lsb 4\lb \frac{m}{b}\rb^1+6\pi\lb \frac{m}{b}\rb^2+96\lb \frac{m}{b}\rb^3 +\frac{315\pi}{2}\lb \frac{m}{b}\rb^4 +\mathcal{O}\lb\frac{m}{b}\rb^{5}\rsb
    +\mathcal{O}(1-v)^2. \label{angschvcinb}\ee
The higher order terms in $\displaystyle \lb \frac{m}{b}\rb$ are given in Eq. \eqref{angschvcinbfull}. Corrections in high orders of $(1-v)^n~(n\geq 2)$ can be easily obtained too.

\subsubsection{Convergence of the series and effect of velocity}

\begin{center}
\begin{figure}[htp!]
\includegraphics[width=0.45\textwidth]{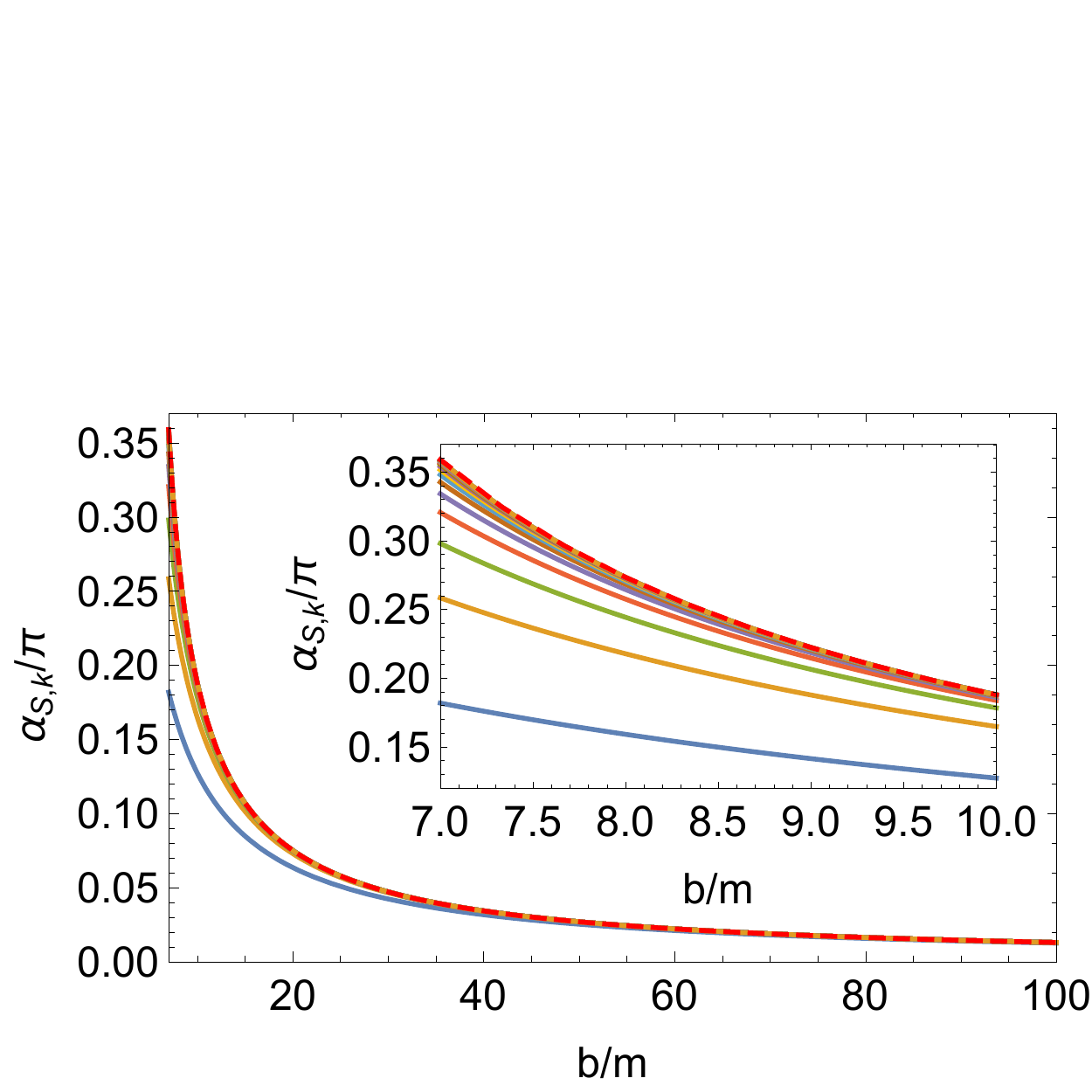}
\hspace{0.5cm}\includegraphics[width=0.45\textwidth]{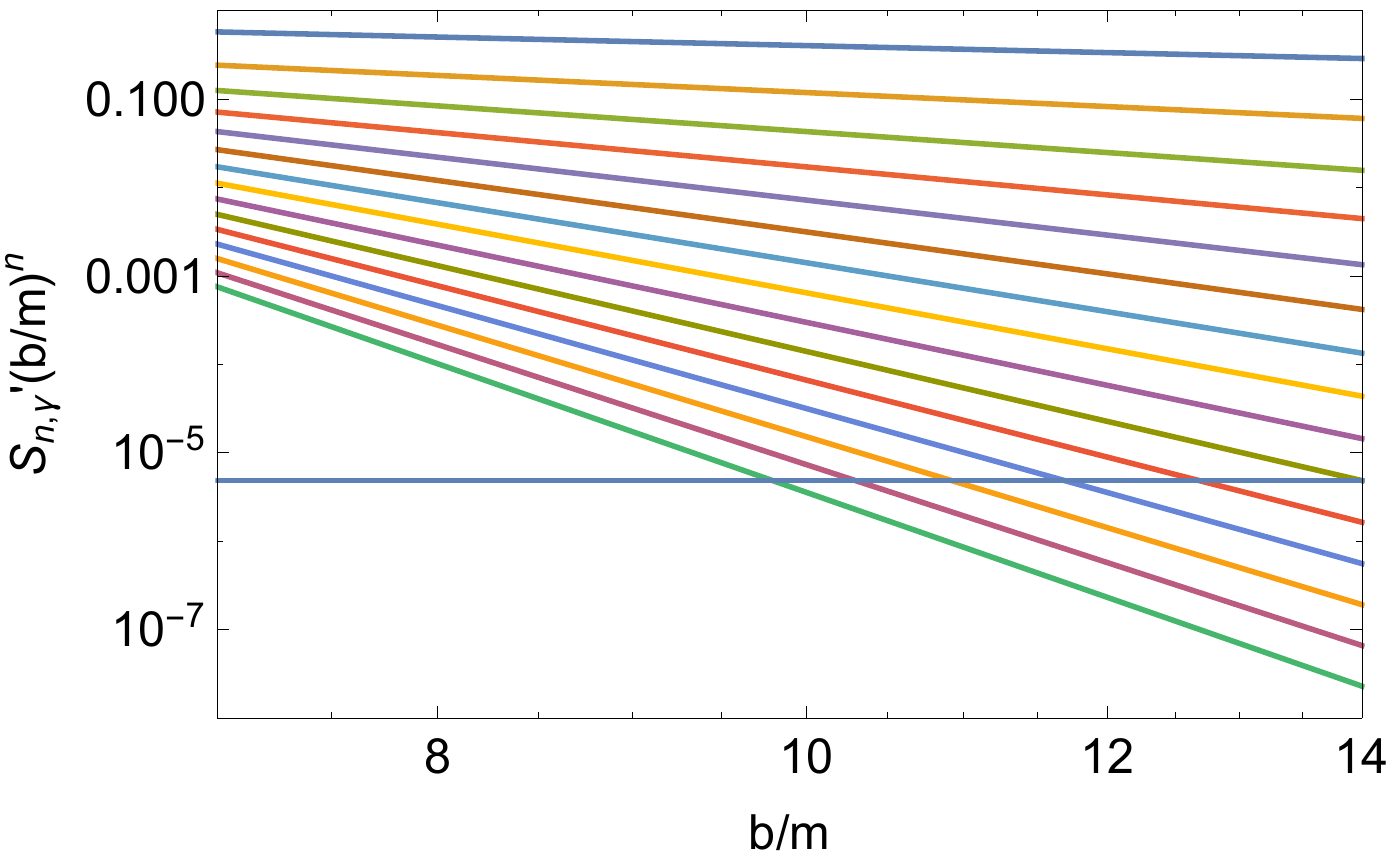}\\
(a)\hspace{8cm}(b)
\caption{The deflection angles in the Schwarzschild spacetime for null rays $(v=1)$. (a) Partial sums \eqref{eq:schpsumdef} for $b/m$ from 7 to 100 and from 7 to 10 in the inset (from bottom to top curve in each plot, $k$ increases from 1 to 17) and the exact deflection angle (dashed red curve). (b) Contribution from each order (from top to bottom curve $n$ increases from 1 to 17) for impact parameter $b/m$ from 7 to 14. In (a), we also illustrated the exact deflection angle obtained using numerical method and in (b) a horizontal line of 1 [as]. \label{fig:sch1}}
\end{figure}
\end{center}

Although the main purpose of this work is not to examine the effect of parameters in each metric, numerical study of the deflection angles such as Eqs. \eqref{angschinbfull}, \eqref{angrnfull} and \eqref{angkrfullinb} can not only reveal their validity but also the regions of converge for $b$ and moreover the effect of velocity. To examine the convergence of the deflection angle \eqref{angschinbfull}, we plot in Fig. \ref{fig:sch1}(a) the partial sums
\be
\alpha_{\mathrm{S},k}=\sum_{n=1}^k S_n^\prime \lb\frac{m}{b}\rb^n~(k=1,\cdots,17), \label{eq:schpsumdef}
\ee
for lightrays and the corresponding exact values obtained using numerical method to an accuracy of $10^{-15}$. We also plot in Fig. \ref{fig:sch1}(b) the contribution from each order $S_n^\prime \lb\frac{m}{b}\rb^n$ for lightrays.

It is seen from Fig. \ref{fig:sch1}(a) that for all fixed $b/m$, the partial sums converges to the exact deflection angle as the number of terms in the partial sum increases. The minimal convergent impact parameter can reach a sub $10m$ level. Indeed, as one can see from the inset of \ref{fig:sch1}(a) that when $b\lesssim 10m$, the deflection angle is already larger than $0.2\pi$. This value certainly exceeds the traditional weak deflection limits. Therefore this nice applicability of the perturbative deflection angle \eqref{angschinbfull} shows clearly the power of the perturbative result, especially when high orders in the expansion can be known.

It is also noticeable from Fig. \ref{fig:sch1}(a) that as $b/m$ decreases from large values, the low order contributions to Eq. \eqref{angschinbfull} becomes less dominate. As $b/m$ approaches the strong field limit which is $b_c/m=3\sqrt{3}$ for lightrays, the high order contributions become more and more important and eventually cause the divergence of the total deflection angle. In order to determine the accuracy of the deflection angle calculated to certain order, we plotted the contributions from each order in Fig. \ref{fig:sch1}(c) and a benchmark line of 1 [as]. Clearly in this plot the slope of contribution $S_n^\prime \lb \frac{m}{b}\rb^n$ should be $-n$. What is important is that for any fixed $b$, as the order increases, the contribution from each order decreases by a factor smaller than 1. Therefore this decrease guarantees that the series converges at all $b$ considered. In particular, one sees that even for $b$ as small as $\sim 9.7m$, the deflection angle expanded to the 17th order is still accurate to the 1 [as] level, which is roughly the limit of GL observations of galaxies and galaxy clusters.

\begin{center}
\begin{figure}[htp!]
\includegraphics[width=0.45\textwidth]{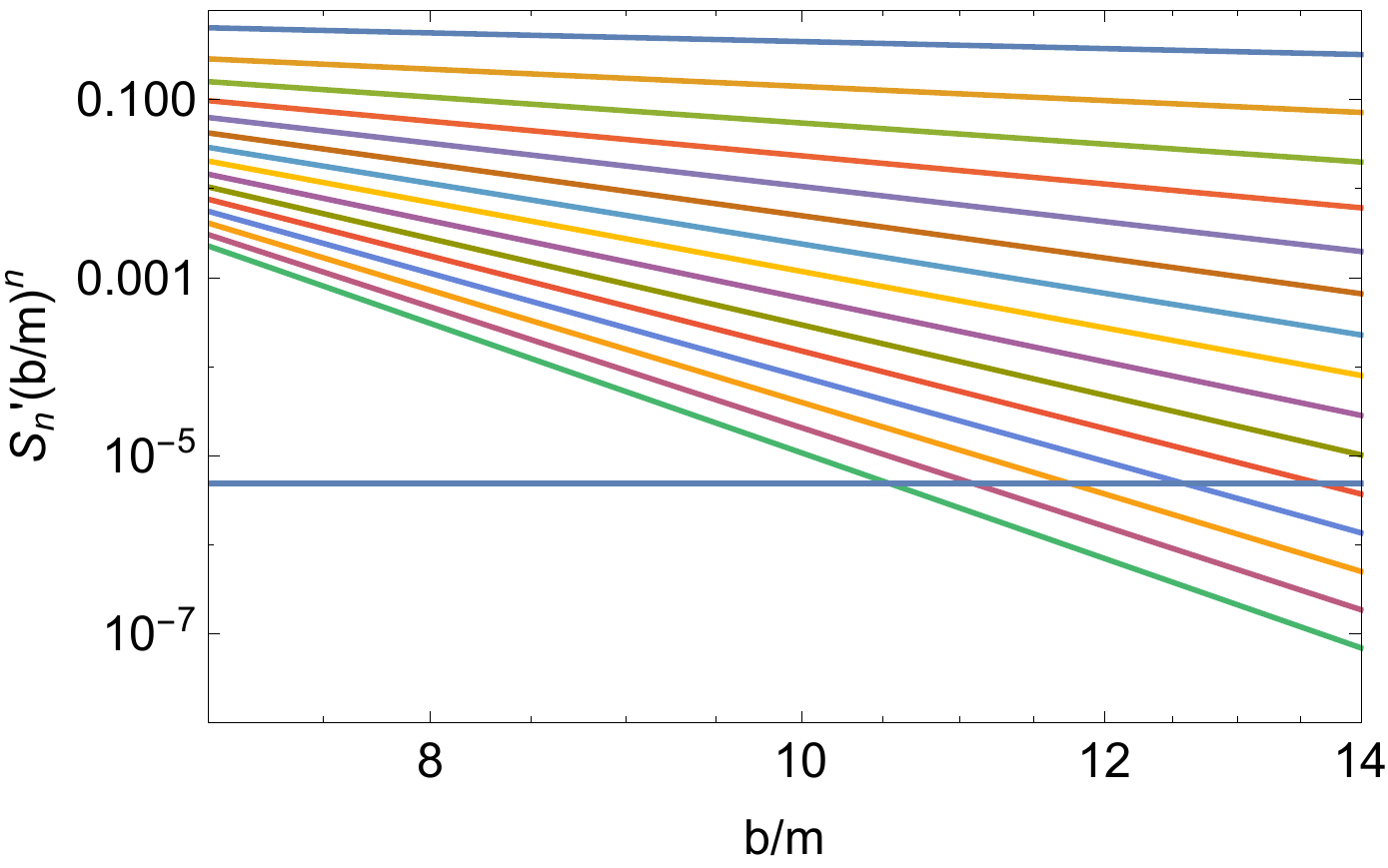}
\hspace{0.5cm}\includegraphics[width=0.45\textwidth]{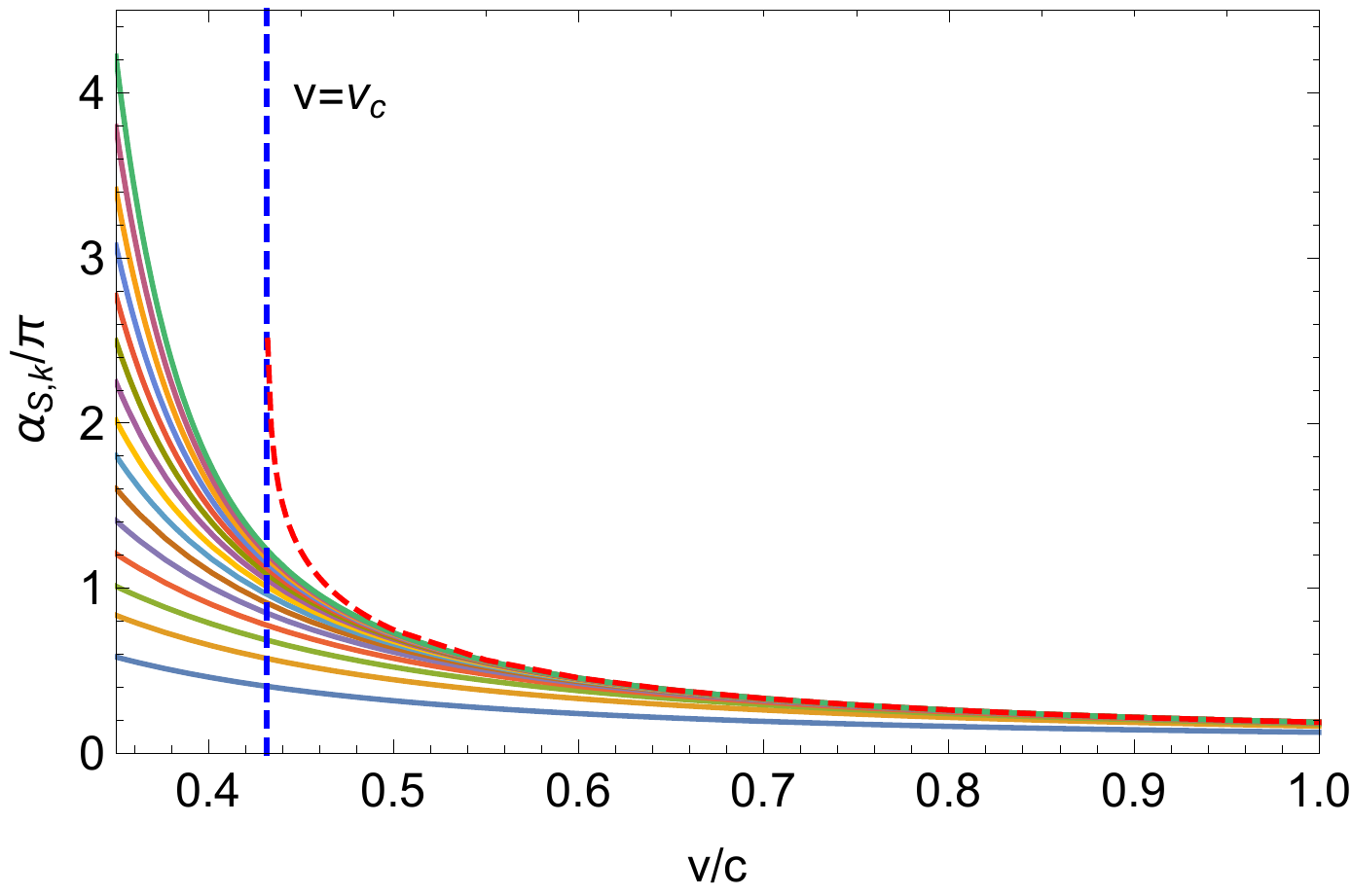}\\
(a)\hspace{8cm}(b)
\caption{The deflection angles in the Schwarzschild spacetime for timelike rays. (a) Contribution from each order of the deflection angle in Eq. \eqref{angschinbfull} (from top to bottom line $n$ increases from 1 to 17) for fixed $v=0.9c$ and $b/m$ from 7 to 14; (b)  Partial sums \eqref{eq:schpsumdef} (from bottom to top curve, $k$ increases from 1 to 17) and exact value of the deflection angle (dashed red curve) for fixed $b=10m$ and $v$ from $0.35c$ to 1$c$. The critical velocity for this value of $b$ is shown by the dashed blue line. \label{fig:sch2}}
\end{figure}
\end{center}

To study the effect of signal velocity on the deflection angle, we plot in Fig. \ref{fig:sch2}(a) the contribution from each order to the deflection angle for $v=0.9c$, and in Fig. \ref{fig:sch2}(b) for fixed $b=10m$ and increasing $v$ from $0.35c$ to $c$.
Comparing Fig. \ref{fig:sch2}(a) to Fig. \ref{fig:sch1}(b), one can see that as $v$ deviates from $c$, the contribution to the deflection angle from each order also increases. This leads to an increase of the impact parameter (from $9.4m$ to $10.5m$) at which the accuracy of the expansion can reach the 1 [as] level. For a fixed $b=10m$, it is seen in Fig. \ref{fig:sch2}(b) that as $v$ decreases, the deflection angle increases as dictated by the series \eqref{angschinbfull}. However, it is known that for massive particle with velocity $v$ at infinity, there exists a critical impact parameter $b_c$ \cite{Jia:2015zon}
\be b_c(v) =\frac{\left[8v^4+20v^2-1+(8v^2+1)^{3/2}\right]^{1/2}m}{\sqrt{2}v^2} \label{critip}\ee
below which the particle will spiral into the black hole. Using this relation, it can be worked out then when $v=v_c=0.43c$, $b_c=10m$. Particles with velocities at or below this value will all have larger $b_c$'s and therefore for these particles if their impact parameter $b=10m$, then they will experience a divergent deflection angle and eventually be captured by the BH. This is also seen from the sharp deviation of the perturbative deflection angle \eqref{angschinbfull} from the exact value starting from $v/c=0.5$. Below this velocity, the deflection angle, Eq. \eqref{angschinbfull}, becomes invalid.

\subsection{Weak deflection angle in RN spacetime}

The RN metric is given by
\be A(x)=1-\frac{2m}{x}+\frac{q^2}{x^2},~B(x)=\frac{1}{A(x)},~C(x)=x^2.
\ee
where $q$ is the total charge of the spacetime.
Substituting this into Eq. \eqref{expabcfull}, simplifying and integrating over $u$ we get the change of the angular coordinate for RN metric to the fifteenth order of $\displaystyle \frac{1}{x_0}$  for general velocity $v$
\be I_\mathrm{RN}(x_0,v)=\sum_{n=0}^{15}R_n\lb \frac{m}{x_0}\rb ^n+ \mathcal{O}\lb \frac{m}{x_0}\rb^{16} ,\label{angrnfull}\ee
where
\begin{subequations}\label{rndetails}
\begin{align}
R_0=
    &
    S_{0},\\
R_1=
    &
    S_{1},\\
R_2=
    &
    S_{2}-\left(\frac{\pi }{4}+\frac{\pi }{2v^2}\right)\hat{q}^2, \label{rndetailssec}\\
R_3=
    &S_{3}-\lsb 2-\lb\frac{\pi }{2} -11\rb\frac{1}{v^2}
     -\lb\pi-1\rb\frac{1}{v^4}\rsb \hat{q}^2,\\
R_4=
    &
    S_{4}-\lsb\frac{45 \pi }{32}+\lb \frac{121 \pi}{8}-10\rb\frac{1}{v^2}
    -\lb 37-\frac{29 \pi }{4}\rb\frac{1}{v^4}+\lb 2\pi- 3\rb\frac{1}{v^6}\rsb \hat{q}^2
    +\left(\frac{9 \pi }{64}+\frac{7 \pi }{8v^2}-\frac{\pi }{8v^4}\right)\hat{q}^4,
\end{align}
\end{subequations}
where $\hat{q}\equiv q/m$, $S_1$ to $S_4$ are the corresponding terms in Schwarzschild spacetime given in Eq. \eqref{sndetails} and $R_5$ to $R_{15}$ are given in Eq. \eqref{rndetails7to15}.
The expansion coefficients \eqref{rndetails} show that the effect of the charge in the RN metric starts to appear in the deflection angle from the second order in $R_2$. Rewriting Eq. \eqref{rndetailssec}, we have
\be R_2=
\frac{3\pi}{4} +(3\pi-2)\frac{1}{v^2} -2\frac{1}{v^4}   -\frac{\pi}{4}\lb 1+\frac{2}{v^2}\rb \hat{q}^2 . \label{eq:rn2order} \ee
This implies that the total deflection angle decreases monotonically as $|\hat{q}|$ increases. This agrees with the observation in Ref. \cite{Pang:2018jpm}.
Similar to the Schwarzschild case, comparison of $R_{n+1}$ and $R_n$ in the large $x_0$ and small $v$ limit also suggests that in order for the expansion \eqref{rndetails} to converge, the velocity of the particles should be bounded within $v\in (\sqrt{m/x_0},1]$.

For lightlike particles, setting $v=1$ we obtain its corresponding $I_{\mathrm{RN},\gamma}(x_0)$
\be
I_\mathrm{RN,\gamma}(x_0)=
\sum_{n=0}^{15} R_{n,\gamma} \lb\frac{m}{x_0}\rb^n +\mathcal{O} \lb\frac{m}{x_0}\rb^{16}
\ee
where the coefficients are given by
\begin{subequations}\label{rngammaexp}
\begin{align}
R_{0,\gamma}=&S_{0,\gamma},\\
R_{1,\gamma}=&S_{1,\gamma},\\
R_{2,\gamma}=&S_{2,\gamma}-\frac{3 \pi  \hat{q}^2}{4},\\
R_{3,\gamma}=&S_{3,\gamma}-\left(14-\frac{3 \pi }{2}\right) \hat{q}^2,\\
R_{4,\gamma}=&S_{4,\gamma}-\left(\frac{825 \pi }{32}-50\right) \hat{q}^2+\frac{57 \pi  \hat{q}^4}{64},
\end{align} \label{eq:rnlightdf}
\end{subequations}
where $S_{n,\gamma}$ are given in Eq. \eqref{sndetailsv1} and $R_{5,\gamma}$ to $R_{15,\gamma}$ are given in Eq. \eqref{rngammaexp7to15}.
The result \eqref{eq:rnlightdf} agrees with the deflection angle in Ref. \cite{Keeton:2005jd}.
For ultrarelativistic particles, the change of the angular coordinates can be expanded around $I_{\mathrm{RN},\gamma}(x_0)$. To the order $(1-v)^1$, this is
\bea
I_\mathrm{RN}(x_0,v\to1)&=&I_{\mathrm{RN},\gamma}(x_0)+(1-v) \lcb  4\lb\frac{m}{x_0}\rb+\lsb (6\pi-12)-\pi\hat{q}^2\rsb \lb \frac{m}{x_0}\rb^2 +\lsb 3(34-9\pi)-(26-5\pi) \hat{q}^2\rsb \lb \frac{m}{x_0}\rb^3\right.\nn\\
&&\left. +\lsb \frac{375 \pi }{2}-542-\lb \frac{285 \pi}{4}- 186
   \rb \hat{q}^2 +\frac{5 \pi}{4}\hat{q}^4\rsb \lb \frac{m}{x_0}\rb^4
+\mathcal{O}\lb \frac{m}{x_0}\rb^5\rcb +\mathcal{O}(1-v)^2,\label{angrnvcinx0}\eea
where for simplicity in the curl bracket the orders higher than $\displaystyle \lb \frac{1}{x_0}\rb^4$  were not shown.

In order to transform the change of angular coordinates \eqref{angrnfull} into a power series of $\displaystyle \frac{1}{b}$, we shall use the relation \eqref{br0rel}
which in this case becomes for $\kappa=1$
\be
\frac{1}{b}=\frac{1}{x_0}\frac{\sqrt{E^2-1}}{\sqrt{E^2-\lb 1-\frac{2m}{x_0}+\frac{q^2}{x_0^2}\rb}}\sqrt{1-\frac{2m}{x_0}+\frac{q^2}{x_0^2}}.\ee
Using this, $\displaystyle \frac{m}{x_0}$ can be expanded as power series of $\displaystyle \frac{m}{b}$
\be
\frac{m}{x_0}=\sum_{n=1}^{15}C_{\mathrm{RN},n}\lb \frac{m}{b}\rb^n +\mathcal{O}\lb \frac{m}{b}\rb^{16}  \label{x0inbrn}
\ee
where the coefficients are
\begin{subequations}\label{x0inbrncoeff}
\begin{align}
C_{\mathrm{RN},1}=
    &
    C_{\mathrm{S},1},\\
C_{\mathrm{RN},2}=
    &
    C_{\mathrm{S},2},\\
C_{\mathrm{RN},3}=
    &
    C_{\mathrm{S},3} -\frac{\hat{q}^2}{2v^2},\\
C_{\mathrm{RN},4}=
    &
    C_{\mathrm{S},4} -\lb\frac{2}{v^2}+\frac{1}{v^4}\rb \hat{q}^2,
\end{align}
\end{subequations}
with $C_{\mathrm{S},1}$ to $C_{\mathrm{S},4}$ present in Eq. \eqref{x0inbsch} and $C_{\mathrm{RN},5}$ to $C_{\mathrm{RN},15}$ given in Eq. \eqref{x0inbrn7to15}.
Substituting this into expansion \eqref{angrnfull}, the change of the angular coordinate becomes
\be I_\mathrm{RN}(b,v)=\sum_{n=0}^{15}R^\prime_n\lb \frac{m}{b}\rb ^n+ \mathcal{O}\lb \frac{m}{b}\rb ^{16} ,\label{angrninbfull}\ee
where
\begin{subequations}\label{angrninb}
\begin{align}
R^\prime_0=
    &
    S^\prime_0,\\
R^\prime_1=
    &
    S^\prime_1,\\
R^\prime_2=
    &
    S^\prime_2-\frac{\pi}{4}\left(1+\frac{2}{v^2}\right) \hat{q}^2,\\
R^\prime_3=
    &
    S^\prime_3-2\lb 1+\frac{6}{v^2}+\frac{1}{v^4}\rb \hat{q}^2,\\
R^\prime_4=
    &
    S^\prime_4-\frac{45\pi}{32}\left(1+\frac{12}{v^2}+\frac{8}{v^4}\right) \hat{q}^2+\frac{3\pi}{64}\left(3+\frac{24}{v^2}+\frac{8}{v^4}\right) \hat{q}^4
    ,
\end{align}
\end{subequations}
and the higher order terms are given in Eq. \eqref{angrninb7to15}.
For lightlike rays, this reduces to
\bea
I_{\mathrm{RN},\gamma}(b)&=&I_{\mathrm{S},\gamma}(b)
    -\frac{3 \pi}{4}\lb\frac{m}{b}\rb^{2}
    -16 \hat{q}^2\lb\frac{m}{b}\rb^{3}
    +\frac{105}{64} \pi \left(-18\hat{q}^2+\hat{q}^4\right)\lb\frac{m}{b}\rb^{4}
+ \mathcal{O}\lb \frac{m}{b}\rb^{5}, \label{angrnlightinb}\eea
where the first three orders were known in Ref. \cite{Keeton:2005jd}.
Again, for the ultrarelativistic particle, the change of the angular coordinate expanded to the $(1-v)^1$ order is
\bea
I_\mathrm{RN}(b,v\to1)&=&I_{\mathrm{RN},\gamma}(b)+(1-v)\lsb 4\lb \frac{m}{b}\rb+ \pi\lb 6-\hat{q}^2\rb  \lb \frac{m}{b}\rb^2 +32\lb 3-\hat{q}^2\rb \lb \frac{m}{b}\rb^3\right.\nn\\
&&\left.+\frac{15\pi}{4}\lb 42 -21\hat{q}^2 +\hat{q}^4\rb \lb \frac{m}{b}\rb ^4 +\mathcal{O}\lb \frac{m}{b}\rb ^5\rsb +\mathcal{O}(1-v)^2 \label{angrnvcinb}\eea
and terms of order $\lb \frac{m}{b}\rb ^5$ and $(1-v)^2$ can be similarly computed but not shown for simplicity reason here.

\subsubsection{Effect of charge on deflection angle}

\begin{center}
\begin{figure}[htp!]
\includegraphics[width=0.3\textwidth]{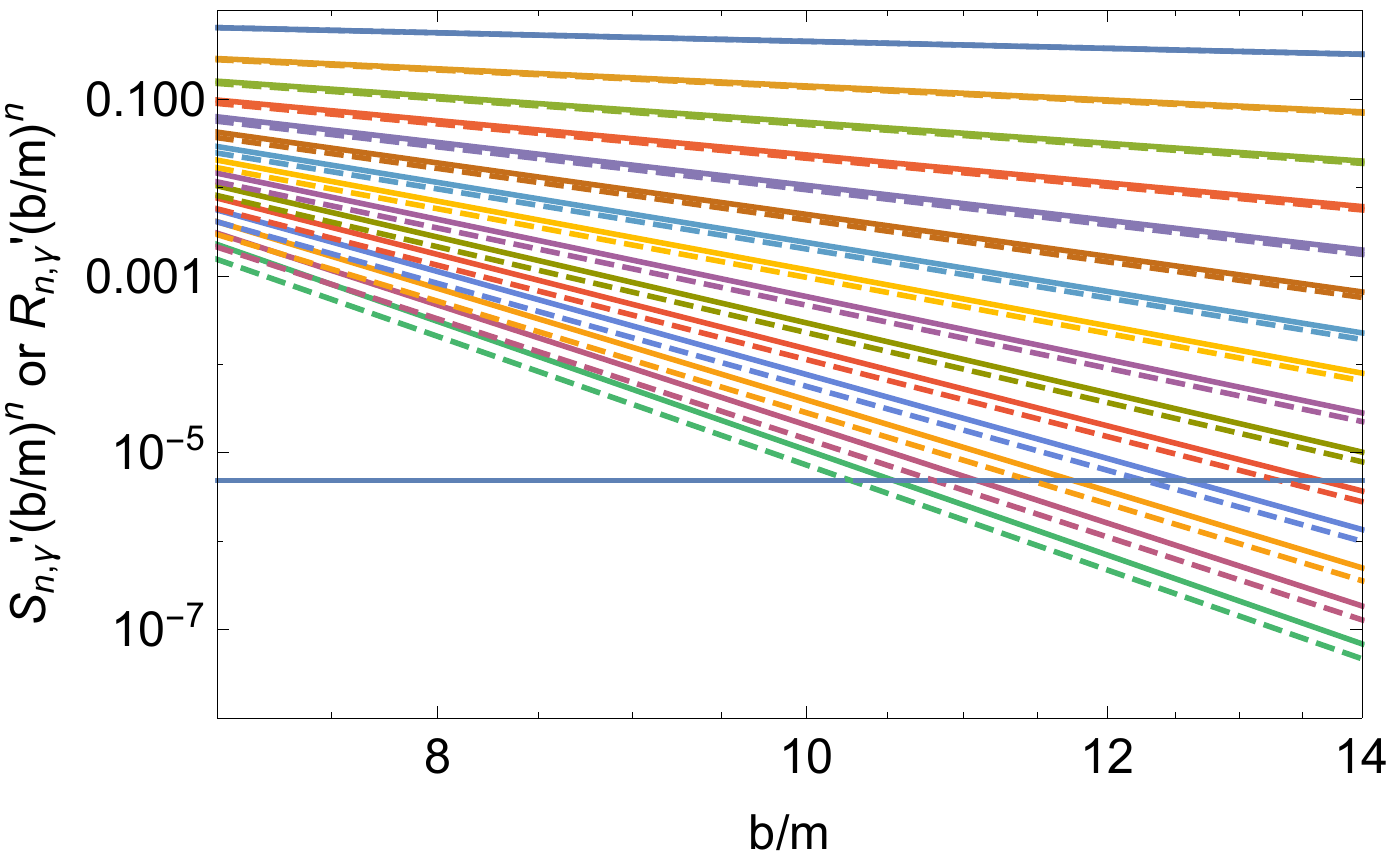}
\hspace{0.5cm}
\includegraphics[width=0.3\textwidth]{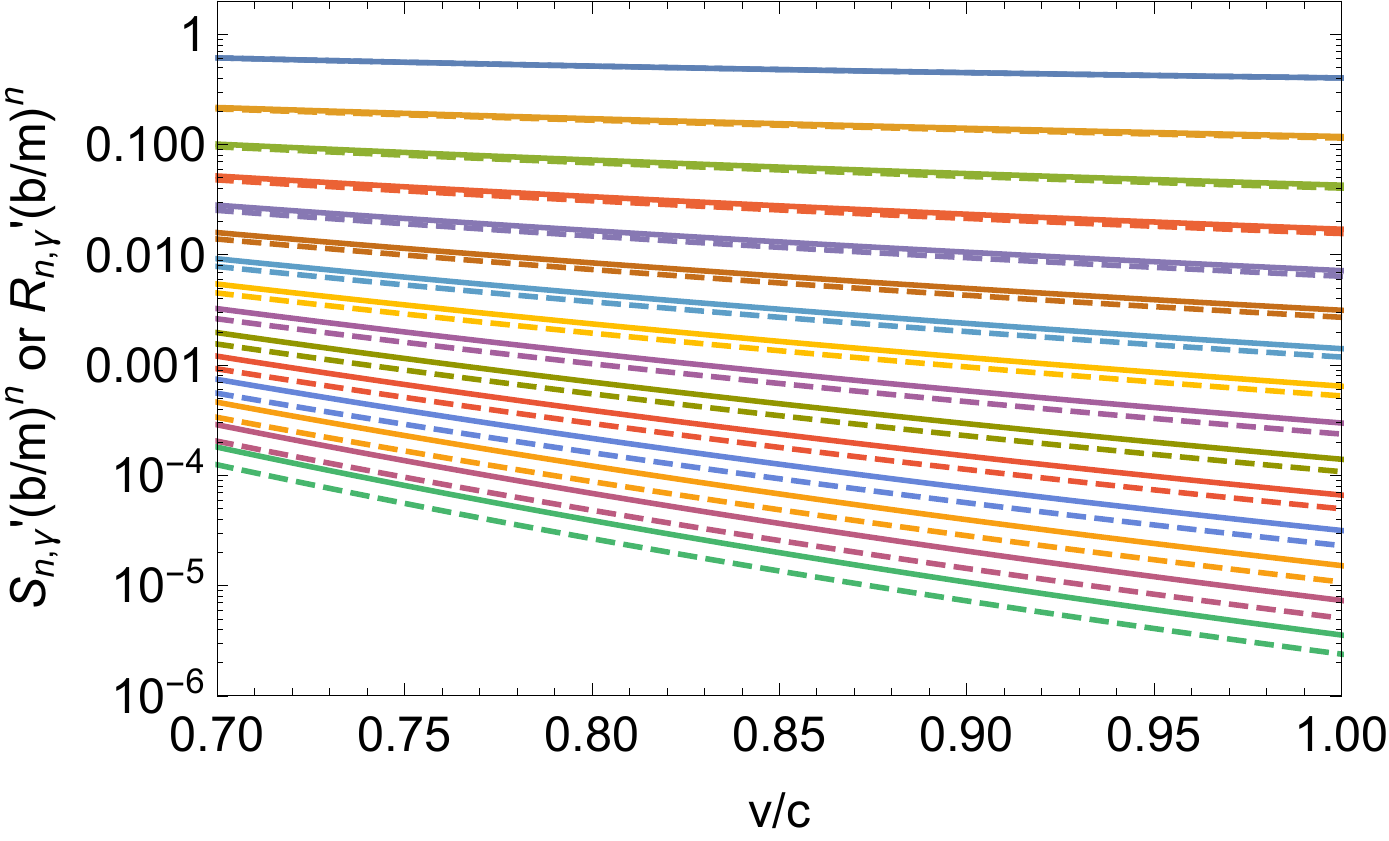}
\hspace{0.5cm}
\includegraphics[width=0.3\textwidth]{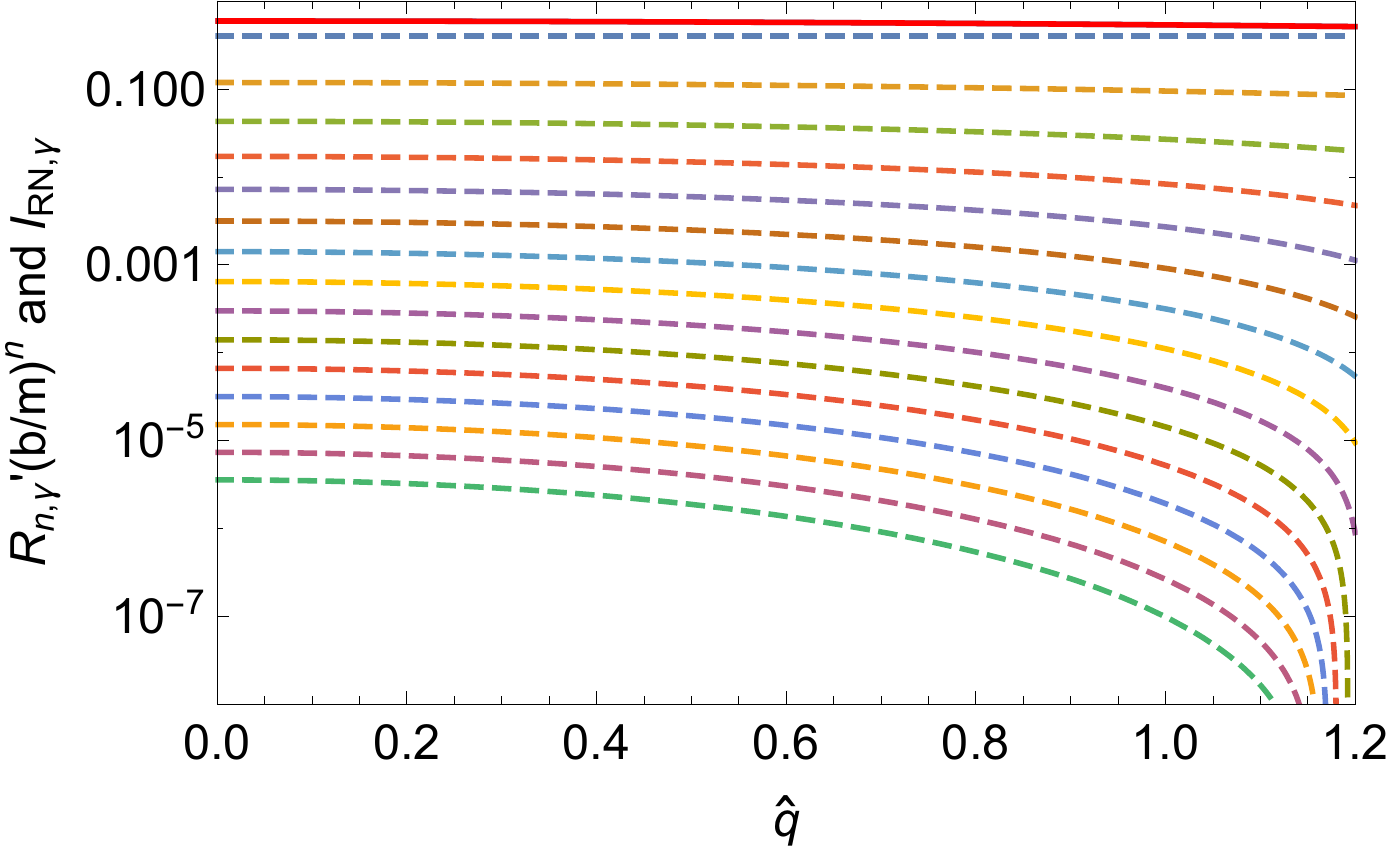}\\
(a)\hspace{5.5cm}(b)\hspace{5.5cm}(c)
\caption{The deflection angles in the RN spacetime. (a) Contribution from each order of the deflection angle in Eq. \eqref{angrninbfull} for fixed $v=0.9c$ and $b/m$ from 7 to 14 (from top to bottom $n$ increases from 1 to 15); (b) Contribution from each order of the deflection angle (from bottom to top $n$ decreases from 15 to 1), the total deflection angle (solid blue line), Eq. \eqref{angrninbfull}, and the exact deflection angle (solid red line) for fixed $b=10m,~v/c=1$ and $\hat{q}$ from 0 to 1.2. The solid blue and red lines overlap. \label{fig:rn1}}
\end{figure}
\end{center}

In Fig. \ref{fig:rn1}(a) we plot contribution of each order $R_n^\prime \lb\frac{b}{m}\rb^n$ to the deflection angle \eqref{angrninbfull} for $\hat{q}=0$ and $\hat{q}=0.4$ respectively and $v=0.9c$, and in Fig. \ref{fig:rn1}(b) the same quantities but for $\hat{q}=0$ and $\hat{q}=0.4$ and $b/m=10$. While in Fig. \ref{fig:rn1}(c) for lightrays with $b=10m$, the contribution from each order in Eq. \eqref{angrninbfull}, the total deflection angle \eqref{angrninbfull} to the 15th order, and the exact deflection angle obtained numerically, are shown.

It is seen from the comparison of Fig. \ref{fig:rn1}(a), (b) with Fig. \ref{fig:sch2}, the effects of $b/m$ and $v/c$ in RN spacetime are similar to the case of Schwarzschild metric. Moreover, in each plot of Fig. \ref{fig:rn1}, a comparison between the $\hat{q}=0$ and $\hat{q}=0.4$  cases shows that as explained from Eq. \eqref{eq:rn2order}, the contribution of each order $R_n^\prime \lb\frac{b}{m}\rb^n$ and therefore the total deflection will all decrease as $q$ increases. Fig. \ref{fig:rn1}(c) further illustrates that this effect indeed is persistent to the region that $\hat{q}>1$, i.e., from an RN BH spacetime to naked singularity case. A comparison of the deflection angle \eqref{angrninbfull} to the exact value also reveals that this expansion approximate the exact deflection angle perfectly in both the RN BH and naked singularity cases.

\subsection{Weak deflection angle in non-explicitly known SSS spacetimes \label{secunknownmetric}}

As explained in Sec. \ref{secdaexp}, the change of the angular coordinate in the large $x_0$ limit shall be determined by the spacetime in the large $x_0$ region. In literatures, there are quite some interesting SSS spacetimes whose metric functions are not known explicitly and/or analytically due to the complexity of the Einstein equations. However, very often their asymptotic behavior can be known from series analysis. We take the colored BH in the SU(2) Yang-Mills-Einstein theory as an example \cite{Bizon:1990sr}. We will show that in this spacetime the change of the angular coordinate can be solved to the order $\displaystyle \lb \frac{1}{x_0}\rb^1$  without having to know the explicit form of the metric functions.

The metric of the colored BH are given at large $x$ by \cite{Bizon:1990sr}
\bea
A(x)&=&\mbox{e}^{-2\delta (x)}\lb 1-\frac{2m(x)}{x}\rb \xrightarrow{x\to \infty} \mbox{e}^{-2\delta_k}\lb 1-\frac{2M_k}{x}\rb,\\
B(x)&=&\lb 1-\frac{2m(x)}{x}\rb^{-1}
\xrightarrow{x\to \infty}
\lb 1-\frac{2 M_k}{x}\rb^{-1},\\
C(x)&=&x^2,
\eea
where $-1<\delta_k<0,~0<M_k<1$ are constants and $k=1,~2,~\cdots$ are indices characterizing colored BHs of different masses.

It is clear that asymptotically this metric is almost identical to the Schwarzschild spacetime at the first order except the factor $\mbox{e}^{-2\delta_k}$ in front of $A(x)$. Substituting the metric into Eq. \eqref{iton}, one finds the change of angular coordinate for general velocity to the order $\displaystyle \lb \frac{1}{x_0}\rb^1$ as
\be I_\mathrm{CB}(x_0)=\pi+ \frac{2\lsb  2-\mbox{e}^{2\delta_k(\delta_k+1)}(1-v^2)^{\delta_k+1}\rsb } {1-\mbox{e}^{2\delta_k(\delta_k+1)}(1-v^2)^{\delta_k+1}} \frac{M_k}{x_0}+\mathcal{O}\lb \frac{M_k}{x_0}\rb^2 . \label{cbhgd}\ee
It is seen that for massive particles, $I_{\mathrm{CB}}(x_0)$ is affected by not only the asymptotic mass $M_k$ but also $\delta_k$. For lightlike rays, the change of the angular coordinate becomes
\be I_\mathrm{CB,\gamma}(x_0)=\pi+ \frac{4M_k}{x_0}+\mathcal{O}\lb \frac{M_k}{x_0}\rb^2 . \ee
To the order $\displaystyle \frac{1}{x_0}$, this equals the changes of the angular coordinate \eqref{angschlight} in the Schwarzschild spacetime and \eqref{angrninbfull} in the RN spacetime.

The expansion of $\displaystyle \frac{M_k}{x_0}$ in term of $\displaystyle \frac{M_k}{b}$ in this metric can be found using the same procedure as in Eq. \eqref{lthm} to be
\be
\frac{M_k}{x_0}=\frac{v}{\sqrt{1-\mbox{e}^{2\delta_k}(1-v^2)}} \frac{M_k}{b} +\mathcal{O}\lb \frac{M_k}{b}\rb^2.\ee
Substituting this into Eq. \eqref{cbhgd}, one can obtain the change of the angular coordinate to the order of $\displaystyle \frac{1}{b}$. Moreover, it is also easy to see that for lightlike rays, this becomes
\be
I_{\mathrm{CB},\gamma}(b)=\pi +\frac{4M_k}{b} +\mathcal{O}\lb \frac{1}{b}\rb^2.\ee

\section{Weak deflection angle in Equatorial plane of SAS spactimes\label{secsas}}

To find the deflection angle in the weak field limit in a non-SSS spacetimes described by metric $g_{\mu\nu}$, in principle one could attempt to derive through the geodesic equations the differential equation that the angular coordinate satisfies
\be \frac{\dd\phi(x)}{\dd x}=f_\mathrm{ns}(x,~E,~p)\label{dphidxdef}\ee
where $f_\mathrm{ns}$ denotes some function derivable from the metric functions $g_{\mu\nu}$. Here the coordinate $x$ should resemble the meaning of distance when it is large, $E$ denotes energy per unit mass of the ray at infinity  and $p$ collectively stands for all other parameters that might appear in the metric, such as spacetime mass $m$, angular momentum per unit mass $a$ etc. Integrating Eq. \eqref{dphidxdef} then yields the change of the angular coordinate for a ray coming and going to spatial infinity
\be I(x_0)=2\int_{x_0}^\infty \frac{\dd\phi(x)}{\dd x} \dd x=2\int_{x_0}^\infty f_\mathrm{ns}(x,~E,~p) \dd x .\label{dagdef}\ee

In order for the above process to make physical sense and the integral to be eventually carried out, usually a few requirements should be satisfied. The first is that there should exist a spatial infinity ($x\to\infty$) at which the spacetime is approximated by a flat spacetime. Only in this circumstance, the interpretation of the integral \eqref{dagdef} subtracting $\pi$ as the deflection angle makes physical sense. The second is that the metric should still possess certain amount of symmetries. Indeed, the geodesic equations are originally second order differential equations while Eq. \eqref{dphidxdef} is already first order. For this to happen, the spacetime will have to have at least the necessary symmetries to allow the first integrals. Moreover, only when the spacetime is static, the function $f_\mathrm{ns}(x,~E,~p)$ in Eq. \eqref{dphidxdef} and so that the change of the angular coordinate  can be static. Finally, from the practical point of view, the function $f_\mathrm{ns}(x,~E,~p)$ should still be simple enough for the integration to be carried out. This can require one or more of the following conditions or techniques, such as simple enough metric functions, specially picked trajectories in the spacetime, or series expansions of the integrand over certain parameters.

The above considerations naturally singles out the SAS spacetimes as next suitable candidates for computation.
We start from the most general SAS metric given by the Weyl-Lewis-Papapetrou line element \cite{Levy:1968,Sloane:1978ne,Stephani:2003tm}
\be
   \dd s^2=-\expe^{2U}(\dd t-\omega \dd \phi)^2+\expe^{-2U}\lsb \expe^{2\gamma}(\dd \rho^2+\dd z^2)+\rho^2\dd \phi^2\rsb\label{metric0}
\ee
where $(t,~\rho,~\phi,~z)$ are the coordinates and $U,~\omega$ and $\gamma$ are arbitrary functions of $\rho$ and $z$ only.
For latter easier reference, we can rewrite this metric into the form
\be
\dd s^2= -A(\rho,z)\dd t^2+B(\rho,z)\dd t\dd \phi +C(\rho,z)\dd \phi^2+D(\rho,z)(\dd \rho^2+\dd z^2) \label{metric}
\ee
where we have set $ A(\rho,z)=\expe^{2U}$, $B(\rho,z)=2 \omega \expe^{2U}$, $C(\rho,z)=\rho^2\expe^{-2U}-\omega^2\expe^{2U}$ and $D(\rho,z)=\expe^{2(\gamma-U)}$. Note there is a relation between these functions
\be C=\frac{\rho^2-B^2/4}{A}. \label{cinab}\ee
We further assume that this spacetime allows a plenary motion for particles in a plane with fixed $z$, which will be called the equatorial plane henceforth. Indeed, without losing any generality, the coordinates can be shifted along the $z$ axis so that this plane becomes $z=0$. The motion in the equatorial plane then becomes effectively 2+1 dimensional whose metric after suppressing the $z$ coordinate in the metric functions becomes
\be
\dd s^2=-A(\rho)\dd t^2+B(\rho)\dd t\dd \phi +C(\rho)\dd \phi^2+D(\rho)\dd \rho^2. \label{metric2}
\ee
We then concentrate on the motion of particles in this plane.

The geodesic equations corresponding to metric \eqref{metric2} can be readily obtained as
\bea
\dot{t}&=&\frac{E}{A(\rho)}\lb 1-\frac{ B(\rho )^2}{4 \rho ^2 }\rb +\frac{L B(\rho )}{2 \rho ^2} , \label{tdotsol}\\
\dot{\phi}&=&\frac{1}{2\rho^2}\lsb 2L A(\rho )-E B(\rho )\rsb ,\label{phidotsol}\\
\dot{\rho}^2&=&\frac{4\rho^2\lsb E^2-\kappa A(\rho)\rsb-\lsb EB(\rho)-2LA(\rho)\rsb^2 }{4 \rho ^2 A  D}. \label{Vdef1}
\eea
where we have substituted \eqref{cinab}.
Here $E$ and $L$ are still energy and angular momentum per unit mass of the particle at infinite $\rho$. Note that because the SAS spacetimes might carry a nonzero angular momentum, the direction of the angular momentum $L$ is important in determining the shape and deflection angle of the geodesics.

Using Eqs. \eqref{phidotsol} and \eqref{Vdef1} to find $\dd \phi/\dd \rho$, it is easy to show that the deflection angle for rays both coming and going to infinite $\rho$ takes the form
\be
\alpha(\rho_0)=|I(\rho_0)|-\pi,\ee
where we add an absolute sign because $I(\rho_0)$, the change of the angular coordinate defined as
\be
I(\rho_0)=2\int_{\rho_0}^\infty
\frac{\lsb 2LA(\rho)-EB(\rho)\rsb \sqrt{A(\rho)D(\rho)}} {\rho\sqrt{4\rho^2\lsb E^2-\kappa A(\rho)\rsb-\lsb EB(\rho)+2LA(\rho)\rsb^2}}\dd \rho, \label{asaiint}
\ee
can be close to $\pi$ or $-\pi$ when $L$ is positive $(s=-1)$ or negative $(s=1)$ respectively. The angular momentum $L$ can be linked to $\rho_0$ by using $\dot{\rho}\big\vert_{\rho=\rho_0}=0$ in Eq. \eqref{Vdef1} to find
\be
L=\frac{-2s\rho_0\sqrt{E^2-\kappa A(\rho_0)}+EB(\rho_0)}{2A(\rho_0)}. \label{lsasdef}
\ee
Here since in the small spacetime spin limit the first term in the numerator dominates the second term, it is clear then the $s=+1$ and $-1$ correspond to the cases that the $L$ are negatively or positively oriented, respectively.

Eq. \eqref{asaiint} usually does not permit integration into closed form in terms of elementary functions. Similar to the procedure in Sec. \ref{secdaexp} which leads $I(x_0)$ to power series of $\displaystyle \frac{1}{x_0}$ and then further to series of $\displaystyle \frac{1}{b}$, one can also expand $I(\rho_0)$ in Eq. \eqref{asaiint} in the weak field, i.e., the large $\rho_0$ limit, and integrate to find the change of the angular coordinate in power series of $\displaystyle \frac{1}{\rho_0}$ and eventually in powers of $\displaystyle \frac{1}{b}$. We will not carry out these formal steps as in Sec. \ref{secdaexp} here because they are exactly the same way and too tedious to show here. Rather we will only illustrate these steps in subsection \eqref{subseckerr} for equatorial motion in Kerr and KN metrics respectively.

\subsection{Weak deflection angle in equatorial plane of the Kerr and KN spacetimes\label{subseckerr}}

With the above consideration, the next spacetimes we will consider is naturally the Kerr and KN spacetimes. We will show that the formalism for expanding the integrand in series of $\displaystyle \frac{1}{x_0}$  can also be used here to find the change of the angular coordinate of both massive and massless particles to high orders. We will only illustrate the procedure and results in the KN spacetime because setting its charge to zero will yield the corresponding results in the Kerr metric. To our knowledge, the deflection angle in Kerr spacetime has been known to fourth order in $\displaystyle \frac{1}{x_0}$ and $\displaystyle \frac{1}{b}$ for lightlike rays  \cite{Iyer:2009hq,Aazami:2011tw}.

The KN spacetime has a metric
\bea
\dd s^2 &=&-\frac{\Delta-a^2 \sin^2\theta}{\Sigma }\dd t^2
+\frac{\lb a^2+x^2\rb^2-a^2 \Delta \sin^2\theta}{\Sigma} \sin^2\theta\dd \phi^2-\frac{2 a\sin^2 \theta\lb a^2-\Delta +x^2\rb}{\Sigma } \dd t \dd \phi +\frac{\Sigma }{\Delta }\dd x^2
+\Sigma \dd \theta^2
\label{metrickerr}
\eea
where functions $\Sigma$ and $\Delta$ are
\be \Sigma(x,\theta)=x^2+a^2\cos^2 \theta ,~\Delta(x,\theta)=x^2-2m x+a^2+q^2,\ee
and $m$, $a$ and $q$ are respectively the mass, angular momentum per unit  mass and charge of the KN spacetime.

In the equatorial plane, $\theta=\frac{\pi}{2}$ and the metric \eqref{metrickerr} reduces to
\be
\dd s^2 =-A_{\mathrm{KN}}(x)\dd t^2+B_{\mathrm{KN}}(x) \dd t \dd \phi
+C_{\mathrm{KN}}(x)\dd \phi^2+D_{\mathrm{KN}}(x)\dd x^2,
\label{metrickerr2p1}
\ee
where
\bea
&&A_{\mathrm{KN}}(x)=1-\frac{2m}{x}+\frac{q^2}{x^2},~
B_{\mathrm{KN}}(x)=\frac{2aq^2}{x^2}- \frac{4am}{x},~\nonumber\\
&&C_{\mathrm{KN}}(x)=x^2+a^2+\frac{2a^2m}{x}-\frac{a^2q^2}{x^2},~
D_{\mathrm{KN}}(x)=\frac{x^2}{x^2-2mx+a^2+q^2}.
\label{abcdprime}
\eea
The coordinate $\rho$ in the standard metric \eqref{metric2} is related to the coordinate $x$ in metric \eqref{metrickerr2p1}
by \cite{Stephani:2003tm, alberto:1989}
\be
\rho(x)=\sqrt{x^2-2mx+a^2+q^2}\sin\theta=\sqrt{x^2-2mx+a^2+q^2}, \label{rhoxrel}
\ee
where $\theta=\frac{\pi}{2}$ was used. The metric functions in \eqref{metrickerr2p1} and \eqref{metric2} are then related by
\be
A(\rho(x))=A_{\mathrm{KN}}(x),~B(\rho(x))=B_{\mathrm{KN}}(x),
~C(\rho(x))=C_{\mathrm{KN}}(x),~D(\rho(x))=D_{\mathrm{KN}}(x)\lb \frac{\dd \rho(x)}{\dd x}\rb^{-2} \label{abcdprimetoabcd}
.\ee

Substituting Eqs. \eqref{abcdprime} to \eqref{abcdprimetoabcd} into  \eqref{asaiint}, the change of angular coordinate can be expressed as
\be
I_\mathrm{KN}(x_0)=2\int_{x_0}^\infty
\frac{\lsb 2LA_{\mathrm{KN}} (x)-EB_{\mathrm{KN}} (x)\rsb \sqrt{A_{\mathrm{KN}} (x)D_{\mathrm{KN}} (x)}} {\rho(x)\sqrt{4\rho(x)^2\lsb E^2-\kappa A_{\mathrm{KN}} (x)\rsb-\lsb EB_{\mathrm{KN}} (x)-2LA_{\mathrm{KN}} (x)\rsb^2}}
\dd x. \label{kerrintform}
\ee
We can get rid of the angular momentum $L$ in this equation by using Eq. \eqref{lsasdef} which in this case becomes
\be
L=\frac{-2s\rho(x_0)\sqrt{E^2-\kappa A_{\mathrm{KN}} (x_0)}+EB_{\mathrm{KN}} (x_0)}{2A_{\mathrm{KN}}(x_0)}.\label{krldef}
\ee
Again, $s=+1$ and $s=-1$ correspond to the cases that $L$ is negative and positive respectively.

The integral \eqref{kerrintform} can not be carried out to find a closed form and therefore an expansion of the integrand for large $x_0$ is needed. After integrating this expansion, we obtain a series approximation of the change of the angular coordinate. Here we skip these tedious middle steps and present the result directly to the order $\displaystyle \lb \frac{m}{x_0}\rb ^6$ as
\be
I_\mathrm{KN}(x_0,v)=\sum_{n=0}^6K_n\lb \frac{m}{x_0}\rb ^n+ \mathcal{O}\lb \frac{m}{x_0}\rb ^7 ,\label{angkrfull}\ee
where
\begin{subequations}\label{krdfres1}
\begin{align}
K_0=&R_0 ,\\
K_1=&R_1,\\
K_2=&R_2+\frac{4 \hat{a} s}{v},\label{krdfres1k2}\\
K_3=&R_3+
\hat{a}s\lcb \lsb(6 \pi-4)\frac{1}{v}+(4 \pi-12) \frac{1}{v^3}\rsb -\frac{\pi }{v}   \hat{q}^2\rcb +\hat{a}^2 \left(1+\frac{1}{v^2}\right),\\
K_4=&R_4+\hat{a}s\lcb \lsb(52-3 \pi)\frac{1 }{v}+(104-30 \pi)\frac{1 }{v^3}-(12 \pi-36) \frac{1}{v^5}\rsb- \lsb(22-\pi)\frac{1}{v}+(18-5 \pi)
\frac{1}{v^3}\rsb \hat{q}^2\rcb\nn\\
&+\hat{a}^2 \lcb \lsb \frac{33 \pi }{16}-2+\lb \frac{21 \pi }{2}-23\rb\frac{1}{v^2}-(5-\frac{3 \pi
   }{2})\frac{1}{v^4}\rsb-\left(\frac{5 \pi }{16}+\frac{\pi }{4 v^2}\right)
   \hat{q}^2\rcb
\\
K_5=&R_5+\hat{a}s\lcb \lsb\lb -36+\frac{93 \pi }{2}\rb\frac{1}{v}+(-436+159 \pi)\frac{1}{v^3}+(-460+141 \pi )\frac{1}{v^5}+(-92+30 \pi )\frac{1}{v^7}\rsb\right.\nonumber\\
&\left.+\lsb\lb 20-\frac{121 \pi
   }{4}\rb \frac{1}{v}+( 184-68 \pi )\frac{1}{v^3}+\lb 84-\frac{51 \pi }{2}\rb\frac{1}{v^5}\rsb \hat{q}^2 +\lb\frac{7 \pi }{4 v}+\frac{\pi }{2 v^3}\rb \hat{q}^4\rcb\nonumber\\
&+\hat{a}^2 \lcb \lsb 19-\frac{3 \pi }{2} +\lb 219-\frac{207 \pi }{4}\rb\frac{1}{v^2}+\lb \frac{453}{2}-72 \pi
   \rb \frac{1}{v^4}+\lb \frac{37}{2}-6 \pi \rb \frac{1}{v^6}\rsb\right.\nonumber\\
&\left.+\lsb -8+\frac{\pi }{2} +\lb -\frac{85}{2}+\frac{35 \pi }{4}\rb\frac{1}{v^2}+\lb -\frac{13}{2}+2 \pi
   \rb\frac{1}{v^4}\rsb \hat{q}^2\rcb \nonumber\\
&+\hat{a}^3s  \lsb(-16+6 \pi )\frac{1}{v}+(-16+6 \pi )\frac{1}{v^3}\rsb +\hat{a}^4
   \left(-\frac{1}{4}-\frac{1}{4 v^2}\right)
\\
K_6=&R_6+\hat{a}s\lcb\lsb\lb 348-\frac{201 \pi }{8}\rb\frac{1}{v}+\lb 2040-\frac{1113 \pi }{2}\rb\frac{1}{v^3} +\lb 3072-\frac{2013 \pi }{2}\rb \frac{1}{v^5}+(1576-495 \pi )\frac{1}{v^7}+\lb \frac{652}{3}-70 \pi
   \rb\frac{1}{v^9}\rsb\right.\nonumber\\
&+\lsb\lb-\frac{946}{3}+\frac{85 \pi }{4}\rb\frac{1}{v}+\lb-\frac{4034}{3}+\frac{1451 \pi }{4}\rb\frac{1}{v^3}+\lb-1330+\frac{873 \pi
   }{2}\rb\frac{1}{v^5}+\lb-\frac{886}{3}+\frac{185 \pi }{2}\rb\frac{1}{v^7}\rsb \hat{q}^2\nonumber\\
&\left.+\lsb\lb 52-\frac{25 \pi }{8}\rb\frac{1}{v}+\lb 112-\frac{109 \pi }{4}\rb\frac{1}{v^3}+\lb 12-\frac{9
   \pi }{2}\rb\frac{1}{v^5}\rsb \hat{q}^4\rcb \nonumber\\
&+\hat{a}^2 \lcb\lsb-18+\frac{1173 \pi }{64} +\lb -747+\frac{2523 \pi
   }{8}\rb\frac{1}{v^2}+\lb-2116+\frac{1365 \pi }{2}\rb\frac{1}{v^4}+\lb-\frac{2149}{2}+339 \pi \rb\frac{1}{v^6}+\lb-\frac{111}{2}+18 \pi \rb\frac{1}{v^8}\rsb\right.\nonumber\\
&\left.+\lsb 10-\frac{379 \pi
   }{32} +\lb 315-\frac{2205 \pi }{16}\rb\frac{1}{v^2}+\lb\frac{885}{2}-144 \pi \rb\frac{1}{v^4}+\lb\frac{75}{2}-\frac{23 \pi }{2}\rb\frac{1}{v^6}\rsb \hat{q}^2+\lb\frac{41
   \pi }{64}+\frac{29 \pi }{16 v^2}+\frac{\pi }{8 v^4}\rb\hat{q}^4\rcb\nonumber\\
&+\hat{a}^3s  \lcb\lsb (136-36 \pi)\frac{1}{v}+\lb \frac{1520}{3}-150 \pi
   \rb\frac{1}{v^3}+(168-54 \pi)\frac{1}{v^5}\rsb+\lsb(-24+6 \pi )\frac{1}{v}+(-24+6 \pi )\frac{1}{v^3}\rsb
   \hat{q}^2\rcb\nonumber\\
&+\hat{a}^4 \lsb  -3+\frac{19 \pi }{32} +\lb-\frac{51}{4}+\frac{33 \pi }{8}\rb\frac{1}{v^2}+\lb-\frac{7}{4}+\frac{3 \pi
   }{4}\rb\frac{1}{v^4} +\left(\frac{3 \pi }{32}+\frac{\pi }{16 v^2}\right)
   \hat{q}^2\rsb
\end{align}
\end{subequations}
where $R_0$ to $R_6$ are given in Eq. \eqref{rndetails}.

For lightlike rays, setting $v=1$ we easily get their change of the angular coordinate as
\begin{subequations}\label{krdfres2}
\begin{align}
K_{0,\gamma}=&R_{0,\gamma},\\
K_{1,\gamma}=&R_{1,\gamma},\\
K_{2,\gamma}=&R_{2,\gamma}+4 \hat{a}s,\\
K_{3,\gamma}=&R_{3,\gamma}+ \hat{a}s \left(10 \pi-16 -\pi  \hat{q}^2\right) +2 \hat{a}^2 , \\
K_{4,\gamma}=&R_{4,\gamma}+ \hat{a}s \lsb 192-45 \pi -(40-6 \pi ) \hat{q}^2\rsb+\hat{a}^2 \left(\frac{225 \pi }{16}-30-\frac{9 \pi  \hat{q}^2}{16}\right) ,\\
K_{5,\gamma}=&R_{5,\gamma}+ a s \lsb \frac{753 \pi }{2}-1024-\left(\frac{495 \pi
   }{4}-288\right) \hat{q}^2+\frac{9 \pi  \hat{q}^4}{4}\rsb +\hat{a}^2 \lsb 483-\frac{525 \pi }{4}-\left(57-\frac{45 \pi }{4}\right) \hat{q}^2\rsb \nn\\
   &+\hat{a}^3 s (12 \pi-32 )-\frac{\hat{a}^4}{2} ,\\
K_{6,\gamma}=&R_{6,\gamma}+ \hat{a}s \lsb \frac{21760}{3}-\frac{17225 \pi
   }{8}-\left(\frac{9856}{3}-913 \pi \right) \hat{q}^2+\left(176-\frac{279 \pi }{8}\right) \hat{q}^4\rsb \nn\\
   &+\hat{a}^2 \lsb -4011+\frac{87885 \pi
   }{64}-\left(\frac{9765 \pi }{32}-805\right) \hat{q}^2+\frac{165 \pi  \hat{q}^4}{64}\rsb \nn\\
   &+\hat{a}^3s \lsb \frac{2432}{3}-240 \pi -(48-12 \pi ) \hat{q}^2\rsb - \hat{a}^4 \left(\frac{35}{2}-\frac{175 \pi }{32}-\frac{5 \pi  \hat{q}^2}{32}\right)
\end{align}
\end{subequations}
And for ultra-relativistic particles, their change of the angular coordinate deviates from the lightlike rays' amount. At order $(1-v)^1$ this deviation is given by
\bea
&&I_{\mathrm{KN}}(x_0,v\to 1)=I_{\mathrm{KN},\gamma}(x_0)+(1-v)\times \nn\\
&&\lsb 4 \hat{a}s\lb\frac{m}{x_0}\rb^2+\lcb\lsb(-40+18 \pi )-\pi  \hat{q}^2\rsb \hat{a}s+2 \hat{a}^2\rcb \lb\frac{m}{x_0}\rb^3\right.\nn\\
&&+\lcb\lsb(544-153 \pi )+(-76+16 \pi ) \hat{q}^2\rsb
   \hat{a}s+\lsb(-66+27 \pi )-\frac{\pi  \hat{q}^2}{2}\rsb \hat{a}^2\rcb \lb\frac{m}{x_0}\rb^4\nn\\
&&+\lcb\lsb\left(-4288+\frac{2877 \pi }{2}\right)+\left(992-\frac{1447 \pi }{4}\right) \hat{q}^2+\frac{13 \pi
   \hat{q}^4}{4}\rsb \hat{a}s\right.\nn\\
&&\left.+\lsb\left(1455-\frac{855 \pi }{2}\right)+\lb-111+\frac{51 \pi }{2}\rb \hat{q}^2\rsb \hat{a}^2+(-64+24 \pi ) \hat{a}^3s-\frac{\hat{a}^4}{2}\rcb
   \lb\frac{m}{x_0}\rb^5\nn\\
&&+\lcb\lsb\left(34816-\frac{86577 \pi }{8}\right)+\left(-\frac{39200}{3}+\frac{7879 \pi }{2}\right) \hat{q}^2+\left(448-\frac{859 \pi }{8}\right) \hat{q}^4\rsb
   \hat{a}s\right.\nn\\
&&+\lsb\left(-16849+\frac{22155 \pi }{4}\right)+\left(2625-\frac{7365 \pi }{8}\right) \hat{q}^2+\frac{33 \pi  \hat{q}^4}{8}\rsb \hat{a}^2+\lsb(2496-756 \pi )+(-96+24 \pi ) \hat{q}^2\rsb
   \hat{a}^3s\nn\\
&&\left.\left.+\lsb\left(-\frac{65}{2}+\frac{45 \pi }{4}\right)+\frac{\pi  \hat{q}^2}{8}\rsb \hat{a}^4\rcb \lb\frac{m}{x_0}\rb^6
+\mathcal{O}\lb \frac{m}{x_0}\rb^7 \rsb +\mathcal{O}\lb 1-v\rb^2.
\label{krdfres3}
\eea

In order to express the change of the angular coordinate in terms of the expansion of the impact parameter, again we use the relation \eqref{blerel} with $L$ given by Eq. \eqref{krldef} to obtain a series expansion of $\frac{1}{x_0}$
\bea
\frac{m}{x_0}&=& \frac{m}{b}+\frac{1}{v^2}\lb \frac{m}{b}\rb^2+\lb \frac{2}{v^2}+\frac{1}{2 v^4} -\frac{1}{2 v^2}  \hat{q}^2 + \frac{2\hat{a}s
   }{v} +\frac{\hat{a}^2}{2}\rb
 \lb\frac{m}{b}\rb^3\nn\\
&&+\lsb \frac{4}{v^2}+\frac{4}{v^4} -\left(\frac{2}{v^2}+ \frac{1}{v^4}\right)
   \hat{q}^2  +\left( \frac{4}{v}+\frac{6}{v^3} -\frac{1}{v} \hat{q}^2\right)
   \hat{a}s+\left(1+\frac{2}{v^2}\right) \hat{a}^2
   \rsb \lb\frac{m}{b}\rb^4\nn\\
&&
+\lcb\frac{8}{v^2}+\frac{18}{v^4}+\frac{3}{v^6}-\frac{1}{8 v^8} -\left(\frac{6}{v^2}+\frac{9}{v^4}+\frac{3}{4 v^6}\right)
   \hat{q}^2+\left(\frac{1}{2 v^2}+\frac{3}{8 v^4}\right)
   \hat{q}^4\right.\nn\\
&&\left.+\lsb \frac{8}{v}+\frac{32}{v^3}+\frac{8}{v^5} -\left(\frac{4}{v}+\frac{8}{v^3}\right)
   \hat{q}^2\rsb \hat{a}s+\lsb 2+\frac{18}{v^2}+\frac{15}{4 v^4}-\left(\frac{1}{2}+\frac{5}{4 v^2}\right) \hat{q}^2\rsb
   \hat{a}^2+ \frac{4}{v} \hat{a}^3s+\frac{3 \hat{a}^4}{8}\rcb
 \lb\frac{m}{b}\rb^5\nn\\
&&+\lcb\frac{16}{v^2}+\frac{64}{v^4}+\frac{32}{v^6}-16\left(\frac{1}{v^2}+\frac{3}{v^4}+\frac{1}{v^6}\right)
   \hat{q}^2+\left(\frac{3}{v^2}+\frac{6}{v^4}+ \frac{1}{v^6}\right)
   \hat{q}^4 \right.\nn\\
&&+\lsb\frac{16}{v}+\frac{120}{v^3}+\frac{90}{v^5}+\frac{5}{v^7}-\left(\frac{12}{v}+\frac{60}{v^3}+\frac{45}{2
   v^5}\right) \hat{q}^2+\left(\frac{1}{v}+\frac{5}{2 v^3}\right) \hat{q}^4\rsb
   \hat{a}s\nn\\
&&\left.+\lsb 4+\frac{76}{v^2}+\frac{76}{v^4}+\frac{4}{v^6}-\left(2+\frac{22}{v^2}+\frac{6}{v^4}\right) \hat{q}^2\rsb
   \hat{a}^2+\left( \frac{20}{v}+\frac{25}{v^3} - \frac{5}{2 v}  \hat{q}^2\right)
   \hat{a}^3s+\left(2+\frac{3}{v^2}\right) \hat{a}^4\rcb \lb\frac{m}{b}\rb^6\nn\\
   &&+\mathcal{O}\lb \frac{m}{b}\rb^7.
\eea
It is seen that when $b>0$ is fixed and $q=0$, in the retrograde case $(sa>0)$ the $x_0$ will be smaller than in the prograde case $(sa<0)$.

Substituting into Eq. \eqref{angkrfull} we obtain
\be
I_\mathrm{KN}(b,v)=\sum_{n=0}^6 K^\prime_n \lb\frac{m}{b}\rb^n + \mathcal{O}\lb \frac{m}{b}\rb ^7 ,\label{angkrfullinb}\ee
where
\begin{subequations}\label{angkerrinb}
\begin{align}
K^\prime_0=& R^\prime_0 ,\\
K^\prime_1=& R^\prime_1,\\
K^\prime_2=& R^\prime_2+\frac{4 \hat{a}s}{v}  ,\label{angkerrinbo2}\\
K^\prime_3=& R^\prime_3+\pi \hat{a}s \left(\frac{4}{v^3}+\frac{6}{v} -\frac{\hat{q}^2}{v}\right) +\hat{a}^2 \left(\frac{2}{v^2}+2\right),\\
K^\prime_4=& R^\prime_4+12 \hat{a}s \lsb \frac{1}{v^5}+\frac{10}{v^3}+\frac{5}{v}-2\hat{q}^2
   \left(\frac{1}{v^3}+\frac{1}{v}\right)\rsb +3\pi  \hat{a}^2 \lsb\frac{1}{2 v^4}+\frac{9}{2 v^2}+\frac{15}{16}-\hat{q}^2 \left(\frac{1}{4 v^2}+\frac{3}{16}\right)\rsb+\frac{4 \hat{a}^3 s}{v}\\
K^\prime_5=& R^\prime_5+3\pi\hat{a}s \lsb 7\lb \frac{4}{v^5}+\frac{10}{v^3}+\frac{5}{2 v}\rb-3\hat{q}^2 \left(\frac{15}{4 v}+\frac{10}{v^3}+\frac{2}{v^5}\right)
+\hat{q}^4 \left(\frac{1}{v^3}+\frac{3}{4 v}\right)\rsb \nonumber\\
&+4\hat{a}^2 \lsb 7+\frac{75}{v^2}+\frac{45}{v^4}+\frac{1}{v^6}- \hat{q}^2
   \left(3+\frac{18}{v^2}+\frac{3}{v^4}\right)\rsb +3\pi  \hat{a}^3 s \left(\frac{5}{v}+\frac{4}{v^3}-\frac{3 \hat{q}^2}{2 v}\right)+2\hat{a}^4 \left(1+\frac{1}{v^2}\right),\\
K^\prime_6=& R^\prime_6+10\hat{a}s \lsb \frac{42}{v}+\frac{280}{v^3}+\frac{252}{v^5}+\frac{24}{v^7}-\frac{2}{3
   v^9}-16\hat{q}^2 \left(\frac{7}{3 v}+\frac{35}{3 v^3}+\frac{7}{v^5}+\frac{1}{3 v^7}\right)+
2\hat{q}^4 \left(\frac{3}{v^5}+\frac{10}{v^3}+\frac{3}{v}\right)\rsb\nonumber\\
&+15\pi  \hat{a}^2 \lsb 7
\lb\frac{15}{64}+\frac{35}{8 v^2}+\frac{25}{4 v^4}+\frac{1}{v^6}\rb
-\hat{q}^2\left(\frac{35}{32}+\frac{225}{16 v^2}+\frac{45}{4 v^4}+\frac{1}{2 v^6}\right)+\hat{q}^4 \left(\frac{5}{64}+\frac{9}{16 v^2}+\frac{1}{8 v^4}\right)
\rsb\nonumber\\
&+40\hat{a}^3 s \lsb \frac{7}{v}+\frac{50}{3 v^3}+\frac{3}{v^5}-2\hat{q}^2
   \left(\frac{1}{v}+\frac{1}{v^3}\right)\rsb +\frac{5\pi}{4}  \hat{a}^4 \lsb \frac{35}{8}+\frac{45}{2 v^2}+\frac{3}{v^4}-\hat{q}^2 \left(\frac{5}{8}+\frac{3}{4 v^2}\right)\rsb+\frac{4 \hat{a}^5 s}{v}
\end{align}
\end{subequations}
where $R^\prime_0$ to $R^\prime_6$ are given in Eq. \eqref{angrninb}.
From Eq. \eqref{angkerrinb} it is seen that the effect of the angular momentum per unit mass $a$ only starts to appear from order $\displaystyle \lb\frac{1}{b}\rb^2$. Writing  Eq. \eqref{angkerrinbo2} out explicitly, we have
\be K_2=
\frac{3\pi}{4}\lb 1+\frac{4}{v^2}\rb-\frac{\pi}{4}\left(1+\frac{2}{v^2}\right) \hat{q}^2+\frac{4 \hat{a}s}{v}
\ee
It is seen that when the geodesic motion is prograde, i.e., $sa<0$ the deflection angle decreases as $|a|$ increases. On the contrary, if the geodesic motion is retrograde, i.e., $sa>0$, then the deflection angle is larger than the prograde case with same $|a|$. Note that when $Q=0$ for Kerr spacetime, the first three orders agree with Eq. (96) of Ref. \cite{Crisnejo:2019ril}.

For lightlike rays, setting $v=1$, this becomes
\be
I_{\mathrm{KN},\gamma}(b)=\sum_{n=0}^6K^\prime_{n,\gamma}\lb\frac{m}{b}\rb^n +\mathcal{O} \lb\frac{m}{b}\rb^7, \ee
where
\begin{subequations}\label{angkerrinblight}
\begin{align}
K^\prime_{0,\gamma}=& R^\prime_{0,\gamma},\\
K^\prime_{1,\gamma}=& R^\prime_{1,\gamma},\\
K^\prime_{2,\gamma}=& R^\prime_{2,\gamma}+4\hat{a}s,\\
K^\prime_{3,\gamma}=& R^\prime_{3,\gamma}+\pi  \hat{a}s (10-\hat{q}^2) +4 \hat{a}^2,\\
K^\prime_{4,\gamma}=& R^\prime_{4,\gamma}+48\hat{a}s (4- \hat{q}^2)
+\frac{3\pi}{16}  \hat{a}^2 (95-7 \hat{q}^2)
+4 \hat{a}^3 s,\\
K^\prime_{5,\gamma}=& R^\prime_{5,\gamma}+\frac{21\pi}{4}  \hat{a}s (66-27\hat{q}^2+ \hat{q}^4)
+32 \hat{a}^2 (16-3 \hat{q}^2)
+3\pi  \hat{a}^3s (9-\frac{ \hat{q}^2}{2})
+4 \hat{a}^4,\\
K^\prime_{6,\gamma}=& R^\prime_{6,\gamma}+\frac{320}{3}\hat{a}s (56-32\hat{q}^2+3 \hat{q}^4 )+\frac{105\pi}{64}\hat{a}^2 (759 - 246 \hat{q}^2 + 7 \hat{q}^4 )+\frac{160}{3}\hat{a}^3s(20-3 \hat{q}^2)\nonumber\\
&+\frac{5}{32}\hat{a}^4 \left(239-11\hat{q}^2 \right)+4 \hat{a}^5 s .
\end{align}
\end{subequations}
The first four orders here agrees with Ref. \cite{Iyer:2009hq,Aazami:2011tw}.


\subsubsection{Effect of angular momentum on deflection angle}

\begin{figure}[htp!]
\begin{center}
\includegraphics[width=0.3\textwidth]{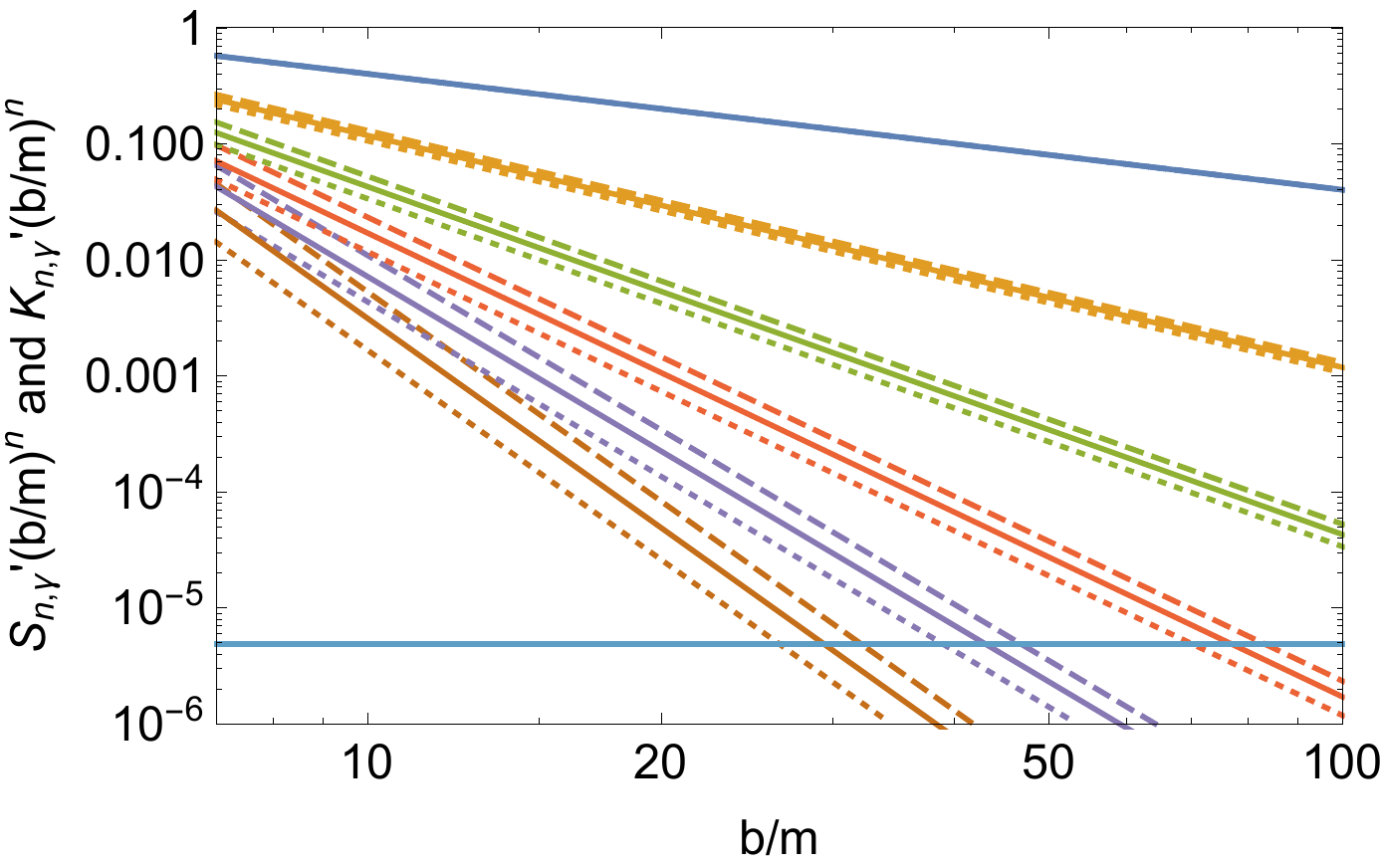}
\hspace{0.5cm}\includegraphics[width=0.3\textwidth]{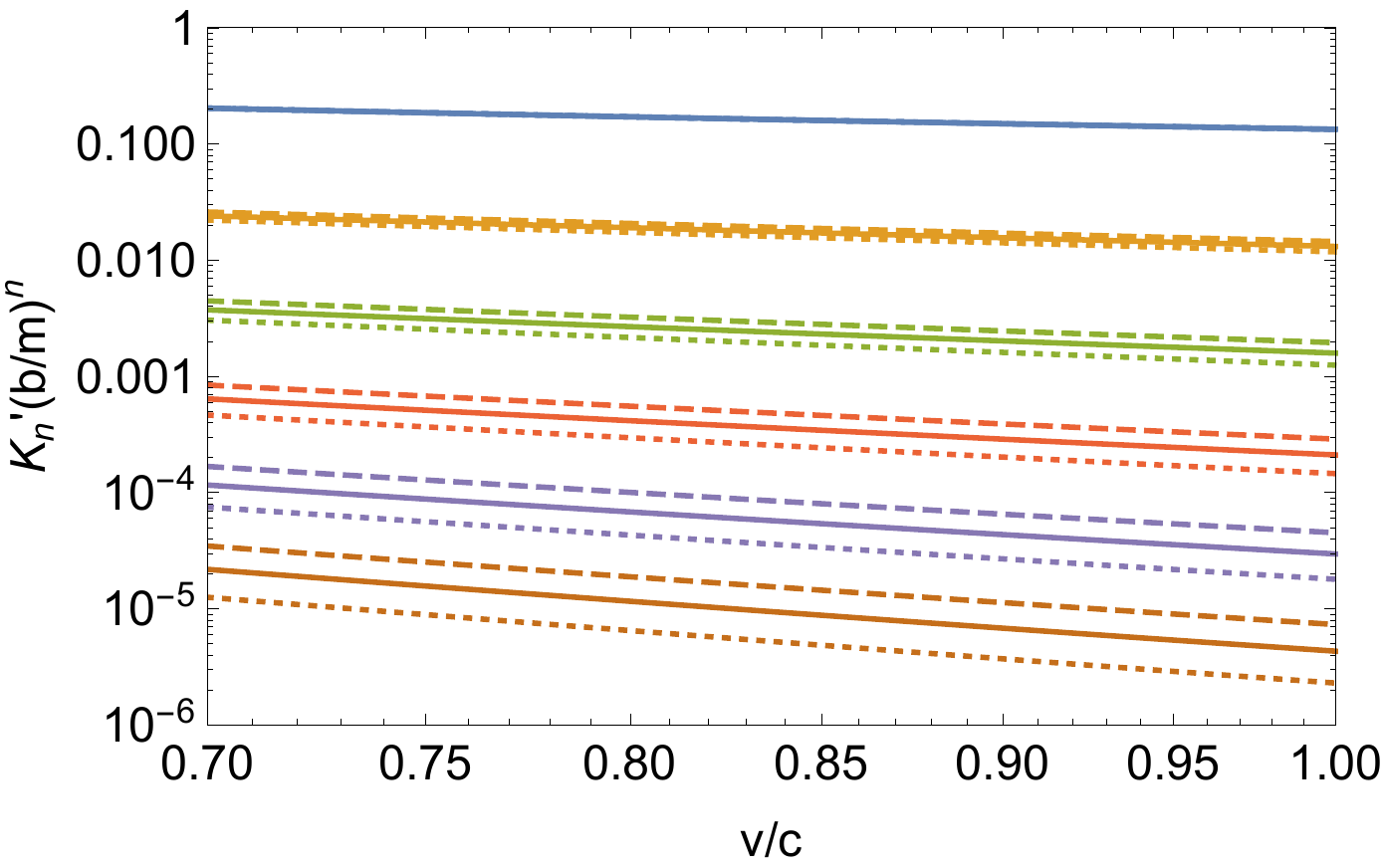}
\hspace{0.5cm}\includegraphics[width=0.3\textwidth]{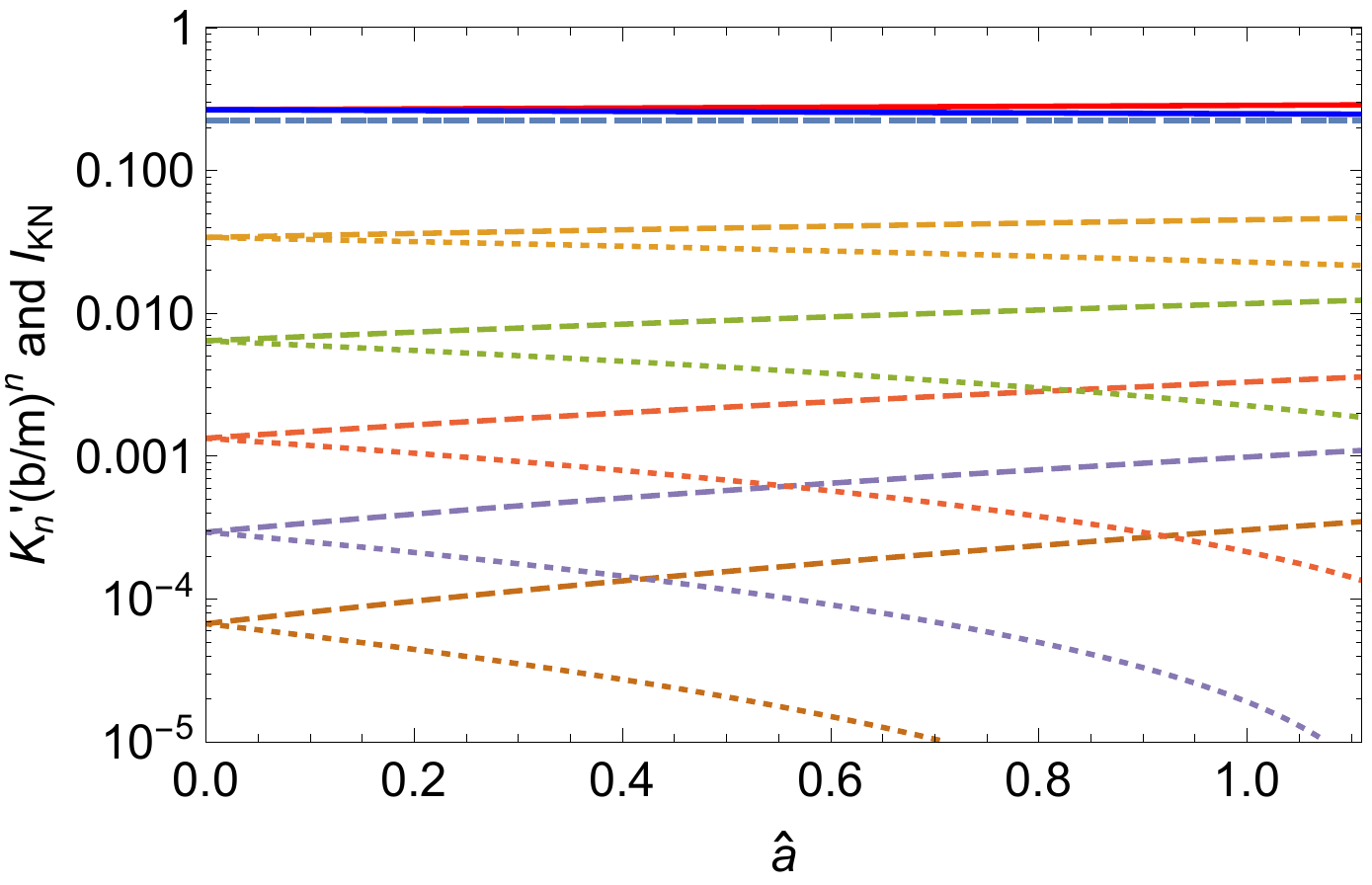}\\
(a)\hspace{5.5cm}(b)\hspace{5.5cm}(c)
\caption{The deflection angles at each order of Eq. \eqref{angkrfullinb} in the KN spacetime. (a) $v=1,~\hat{q}=0,~\hat{a}=0.3$; (b) $b/m=30,~\hat{q}=0,~\hat{a}=0.3$; The horizontal line in (a) is the 1 [as] line. (c) $b/m=100,~v/c=0.9,~\hat{q}=0.4$. For each order in these plots, solid, dashed and dotted lines correspond to $\hat{a}=0$, prograde motion and retrograde motion respectively. \label{figkn}}
\end{center}
\end{figure}

In Fig. \ref{figkn}(a), we plot each order in the deflection angle \eqref{angkrfullinb} for $v=1$ and $\hat{a}=0$, $\hat{a}=0.3$ for both the retrograde and prograde motions. It is seen for the first nontrivial order of the deflection angles, the lines completely overlap for different $\hat{a}$. This is indeed a consequence of Eq. \eqref{angkrfullinb}, where the effect of $\hat{a}$ only starts to appear from the third order (the second nontrivial order). For third or higher orders, it is seen that the retrograde motion has a larger deflection angle than in the Schwarzschild spacetime which is yet larger than the prograde motion case. This is indeed a manifestation of the frame-dragging effect of a rotating spacetime. In Fig. \ref{figkn}(b), the impact parameter is fixed at $b=10m$ and the velocity varies from $0.7c$ to $1.0c$ with two spacetime rotation directions with $\hat{a}=0.3$. It is seen that as $v$ decreases, the deflection angles at all orders increase for both rotation directions. This can be understand from the observation that slower particles (keeping $b$ fixed) tend to pass by the BH with a closer distance and therefore are more influenced by it, leading to a larger deflection angle. Moreover,  the deflection angle is increased regardless whether the BH is rotating or not, or the direction of the rotation. In Fig. \ref{figkn}(c), the dependence of the deflection angles on the angular momentum is shown. It is seen that as $\hat{a}$ increases, i.e., the rotation of the spacetime becomes faster, the deflection angle increases for retrograde motion and decreases for prograde motion. Moreover, similar to the case of RN spacetime, it is seen that the total deflection angle computed using Eq. \eqref{angkrfullinb} works even when the angular parameter is beyond its critical value $\hat{a}=\sqrt{1-\hat{q}^2}$ for the extreme KN spacetime. In other words, the deflection angle \eqref{angkrfullinb} is valid for both the KN BH and naked singularity spacetimes.

\subsection{Universal weak deflection angle to the lowest non-trivial order}

In previous sections/subsections we see that the change of the angular coordinate can be calculated for all known SSS spacetimes and for SAS spacetimes in the equatorial plane. Changes of the angular coordinate at high orders of $\displaystyle \frac{1}{x_0}$ or $\displaystyle \frac{1}{b}$  depends on the parameters in different ways. But through the examples, we do see that to the lowest order,  all spacetimes have a change of the angular coordinate  $\pi$. Now we will show that this feature is universal to at least geodesics on the equatorial plane of general SAS spacetimes which are also asymptotic
ally flat. These spacetime certainly include all the spacetimes in previous sections, in particular the SSS spacetimes specified by metric \eqref{sss1}. This value although sounds intuitively trivial, was never proven rigorously before.

Furthermore and more importantly, we will also show that for geodesics in the equatorial plane of such spacetimes, to the next (the lowest nontrivial) order, the change of the angular coordinate always takes the form
\be
I(b)=\pi+2\lb 1+\frac{1}{v^2}\rb \lb \frac{m}{b}\rb +\mathcal{O}\lb \frac{m}{b}\rb^2, \label{eqgcac}
\ee
where $m$ is the ADM mass of the spacetime.
Recently, Ref. \cite{Crisnejo:2019xtp} derived a deflection angle to the first nontrivial order in terms of an energy-momentum distribution of the SSS spacetime. From that point of view, our result here further proves that to the first non-trivial order, ADM mass is the only parameter of an asymptotically flat spacetime that will affect the change of the angular coordinate. Other parameters in these spacetimes such as charge or angular momentum can only appear at higher orders.
In addition, Eq. \eqref{eqgcac} also dictates how the particle velocity will affect this angle. Expanding Eq. \eqref{eqgcac} around $v=1$ yield the correction of the deflection angle of ultrarelativistic particle with respect to lightlike rays to the lowest nontrivial order of both $\frac{1}{b}$ and $(1-v)$.
\be
\alpha(b,v\to 1)=\alpha_\gamma(b)+4(1-v)\lb \frac{m}{b}\rb+\mathcal{O}\lb (1-v)^i\lb\frac{m}{b}\rb^j\rb,
\ee
where $i+j\geq 2$.

We now show \eqref{eqgcac}. In Eq. \eqref{asaiint} we have calculated the change of the angular coordinate in terms of general metric functions in Eq. \eqref{metric2}.
In order for this spacetime to be asymptotically flat, these metric functions should asymptotically satisfy the following conditions \cite{Bardeen:1973}
\begin{subequations}\label{abcdexp}
\begin{align}
A(\rho)=&1-\frac{2m}{\rho}+\mathcal{O}(\rho)^{-2}, \\ B(\rho)=&-\frac{4am}{\rho}+\mathcal{O}(\rho)^{-2},\\
C(\rho)=&\rho^2+2m\rho+\mathcal{O}(\rho)^0,\\
D(\rho)=&1+\frac{2m}{\rho}+\mathcal{O}(\rho)^{-2},
\end{align}
\end{subequations}
where $m$ is the effective total ADM mass of the spacetime. The effective parameters of charge and angular momentum of the spacetime will only appear in the higher orders of this expansion.
Substituting Eq. \eqref{abcdexp} into Eq. \eqref{asaiint}, changing the integration variables from $\rho$ to $u=\rho_0/\rho$, series expanding the integrand in the large $\rho_0$ limit and then carrying out the integration, the result of the change of angular coordinate is found to be
\be
I(\rho_0)=\pi+
2\lb 1+\frac{1}{v^2}\rb \lb \frac{m}{\rho_0}\rb +\mathcal{O}\lb \frac{m}{\rho_0}\rb^2. \label{sascares}\ee

To express Eq. \eqref{sascares}  in terms of the impact parameter $b$, we can again using Eq. \eqref{blerel} and \eqref{lsasdef} to find to the lowest order
\be
\frac{M}{\rho_0}=\frac{M}{b}+\mathcal{O}\left(\frac{1}{b^2}\right).
\ee
Substituting this into Eq. \eqref{sascares}, we immediately have to the first non-trivial order
\be
I(b)=\pi+
2\lb 1+\frac{1}{v^2}\rb \lb \frac{m}{b}\rb +\mathcal{O}\lb \frac{m}{b}\rb^2, \label{sascaresinb}\ee
which is the desired Eq. \eqref{eqgcac}.

\section{Gravitational Lensing in the weak field limit\label{secgl}}

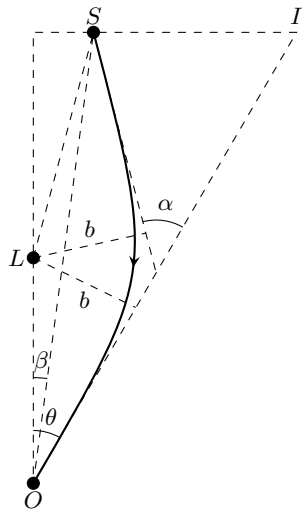
\begin{figure}[htp]
\tikzmath{
\ixvalue=3.5;
\cyvalue=-0.2;
\myvalue=3;
\cxvalue=(1-(\myvalue-\cyvalue)/(2*\myvalue))*\ixvalue;
\sxvalue=0.8;
}
\begin{tikzpicture}
\coordinate (L) at (0,0);
\coordinate (O) at (0,-\myvalue);
\coordinate (S) at (\sxvalue,\myvalue);
\coordinate (M) at (0,\myvalue);
\coordinate (C) at (\cxvalue,\cyvalue);
\coordinate (I) at (\ixvalue,\myvalue);
\coordinate (P) at (1.36,-0.66);
\coordinate (Q) at (1.5,0.33);
\draw[dashed] (L)--(S)--(O)--(M) --(I) node[above] {$I$} --(O);
\draw[thick,decoration={markings, mark=at position 0.5 with {\arrow{stealth}}},postaction={decorate}] (S) .. controls (C) ..(O);
\draw[dashed] (S) --(C);
\draw[fill] (L) circle[radius=0.08] node[left] {$L$};
\draw[fill] (S) circle[radius=0.08] node[above] {$S$};
\draw[fill] (O) circle[radius=0.08] node[below] {$O$};
\pic[draw,-, "$\theta$", angle eccentricity=1.3, angle radius=20] {angle=C--O--L};
\pic[draw,-, "$\beta$", angle eccentricity=1.15, angle radius=40] {angle=S--O--L};
\pic[draw,-, "$\alpha$", angle eccentricity=1.3, angle radius=20] {angle=I--C--S};
\draw[dashed] (L) --(P) node[midway,color=black,below] {$b$};
\draw[dashed] (L) --(Q) node[midway,color=black,above] {$b$};

\end{tikzpicture}
\caption{GL in the weak lensing limit. $O,~L,~S$ and $I$ stand for the observer, lens, source and image respectively. $\alpha,~\beta,~\theta$ are the deflection angle, the source and image angular positions respectively.
\label{fig:lensreg} }
\end{figure}

One main application of the deflection angles is in the GL in the weak field limit. The configuration of the GL in the weak field limit in the SSS or equatorial plane of SAS spacetime is illustrated in Fig. \ref{fig:lensreg}. We will denote the distance between the lens and source as $d_{\rm ls}$ and observer and lens as $d_{\rm ol}$, and the angular position of the source and its image against the observer-lens axis as $\beta$ and $\theta$ respectively. These angles as well as the deflection angle $\alpha$ are all small in the weak field limit. A lens equation linking these lengths and angles can be established, from which the apparent angle $\theta$ of the image are usually solved.

In the following, we will start from an exact lens equation derived directly from the geometric relations in GL but without using small angle or large length approximation \cite{Bozza:2008ev}
\be
d_{\rm os} \sin\beta=d_{\rm ol} \sin\theta\cos(\alpha-\theta)-\sqrt{d_{\rm ls}^2-d_{\rm ol}^2\sin^2\theta }\sin(\alpha-\theta). \label{fullleq}
\ee
The reason that we use the exact length equation but not the usual first order equation
\be
  \beta= \theta-\frac{d_{\rm ls}}{d_{\rm ls}+d_{\rm ol}}\alpha. \label{eq:lensingnormalsmall}
\ee
is that we would like to compute the apparent angle $\theta$ using perturbative method to higher orders and therefore this requires a lens equation accurate to high orders too. Eq. \eqref{eq:lensingnormalsmall} although can be derived using different approaches (see Ref. \cite{Bozza:2008ev} for a review and comparison of different approaches) and were used commonly in most GL computations, more exact equation such as Eq. \eqref{fullleq} has to be used when the error induced by the Eq. \eqref{eq:lensingnormalsmall} exceeds observational sensitivity.

Using Eq. \eqref{fullleq}, we will then illustrate how the various orders in the deflection angle $\alpha$ (e.g., Eqs. \eqref{angschinbfull}, \eqref{angrninbfull} and \eqref{angkrfullinb} after subtracting $\pi$) will perturbatively determine the apparent angle $\theta$ to the corresponding order. Using this perturbative method, the apparent angle $\theta$ to any desired accuracy can be achieved, provided that the deflection angle $\alpha$ can be calculated in prior to that order. Note that our method and result here will apply to not only null signal but signal with general velocity.

For the solution process, we first substitute the deflection angle $\alpha$ whose general form is a power series of $\displaystyle \frac{1}{b}$
\be
\alpha=\sum_{i=1}^\infty \frac{\alpha_i \epsilon^i}{b^i} \label{alphaexp}
\ee
into the Eq. \eqref{fullleq}.
Here $\alpha_i$ for Schwarzschild, RN, Kerr and KN spacetimes can be read off from Eqs. \eqref{angschinb}, \eqref{angrninb}, \eqref{angkerrinb}, \eqref{angschinbsupp} and \eqref{angrninb7to15} respectively. The small parameter $\epsilon$ is supplemented to track the order of large distances such as $b$ or $d_{\rm os}$ and it shall be set to 1 in final results.
Next, the $\displaystyle \frac{1}{b}$ in $\alpha$ can be replaced by the geometric relation
\be
\frac{1}{b}=\frac{1}{d_{\rm ol}\sin\theta}\ee
so that only measurable lengths appear. In the third step, we can substitute the ansatz for the series solution of $\theta$
\be
\theta=\sum_{i=1}\theta_i\epsilon^i
\ee
into the resultant equation. Finally, expanding both sides of the equation as power series of $\epsilon$ and collecting the coefficient of each order, one finds a system of algebraic equations of $\theta_i$. The first three of these are
\begin{subequations}\label{epsilonpower}
\begin{align}
0=&\frac{d_{\rm ls}+d_{\rm ol}}{d_{\rm os}}\theta_1-\frac{d_{\rm ls}\alpha_1}{d_{\rm ol}}\frac{1}{\theta_1}-\sin\beta,\label{epsone}\\
0=&d_{\rm ol}\lsb d_{\rm ls}\alpha_1+d_{\rm ol}(d_{\rm ls}+d_{\rm ol})\theta_1^2\rsb \theta_2-d_{\rm ls}\alpha_2,\\
0=&  6d_{\rm ls}d_{\rm ol}^2\theta_1 \lsb d_{\rm ls}\alpha_1+d_{\rm ol}\lb d_{\rm ls}+d_{\rm ol}\rb \theta_1^2\rsb  \theta_3-
\lsb d_{\rm ls}^2\lb 6\alpha_3-12d_{\rm ol}\alpha_2 \theta_2+6d_{\rm ol}^2\alpha_1\theta_2^2-\alpha_1^3\rb\right.\nn\\
&\left.+ 3d_{\rm ls}d_{\rm ol}(d_{\rm ls}+d_{\rm ol})\alpha_1^2\theta_1^2-d_{\rm ol}^2\lb 2d_{\rm ls}^2+6d_{\rm ls}d_{\rm ol}+3d_{\rm ol}^2\rb\alpha_1\theta_1^4+ d_{\rm ol}^3\lb d_{\rm ls}^2+4d_{\rm ls}d_{\rm ol}+3d_{\rm ol}^2\rb \theta_1^6  \rsb .
\end{align}
\end{subequations}

After expanding $\sin\beta$, the first equation \eqref{epsone} coincides with the leading order lens equation \eqref{eq:lensingnormalsmall}. Note that in this equation system, for each $i$, only $\theta_i$ is the unknown in the $i$-th equation. $\theta_j$ with $j<i$ in the $i$-th equation can always be solved in the equations prior to the $i$-th equation. Moreover, except equation \eqref{epsone} which is not a linear equation of $\theta_1$ but still solvable, all other equations are linear in its unknown $\theta_i$. Therefore the perturbative method guarantees that the equation system is iteratively solvable so that the final apparent angle can be obtained. This solvability is one of the advantages of the method.

Solving the system \eqref{epsilonpower}, we find the solution to $\theta_i$ as
\begin{subequations}\label{thetasol}
\begin{align}
\theta_1=&\frac{d_{\rm os}d_{\rm ol}\sin\beta\pm \sqrt{(d_{\rm os}d_{\rm ol}\sin\beta)^2+4d_{\rm ls}d_{\rm ol}(d_{\rm ls}+d_{\rm ol})\alpha_1}}{2d_{\rm ol}(d_{\rm ls}+d_{\rm ol})},\\
\theta_2=&\frac{d_{\rm ls}\alpha_2}{d_{\rm ol}\lsb d_{\rm ls}\alpha_1+d_{\rm ol}(d_{\rm ls}+d_{\rm ol})\theta_1^2\rsb},\\
\theta_3=&\frac{1}{ 6d_{\rm ls}d_{\rm ol}^2\theta_1 \lsb d_{\rm ls}\alpha_1+d_{\rm ol}\lb d_{\rm ls}+d_{\rm ol}\rb \theta_1^2\rsb }
\lsb d_{\rm ls}^2\lb 6\alpha_3-12d_{\rm ol}\alpha_2 \theta_2+6d_{\rm ol}^2\alpha_1\theta_2^2-\alpha_1^3\rb\right.\nn\\
&\left.+ 3d_{\rm ls}d_{\rm ol}(d_{\rm ls}+d_{\rm ol})\alpha_1^2\theta_1^2-d_{\rm ol}^2\lb 2d_{\rm ls}^2+6d_{\rm ls}d_{\rm ol}+3d_{\rm ol}^2\rb\alpha_1\theta_1^4+ d_{\rm ol}^3\lb d_{\rm ls}^2+4d_{\rm ls}d_{\rm ol}+3d_{\rm ol}^2\rb \theta_1^6  \rsb.
\end{align}
\end{subequations}
The higher order equations and solutions can be similarly obtained.

It is seen from Eq. \eqref{thetasol} that the solution to $\theta_m~(m\geq2)$ depends on the lowest order apparent angle $\theta_1$ and the deflection angle from the 1-st to the $m$-th orders, $\alpha_n~(1\leq n\leq m)$, as a power law function, except a common denominator which only depend on $\alpha_1$ and $\theta_1$ but not higher order ones. Moreover, counting the orders of distances $d_{\rm ol}$ and $d_{\rm ls}$ and the small angles $\theta_1$, one finds that $\theta_{m+1}$ is smaller than $\theta_m$ by an order of $d_{\rm ol}\theta_1$. This implies that if a certain calculation accuracy of the apparent angle $\theta$ is desired, one can work out from Eq. \eqref{thetasol} to what order the deflection angle should be used. This tractability is another advantage of the perturbative method. With the series expansion of $\theta$ known, we can find the magnification of the images using
\be
\mu =\frac{\sin\theta}{\sin\beta}\frac{\dd \theta}{\dd\beta}
\ee
and expand around small angles of $\theta$. Again, the result will be of a series form. For simplicity reason, we will not show these results explicitly.

\section{Discussions\label{secdiscussion}}

We have studied a perturbative method to compute the deflection angle in general SSS and equatorial plane of SAS spacetimes for arbitrary velocity $v$. It was shown that the involved integral in this method can always be carried out and therefore a series in either the closest radial coordinate $x_0$ or the impact parameter $b$ can always be found
\be
I=\sum_{n=0}^{\infty} C_n(v,p)\left(\frac{1}{x_0}\right)^n=\sum_{n=0}^{\infty}C^\prime_n(v,p)\left(\frac{1}{b}\right)^n
\ee
where we used $p$ to collectively denote any parameter of the spacetime, and $C_n(v,p)$ and $C^\prime_n(v,p)$ are two sets of coefficient functions.
Using this method, we were able to compute the deflection angle to the 17-th order in the Schwarzschild spacetime, 15-th order in the RN spacetime and 6-th order in the KN and consequently Kerr spacetime. Using these results, we studied how $v,~b$ and various parameters of the spacetime affect the total deflection angle and the deflection angle at each order. Two general features are particularly worth mentioning.
The first is that although the deflection angles are obtained perturbatively as a series of $1/b$ when its large, the valid range of the found deflection angle can extend to much smaller $b$, even when the deflection angle is not small anymore. The second is that the found deflection angles in the RN and KN metrics describe accurately the deflection angle not only for their BH spacetimes but also for their naked singularity cases. This last point here lays the foundation to apply these deflection angle results in the corresponding GLs to reveal the relevant features, if any, of these naked singularity spacetimes.

Using this perturbative method, we have shown that the deflection angle in the weak field limit in the asymptotically flat spacetimes depends only on the asymptotical behavior of the metric functions but not their values at small $b$. In particular, it was shown that the deflection angle of particles with general velocity in an EYM spacetime whose metric is only asymptotically known, can be computed. Moreover, we also illustrated that for equatorial motion in general SAS (including SSS) spacetimes, the deflection angle to the first nontrivial order depends only on the ADM mass of the spacetime and the asymptotical velocity of the particle in the specific way given by Eq. \eqref{sascaresinb}.

Regarding the extension of this method, the first and most apparent is to apply this method in other interesting SSS and SAS spacetimes to compute their deflection angles and study the GL effects in the weak field limit \cite{Duan:2020tsq}. Results found for these spacetimes are expected to not only reveal effect of spacetime parameters to signal deflection or GL, but also properties of the messenger itself, such as the neutrino mass/mass hierarchy and massive GWs.
Secondly, we also expect that this perturbative method is applicable to other computations involving integration of the geodesic equations, e.g., in the computation of time delay in GLs. We are currently working along this direction.

\begin{acknowledgments}

This work is supported in part by the NNSF China 11504276 and MOST China 2014GB109004. The author greatly appreciate discussion with Dr. Nan Yang, Haotian Liu and Ke Huang.
\end{acknowledgments}

\appendix

\section{High order terms of the deflection angles in Schwarzschild spacetime}
For Schwarzschild metric, the $S_5$ to $S_{17}$ in Eq. \eqref{angschfull} are
\begin{subequations}\label{sndetailssupp}
\begin{align}
S_5=&\frac{42}{5}-\lb\frac{201\pi}{16}-174\rb\frac{1}{v^2}-(117\pi-285)\frac{1}{v^4} -(63\pi-255)\frac{1}{v^6}-\lb24\pi-\frac{207}{4}\rb\frac{1}{v^8} +\frac{83}{20}\frac{1}{v^{10}},\\
S_6=&\frac{1155\pi}{256}+\lb\frac{8787\pi}{64}-114\rb\frac{1}{v^2}+ \lb\frac{10851\pi}{32}-1246\rb\frac{1}{v^4}+\lb \frac{2007\pi}{4}-\frac{4207}{3}\rb\frac{1}{v^6}+(183\pi-687)\frac{1}{v^8}\nn\\
&+\lb48\pi-\frac{443}{4}\rb\frac{1}{v^{10}} -\frac{73}{12}\frac{1}{v^{12}},\\
S_7=&
\frac{858}{35}+\lb \frac{4866}{5}-\frac{9897 \pi }{128}\rb
   \frac{1}{v^2}+\lb 3809-\frac{42105 \pi }{32}\rb  \frac{1}{v^4}+\lb 7431-\frac{35535 \pi }{16}\rb  \frac{1}{v^6}+\lb \frac{20861}{4}-\frac{3585 \pi }{2}\rb
   \frac{1}{v^8}\nn\\
&+\lb \frac{34731}{20}-480 \pi \rb  \frac{1}{v^{10}}+\lb \frac{9253}{40}-96 \pi \rb
   \frac{1}{v^{12}}+\frac{523}{56}\frac{1}{v^{14}},\\
S_8=&
\frac{225225 \pi }{16384}+\lb \frac{185637 \pi }{256}-\frac{3138}{5}\rb\frac{1}{v^2}+\lb\frac{963063 \pi }{256}-\frac{61062}{5}\rb\frac{1}{v^4}+\lb\frac{331515 \pi }{32}-31487\rb\frac{1}{v^6}\nonumber\\
   &+\lb \frac{670665 \pi
   }{64}-34265\rb\frac{1}{v^8}+\lb 5670 \pi -\frac{67199}{4}\rb\frac{1}{v^{10}}+\lb 1188 \pi -\frac{84097}{20}\rb\frac{1}{v^{12}}+\lb 192 \pi -\frac{19071}{40}\rb\frac{1}{v^{14}}-\frac{119}{8}\frac{1}{v^{16}},\\
S_9=&
\frac{4862}{63}+\lb\frac{174858}{35}-\frac{858633 \pi }{2048}\rb\frac{1}{v^2}+\lb\frac{179357}{5}-\frac{363561 \pi }{32}\rb\frac{1}{v^4}+\lb\frac{1873517}{15}-\frac{1248837 \pi
   }{32}\rb\frac{1}{v^6}\nonumber\\
   &+\lb\frac{733511}{4}-\frac{119295 \pi }{2}\rb\frac{1}{v^8}+\lb\frac{533009}{4}-\frac{329445 \pi }{8}\rb\frac{1}{v^{10}}+\lb\frac{1979077}{40}-16524 \pi \rb\frac{1}{v^{12}}\nonumber\\
   &+\lb \frac{2765857}{280}-2832 \pi
   \rb\frac{1}{v^{14}}+\lb\frac{2183697}{2240}-384 \pi \rb\frac{1}{v^{16}}+\frac{14051}{576}\frac{1}{ v^{18}},\\
S_{10}=
       &\frac{2909907 \pi }{65536}+\left(\frac{59009547 \pi }{16384}-\frac{112174}{35}\right)\frac{1}{v^2}+\left(\frac{131274411 \pi }{4096}-\frac{3405538}{35}\right)\frac{1}{v^4}+\left(\frac{37060149 \pi }{256}-\frac{2249681}{5}\right)\frac{1}{v^6}\nonumber\\
       &+\left(\frac{71222277 \pi }{256}-\frac{4427723}{5}\right)\frac{1}{v^8}+\left(\frac{17880387 \pi }{64}-\frac{17282663}{20}\right)\frac{1}{v^{10}}+\left(\frac{1148721 \pi }{8}-\frac{9235633}{20}\right)\frac{1}{v^{12}}\nonumber\\
       &+\left(45444 \pi -\frac{5480777}{40}\right) \frac{1}{v^{14}}+\left(6576 \pi -\frac{6354781}{280}\right)\frac{1}{v^{16}}+\left(768 \pi -\frac{4442329}{2240}\right)\frac{1}{v^{18}}-\frac{13103}{320} \frac{1}{v^{20}},\\
S_{11}=
       &\frac{8398}{33}+\left(\frac{2560186}{105}-\frac{69502047 \pi
   }{32768}\right)
   \frac{1}{v^2}+\left(\frac{9762087}{35}-\frac{687200823
   \pi }{8192}\right)
   \frac{1}{v^4}+\left(\frac{53657057}{35}-\frac{992311497
   \pi }{2048}\right)
   \frac{1}{v^6}\nn\\
   &+\left(\frac{76784969}{20}-\frac{157756977
   \pi }{128}\right)
   \frac{1}{v^8}+\left(\frac{98956459}{20}-\frac{199770417
   \pi }{128}\right)
   \frac{1}{v^{10}}+\left(\frac{141049413}{40}-\frac{36346
   527 \pi }{32}\right)
   \frac{1}{v^{12}}\nn\\
   &+\left(\frac{58958723}{40}-\frac{920367
   \pi }{2}\right)
   \frac{1}{v^{14}}+\left(\frac{812709823}{2240}-119736
   \pi \right)
   \frac{1}{v^{16}}+\left(\frac{344536399}{6720}-14976 \pi
   \right)
   \frac{1}{v^{18}}\nn\\
   &+\left(\frac{54036263}{13440}-1536 \pi
   \right)\frac{1}{v^{20}}+\frac{98601
   }{1408}\frac{1}{v^{22}},\\
S_{12}=
   &\frac{156165009 \pi }{1048576}+\left(\frac{1127476893 \pi
   }{65536}-\frac{1640018}{105}\right)
   \frac{1}{v^2}+\left(\frac{15337199865 \pi
   }{65536}-\frac{14309566}{21}\right)
   \frac{1}{v^4}\nn\\
   &+\left(\frac{811056015 \pi
   }{512}-4948511\right)
   \frac{1}{v^6}+\left(\frac{40234680075 \pi
   }{8192}-\frac{108659839}{7}\right)
   \frac{1}{v^8}+\left(\frac{2069467071 \pi
   }{256}-\frac{505097751}{20}\right)
   \frac{1}{v^{10}}\nn\\
   &+\left(\frac{1897384125 \pi
   }{256}-\frac{1406267327}{60}\right)
   \frac{1}{v^{12}}+\left(\frac{66690645 \pi
   }{16}-\frac{103772803}{8}\right)
   \frac{1}{v^{14}}+\left(\frac{5541675 \pi
   }{4}-\frac{247796467}{56}\right)
   \frac{1}{v^{16}}\nn\\
   &+\left(305280 \pi
   -\frac{1247909311}{1344}\right)
   \frac{1}{v^{18}}+\left(33600 \pi
   -\frac{109722263}{960}\right)
   \frac{1}{v^{20}}+\left(3072 \pi
   -\frac{109282609}{13440}\right)
   \frac{1}{v^{22}}-\frac{15565}{128}
   \frac{1}{v^{24}},\\
S_{13}=
   &\frac{371450}{429}+\left(\frac{132929354}{1155}-\frac{2692560531 \pi
   }{262144}\right)
   \frac{1}{v^{2}}+\left(\frac{201718541}{105}-\frac{9148875
   279 \pi }{16384}\right)
   \frac{1}{v^{4}}\nn\\
   &+\left(\frac{322639187}{21}-\frac{79680420
   705 \pi }{16384}\right)
   \frac{1}{v^{6}}+\left(\frac{1640587857}{28}-\frac{3836959
   9155 \pi }{2048}\right)
   \frac{1}{v^{8}}\nn\\
   &+\left(\frac{3320156743}{28}-\frac{7701893
   7885 \pi }{2048}\right)
   \frac{1}{v^{10}}+\left(\frac{5524434529}{40}-\frac{2824
   294095 \pi }{64}\right)
   \frac{1}{v^{12}}\nn\\
   &+\left(\frac{3935582091}{40}-\frac{1994
   399505 \pi }{64}\right)
   \frac{1}{v^{14}}+\left(\frac{19772726781}{448}-\frac{11
   3269545 \pi }{8}\right)
   \frac{1}{v^{16}}\nn\\
   &+\left(\frac{17027542177}{1344}-3974940
   \pi \right)
   \frac{1}{v^{18}}+\left(\frac{6218405311}{2688}-758400
   \pi \right)
   \frac{1}{v^{20}}+\left(\frac{37267872227}{147840}-74496
   \pi \right)
   \frac{1}{v^{22}}\nn\\
   &+\left(\frac{9707157937}{591360}-6144
   \pi \right) \frac{1}{v^{24}}+\frac{1423159}{6656}\frac{1}{v^{26}},\\
S_{14}=
   &\frac{2151252675 \pi }{4194304}+\left(\frac{83870896827 \pi
   }{1048576}-\frac{85161542}{1155}\right)
   \frac{1}{v^{2}}+\left(\frac{809464836033 \pi
   }{524288}-\frac{240298306}{55}\right)
   \frac{1}{v^{4}}\nn\\
   &+\left(\frac{962194187907 \pi
   }{65536}-\frac{1604380399}{35}\right)
   \frac{1}{v^{6}}+\left(\frac{2189788132125 \pi
   }{32768}-\frac{4424711827}{21}\right)
   \frac{1}{v^{8}}\nn\\
   &+\left(\frac{1349158008105 \pi
   }{8192}-\frac{14451331681}{28}\right)
   \frac{1}{v^{10}}+\left(\frac{964210959135 \pi
   }{4096}-\frac{103776190227}{140}\right)
   \frac{1}{v^{12}}\nn\\
   &+\left(\frac{53821800135 \pi
   }{256}-\frac{184429954727}{280}\right)
   \frac{1}{v^{14}}+\left(\frac{7644473325 \pi
   }{64}-\frac{105485967949}{280}\right)
   \frac{1}{v^{16}}\nn\\
   &+\left(\frac{90569745 \pi
   }{2}-\frac{9045013899}{64}\right)
   \frac{1}{v^{18}}+\left(10977900 \pi
   -\frac{46951853681}{1344}\right)
   \frac{1}{v^{20}}+\left(1844928 \pi
   -\frac{25271133577}{4480}\right)
   \frac{1}{v^{22}}\nn\\
   &+\left(163584 \pi
   -\frac{27159543721}{49280}\right)
   \frac{1}{v^{24}}+\left(12288 \pi
   -\frac{19569983309}{591360}\right)
   \frac{1}{v^{26}}-\frac{1361617}{3584}\frac{1}{v^{28}},\\
S_{15}=
   &\frac{430882}{143}+\left(\frac{7965387158}{15015}-\frac{101220705717
   \pi }{2097152}\right)
   \frac{1}{v^{2}}+\left(\frac{14056313867}{1155}-\frac{1810
   356103581 \pi }{524288}\right)
   \frac{1}{v^{4}}\nn\\
   &+\left(\frac{154414032013}{1155}-\frac{110
   74290444657 \pi }{262144}\right)
   \frac{1}{v^{6}}+\left(\frac{303126382501}{420}-\frac{7550
   111290137 \pi }{32768}\right)
   \frac{1}{v^{8}}\nn\\
   &+\left(\frac{127020227977}{60}-\frac{11021
   492762301 \pi }{16384}\right)
   \frac{1}{v^{10}}+\left(\frac{1026649289287}{280}-\frac{
   4787507625975 \pi }{4096}\right)
   \frac{1}{v^{12}}\nn\\
   &+\left(\frac{1116601325089}{280}-\frac{
   2595547784295 \pi }{2048}\right)
   \frac{1}{v^{14}}+\left(\frac{6334749108503}{2240}-\frac
   {115465265325 \pi }{128}\right)
   \frac{1}{v^{16}}\nn\\
   &+\left(\frac{3006211349669}{2240}-\frac
   {851607585 \pi }{2}\right)
   \frac{1}{v^{18}}+\left(\frac{5794564149407}{13440}-1380
   28101 \pi \right)
   \frac{1}{v^{20}}\nn\\
   &+\left(\frac{13812175825843}{147840}-29
   395104 \pi \right)
   \frac{1}{v^{22}}+\left(\frac{1141544032877}{84480}-4411
   008 \pi \right)
   \frac{1}{v^{24}}\nn\\
   &+\left(\frac{9195429924133}{7687680}-35
   6352 \pi \right)
   \frac{1}{v^{26}}+\left(\frac{1024712963063}{15375360}-2
   4576 \pi \right) \frac{1}{v^{28}}+\frac{10461043}{15360}\frac{1}{v^{30}},\\
S_{16}=
   &\frac{1933976154825 \pi }{1073741824}+\left(\frac{1529093331957 \pi
   }{4194304}-\frac{5107039334}{15015}\right)
   \frac{1}{v^{2}}+\left(\frac{39753044858427 \pi
   }{4194304}-\frac{394482990514}{15015}\right)
   \frac{1}{v^{4}}\nn\\
   &+\left(\frac{63300442855389 \pi
   }{524288}-\frac{434319372413}{1155}\right)
   \frac{1}{v^{6}}+\left(\frac{794767425731583 \pi
   }{1048576}-\frac{2756992108363}{1155}\right)
   \frac{1}{v^{8}}\nn\\
   &+\left(\frac{85917453284949 \pi
   }{32768}-\frac{3455378035567}{420}\right)
   \frac{1}{v^{10}}+\left(\frac{176717454948705 \pi
   }{32768}-\frac{7122654069149}{420}\right)
   \frac{1}{v^{12}}\nn\\
   &+\left(\frac{28896972701355 \pi
   }{4096}-\frac{6199928952201}{280}\right)
   \frac{1}{v^{14}}+\left(\frac{99481202698185 \pi
   }{16384}-\frac{763897992473}{40}\right)
   \frac{1}{v^{16}}\nn\\
   &+\left(\frac{57006172575 \pi
   }{16}-\frac{25030972225933}{2240}\right)
   \frac{1}{v^{18}}+\left(\frac{11456023755 \pi
   }{8}-\frac{30314838647029}{6720}\right)
   \frac{1}{v^{20}}\nn\\
   &+\left(404474862 \pi
   -\frac{16991569011289}{13440}\right)
   \frac{1}{v^{22}}+\left(76722612 \pi
   -\frac{36014522343589}{147840}\right)
   \frac{1}{v^{24}}\nn\\
   &+\left(10394112 \pi
   -\frac{18859803514961}{591360}\right)
   \frac{1}{v^{26}}+\left(771072 \pi
   -\frac{19833376444423}{7687680}\right)
   \frac{1}{v^{28}}\nn\\
   &+\left(49152 \pi
   -\frac{2061930347341}{15375360}\right)
   \frac{1}{v^{30}}-\frac{1259743}{1024}\frac{1}{v^{32}},\\
S_{17}=
   &\frac{2357178}{221}+\left(\frac{36069111362}{15015}-\frac{1489022648660
   1 \pi }{67108864}\right)
   \frac{1}{v^{2}}+\left(\frac{72583108089}{1001}-\frac{5297
   294087775 \pi }{262144}\right)
   \frac{1}{v^{4}}\nn\\
   &+\left(\frac{149950442373}{143}-\frac{8661
   4899454125 \pi }{262144}\right)
   \frac{1}{v^{6}}+\left(\frac{1004889697831}{132}-\frac{497
   2861857705 \pi }{2048}\right)
   \frac{1}{v^{8}}\nn\\
   &+\left(\frac{6729314550183}{220}-\frac{637
   518587407443 \pi }{65536}\right)
   \frac{1}{v^{10}}+\left(\frac{2956925896701}{40}-\frac{2
   4110437475817 \pi }{1024}\right)
   \frac{1}{v^{12}}\nn\\
   &+\left(\frac{2743131628633}{24}-\frac{7
   4466297264975 \pi }{2048}\right)
   \frac{1}{v^{14}}+\left(\frac{52773568983405}{448}-\frac
   {2401364359845 \pi }{64}\right)
   \frac{1}{v^{16}}\nn\\
   &+\left(\frac{37347372420843}{448}-\frac
   {27148502564865 \pi }{1024}\right)
   \frac{1}{v^{18}}+\left(\frac{79271572704617}{1920}-\frac{105280583331 \pi }{8}\right)
   \frac{1}{v^{20}}\nn\\
   &+\left(\frac{9252941772369}{640}-\frac{
   9182292273 \pi }{2}\right)
   \frac{1}{v^{22}}+\left(\frac{20203810510113}{5632}-1147
   100220 \pi \right)
   \frac{1}{v^{24}}\nn\\
   &+\left(\frac{136575742403537}{219648}-1
   95992160 \pi \right)
   \frac{1}{v^{26}}+\left(\frac{10884221610983}{146432}-24
   192000 \pi \right)
   \frac{1}{v^{28}}\nn\\
   &+\left(\frac{28366854891847}{5125120}-1
   658880 \pi \right)
   \frac{1}{v^{30}}+\left(\frac{66338835211301}{246005760}
   -98304 \pi \right)\frac{1}{v^{32}}+\frac{623034403}{278528}\frac{1}{v^{34}}.
\end{align}
\end{subequations}
and for lightlike rays these become
\begin{subequations}\label{sndetailsv1supp}
\begin{align}
S_{5,\gamma}=&\frac{7783}{10}-\frac{3465\pi}{16},\\
S_{6,\gamma}=&\frac{310695\pi}{256}-\frac{21397}{6},\\
S_{7,\gamma}=&\frac{544045}{28}-\frac{765765\pi }{128},\\
S_{8,\gamma}=&\frac{530675145 \pi}{16384}-\frac{400353}{4},\\
S_{9,\gamma}=&\frac{1094345069}{2016}-\frac{350975625 \pi }{2048},\\
S_{10,\gamma}=&\frac{61238992815 \pi }{65536}-\frac{3274477761}{1120},\\
S_{11,\gamma}=&\frac{33880841953}{2112}-\frac{166985013195 \pi }{32768},\\
S_{12,\gamma}=&\frac{819 (36025412555 \pi -113062658048)}{1048576},\\
S_{13,\gamma}=&\frac{17954674772417}{36608}-\frac{40904635446675 \pi }{262144},\\
S_{14,\gamma}=&\frac{3654071183280375 \pi }{4194304}-\frac{53937207017735}{19712},\\
S_{15,\gamma}=&\frac{1532445398265737}{99840}-\frac{10244859509950065 \pi }{2097152},\\
S_{16,\gamma}=&\frac{29546784214957377225 \pi }{1073741824}-\frac{4027582104301883}{46592},\\
S_{17,\gamma}=&\frac{6193832627891384481}{12673024}-\frac{10439883710287799625 \pi }{67108864}.
\end{align}
\end{subequations}

The full form of the expansion \eqref{angschvcinx0} is
\bea
&&I_\mathrm{S}(x_0,v\to 1)=I_\mathrm{S,\gamma}(x_0)+(1-v)\times \nn\\
&&
\lsb
  4\lb\frac {m} {x_ 0} \rb^{1}
  +\lb 6 \pi -12\rb \lb\frac {m} {x_ 0} \rb^{2}
  +\lb 102-27 \pi\rb \lb\frac {m} {x_ 0} \rb^{3}
  +\lb\frac{375 \pi }{2}-542\rb \lb\frac {m} {x_ 0} \rb^{4}
  \right.\nn\\
&&
  +\lb\frac{6947}{2}-\frac{8505 \pi }{8}\rb \lb\frac {m} {x_ 0} \rb^{5}
  +\lb\frac{210735 \pi }{32}-\frac{40605}{2}\rb \lb\frac {m} {x_ 0} \rb^{6}
  +\lb\frac{2475251}{20}-\frac{2498265 \pi }{64}\rb \lb\frac {m} {x_ 0} \rb^{7}
  \nn\\
&&
  +\lb\frac{30225195 \pi }{128}-\frac{14770307}{20}\rb \lb\frac {m} {x_ 0} \rb^{8}
  +\lb\frac{996907893}{224}-\frac{1447430985 \pi }{1024}\rb \lb\frac {m} {x_ 0} \rb^{9}
  \nn\\
&&
  +\lb\frac{69647092515 \pi }{8192}-\frac{5976368855}{224}\rb \lb\frac {m} {x_ 0} \rb^{10}
  +\lb\frac{215463718145}{1344}-\frac{835623443655 \pi }{16384}\rb \lb\frac {m} {x_ 0} \rb^{11}
  \nn\\
&&
  +\lb\frac {10036865773935\pi} {32768} - \frac {307844832381} {320} \rb\lb\frac {m} {x_0} \rb^{12}\nn\\
&&
  +\lb\frac {243970659948739}{42240} - \frac {240943980632355\pi}{131072}\rb \lb\frac{m}{x_0}\rb^{13}\nn\\
&&
  +\lb\frac {5785423336868175\pi}{524288}-\frac {97615569852587}{2816}\rb \lb\frac{m}{x_0}\rb^{14}\nn\\
&&
  +\lb\frac{15233751847277841}{73216}-\frac {69444310090079925\pi}{1048576} \rb\lb\frac {m} {x_ 0} \rb^{15}\nn\\
&&
  +\lb\frac{833588786453526675\pi} {2097152} - \frac{639983494401301615} {512512} \rb\lb\frac {m} {x_ 0}\rb^{16}\nn\\
&&
  \left.+\lb\frac{307272383850733443143}{41000960} - \frac{80043485420373397065\pi} {33554432}\rb\lb\frac {m} {x_ 0} \rb^{17}
+ \mathcal{O}\lb \frac{m}{x_0}\rb ^{18}\rsb +\mathcal{O}(1-v)^2.\label{eq:isvto1full}\eea

For Schwarzschild metric, when $x_0$ is expressed in terms of $b$ in Eq. \eqref{eq:x0inbschgeneral}, the high order terms are
\begin{subequations}\label{x0inbschsupp}
\begin{align}
C_{\mathrm{S},5}=&\lb\frac{8}{v^2}+\frac{18}{v^4}+\frac{3}{v^6}-\frac{1}{8v^8}\rb,\\
C_{\mathrm{S},6}=&16\lb\frac{1}{v^2}+\frac{4}{v^4}+\frac{2}{v^6}\rb,\\
C_{\mathrm{S},7}=&\lb \frac{32}{v^2}+\frac{200}{v^4}+\frac{200}{v^6}+\frac{25}{v^8} -\frac{5}{4v^{10}}+\frac{1}{16v^{12}}\rb,\\
C_{\mathrm{S},8}=&64\lb \frac{1}{v^2}+\frac{9}{v^4}+\frac{15}{v^6}+\frac{5}{v^8}\rb,\\
C_{\mathrm{S},9}=& \lb\frac{128}{v^2}+\frac{1568}{v^4}+\frac{3920}{v^6}+\frac{2450}{v^8}+\frac{245}{v^{10}}-\frac{49}{4 v^{12}}+\frac{7}{8 v^{14}}-\frac{5}{128 v^{16}}\rb ,\\
C_{\mathrm{S},10}=&256\lb
\frac{1}{v^2}+\frac{16}{v^4}+\frac{56}{v^6}+\frac{56}{v^8}
  +\frac{14}{v^{10}}\rb,\\
C_{\mathrm{S},11}=&\frac{512}{v^2}+\frac{10368}{v^4}
  +\frac{48384}{v^6}+\frac{70560}{v^8}
  +\frac{31752}{v^{10}}+\frac{2646}{v^{12}}-\frac{126}{v^{14}}+\frac{81}{8 v^{16}}
  -\frac{45}{64 v^{18}}+\frac{7}{256 v^{20}},\\
C_{\mathrm{S},12}=&1024\lb \frac{1}{v^2}+\frac{25}{v^4}+\frac{150
  }{v^6}+\frac{300}{v^8}+\frac{210}{v^{10}}+\frac{42}{v^{12}}\rb,\\
C_{\mathrm{S},13}=&\frac{2048 }{v^2}+\frac{61952 }{v^4}+\frac{464640
   }{v^6}+\frac{1219680}{v^8}+\frac{1219680
   }{v^{10}}+\frac{426888}{v^{12}}+\frac{30492}{v^{14}}-\frac{5445}{4 v^{16}}+\frac{1815}{16
   v^{18}}-\frac{605}{64 v^{20}}\nn\\
   &+\frac{77}{128
  v^{22}}-\frac{21}{1024 v^{24}},\\
C_{\mathrm{S},14}=&4096\lb \frac{1}{v^2}+\frac{36 }{v^4}+\frac{330
  }{v^6}+\frac{1100}{v^8}+\frac{1485
  }{v^{10}}+\frac{792}{v^{12}}+\frac{132}{v^{14}}\rb,\\
C_{\mathrm{S},15}=&\frac{8192}{v^2}+\frac{346112 }{v^4}+\frac{3807232
   }{v^6}+\frac{15704832}{v^8}+\frac{27483456
   }{v^{10}}+\frac{20612592}{v^{12}}+\frac{5889312
   }{v^{14}}+\frac{368082 }{v^{16}}-\frac{61347 }{4
   v^{18}}+\frac{20449 }{16 v^{20}}\nn\\
&  -\frac{1859 }{16
   v^{22}}+\frac{1183 }{128 v^{24}}-\frac{273}{512
   v^{26}}+\frac{33 }{2048 v^{28}},\\
C_{\mathrm{S},16}=&16384\lb\frac{1}{v^2}+\frac{49}{v^4}+\frac{637
   }{v^6}+\frac{3185}{v^8}+\frac{7007
   }{v^{10}}+\frac{7007}{v^{12}}+\frac{3003
   }{v^{14}}+\frac{429}{v^{16}}\rb,\\
C_{\mathrm{S},17}=&\frac{32768}{v^2}+\frac{1843200}{v^4}+\frac{27955200
  }{v^6}+\frac{166566400}{v^8}+\frac{449729280
   }{v^{10}} +\frac{577152576}{v^{12}}+\frac{343543200
   }{v^{14}}+\frac{82818450}{v^{16}},\nn\\
   &+\frac{4601025
   }{v^{18}}-\frac{715715}{4 v^{20}}+\frac{117117
  }{8 v^{22}}-\frac{88725}{64 v^{24}}+\frac{15925
  }{128 v^{26}}-\frac{4725 }{512 v^{28}}+\frac{495
  }{1024 v^{30}}-\frac{429}{32768
  v^{32}}.
\end{align}
\end{subequations}

The $S^\prime_5$ to $S^\prime_{17}$ in Eq. \eqref{angschinbfull} are given by
\begin{subequations}\label{angschinbsupp}
\begin{align}
S^\prime_5=&2\lb \frac{21}{5}+\frac{105}{v^2}+\frac{210}{v^4}+\frac{42}{v^6}-\frac{3}{v^8}+\frac{1}{5 v^{10}} \rb ,\\
S^\prime_6=&\frac{1155\pi}{256}\lb 1+\frac{36}{v^2} +\frac{30}{v^4} +\frac{4}{v^6}\rb,\\
S^\prime_7=&2\lb \frac{429}{35}+\frac{3003}{5v^2}+\frac{3003}{v^4}+\frac{3003}{ v^6}+\frac{429}{v^8}-\frac{143}{5v^{10}}+\frac{13}{5 v^{12}}-\frac{1}{7v^{14}}\rb,\\
S^\prime_8=&\frac{45045\pi}{16384}\lb 5+\frac{320}{v^2}+\frac{2240}{v^4}+\frac{3584}{v^6}+\frac{1280}{v^8}\rb ,\\
S^\prime_9=&2 \lb \frac{2431}{63}+\frac{21879}{7 v^2}+\frac{29172}{v^4}+\frac{68068}{v^6}+\frac{43758}{v^8}+\frac{4862}{v^{10}}-\frac{884}{3 v^{12}}+\frac{204}{7 v^{14}}-\frac{17}{7 v^{16}}+\frac{1}{9  v^{18}}\rb,\\
S^\prime_{10}=&\frac{2909907 \pi}{65536}\lb 1+\frac{100}{v^{2}}+\frac{1200}{v^{4}}
   +\frac{3840}{v^{6}}+\frac{3840}{v^{8}}+\frac{1024}{v^{10}}\rb,\\
S^\prime_{11}=&
   2\lb\frac{4199}{33}+\frac{46189}{3 v^2}+\frac{230945}{v^4}+\frac{969969 }{v^6}+\frac{1385670
   }{v^8}+\frac{646646}{v^{10}}+\frac{58786
   }{v^{12}}-\frac{3230 }{v^{14}}+\frac{323
   }{v^{16}}-\frac{95}{3 v^{18}}\right.\nn\\
   &
   \left.+\frac{7 }{3
   v^{20}}-\frac{1}{11 v^{22}}\rb,\\
S^\prime_{12}=&\frac{22309287 \pi}{1048576}\lb 7+\frac{1008}{v^{2}}+\frac{18480}{v^{4}}
    +\frac{98560}{v^{6}}+\frac{190080}{v^{8}}+\frac{135168}{v^{10}}+\frac{28672}{v^{12}}\rb,\\
S^\prime_{13}=&2\lb\frac{185725}{429}+\frac{2414425}{33 v^2}+\frac{4828850}{3
   v^4}+\frac{10623470}{v^6}+\frac{26558675
   }{v^8}+\frac{26558675}{v^{10}}+\frac{9657700
   }{v^{12}}+\frac{742900}{v^{14}}\right.\nn\\
   &
   \left.-\frac{37145
   }{v^{16}}+\frac{10925}{3 v^{18}}-\frac{1150}{3
   v^{20}}+\frac{1150}{33 v^{22}}-\frac{25}{11
   v^{24}}+\frac{1}{13 v^{26}}\rb,\\
S^\prime_{14}=&\frac{717084225 \pi}{4194304}\lb 3+\frac{588}{v^{2}}
    +\frac{15288}{v^{4}}+\frac{122304}{v^{6}}+\frac{384384}{v^{8}}
    +\frac{512512}{v^{10}}+\frac{279552}{v^{12}}+\frac{49152}{v^{14}}\rb,\\
S^\prime_{15}=&2\lb\frac{215441}{143}+\frac{48474225}{143 v^2}+\frac{113106525}{11
   v^4}+\frac{98025655}{v^6}+\frac{378098955
   }{v^8}+\frac{646969323}{v^{10}}+\frac{490128275
   }{v^{12}}\right.\nn\\
   &
   \left.+\frac{145422675}{v^{14}}+\frac{9694845
   }{v^{16}}-\frac{443555 }{v^{18}}+\frac{42021
   }{v^{20}}-\frac{50025}{11 v^{22}}+\frac{5075}{11
   v^{24}}-\frac{5481}{143 v^{26}}+\frac{29}{13
   v^{28}}-\frac{1}{15 v^{30}}\rb,\\
S^\prime_{16}=&\frac{644658718275 \pi}{1073741824}\lb 3+\frac{768}{v^{2}}+\frac{26880}{v^{4}}
    +\frac{301056}{v^{6}}+\frac{1397760}{v^{8}}+\frac{2981888}{v^{10}}
    +\frac{2981888}{v^{12}}+\frac{1310720}{v^{14}}+\frac{196608}{v^{16}}\rb,\\
S^\prime_{17}=&2\lb\frac{1178589}{221}+\frac{20036013}{13 v^2}+\frac{801440520}{13
   v^4}+\frac{801440520}{v^6}+\frac{4407922860
   }{v^8}+\frac{11460599436}{v^{10}}+\frac{14586217464
   }{v^{12}}\right.\nn\\
   &
   +\frac{8815845720}{v^{14}}+\frac{2203961430
   }{v^{16}}+\frac{129644790}{v^{18}}-\frac{5458728
   }{v^{20}}+\frac{496248 }{v^{22}}-\frac{53940
   }{v^{24}}+\frac{75516 }{13 v^{26}}-\frac{7192}{13
   v^{28}}+\frac{2728}{65 v^{30}}\nn\\
   &
   \left.-\frac{11}{5
   v^{32}}+\frac{1}{17 v^{34}}\rb.
\end{align}
\end{subequations}

The high order terms in Eq. \eqref{angschlightinbgeneral} are given by
\begin{subequations}\label{angschinblightsupp}
\begin{align}
S^\prime_{5,\gamma}=&\frac{3584}{5} ,\\
S^\prime_{6,\gamma}=&\frac{255255\pi}{256},\\
S^\prime_{7,\gamma}=&\frac{98304}{7} ,\\
S^\prime_{8,\gamma}=&\frac{334639305 \pi}{16384} ,\\
S^\prime_{9,\gamma}=&\frac{18743296}{63},\\
S^\prime_{10,\gamma}=&\frac{29113619535 \pi}{65536},\\
S^\prime_{11,\gamma}=&\frac{218103808}{33},\\
S^\prime_{12,\gamma}=&\frac{10529425731825 \pi}{1048576},\\
S^\prime_{13,\gamma}=&\frac{21676163072}{143},\\
S^\prime_{14,\gamma}=&\frac{977947275623175 \pi}{4194304},\\
S^\prime_{15,\gamma}=&\frac{693637218304}{195},\\
S^\prime_{16,\gamma}=&\frac{5929294332103310025 \pi}{1073741824},\\
S^\prime_{17,\gamma}=&\frac{18769007083520}{221}.
\end{align}
\end{subequations}

For the fast speed expansion of the deflection angle in terms of $b$, the full results to 17-th order is
\bea
&&I_\mathrm{S}(b,v\to1)=I_\mathrm{S,\gamma}(b)+(1-v)\lsb 4\lb \frac{m}{b}\rb^1+6\pi\lb \frac{m}{b}\rb^2+96\lb \frac{m}{b}\rb^3 +\frac{315\pi}{2}\lb \frac{m}{b}\rb^4 \right.\nn\\
&&+2560\lb \frac{m}{b}\rb^5 +\frac{135135\pi}{32}\lb \frac{m}{b}\rb^6+\frac{344064\pi}{5}\lb \frac{m}{b}\rb^7+\frac{14549535 \pi}{128}\lb \frac{m}{b}\rb^8+\frac{12976128}{7}\lb \frac{m}{b}\rb^9\nn\\
&&
  +\frac{25097947875 \pi}{8192}\lb\frac{m}{b}\rb^{10}
  +\frac{149946368}{3}\lb\frac{m}{b}\rb^{11}
  +\frac{2707566616755 \pi}{32768}\lb\frac{m}{b}\rb^{12}+\frac{14831058944}{11}\lb\frac{m}{b}\rb^{13}\nn\\
&&
  +\frac{1168766256232575\pi}{524288}\lb\frac{m}{b}\rb^{14}
  +\frac{5202279137280}{143}\lb\frac{m}{b}\rb^{15}
  +\frac{126155198555389575 \pi}{2097152}\lb\frac{m}{b}\rb^{16}\nn\\
&&
  \left.+\frac{63814624083968}{65}\lb\frac{m}{b}\rb^{17}
  +\mathcal{O}\lb\frac{m}{b}\rb^{18}\rsb
    +\mathcal{O}(1-v)^2. \label{angschvcinbfull}\eea

\section{High order terms of the deflection angles in RN spacetime}

For $R_5$ to $R_{15}$ in Eq. \eqref{rndetails}, they are given by
\begin{subequations}\label{rndetails7to15}
\begin{align}
R_5=
    &
    S_{5}-\left[\frac{28}{3}+\lb \frac{473}{3}-\frac{85 \pi}{8} \rb\frac{1}{v^2}
    -\lb \frac{153 \pi}{2}-184 \rb\frac{1}{v^4}
    +\lb \frac{665}{6}-27 \pi \rb\frac{1}{v^6}
    -\lb 4 \pi-\frac{43}{6} \rb\frac{1}{v^8}\right]\hat{q}^2\nn\\
    &
    +\left[2+\lb 26-\frac{25 \pi}{16}\rb\frac{1}{v^2}
    -\lb \frac{11 \pi}{2}-\frac{47}{4}\rb\frac{1}{v^4}
    +\lb \frac{\pi}{2}-\frac{1}{4} \rb\frac{1}{v^6}\right]\hat{q}^4,\\
R_6=
    &
    S_{6}-\left[\frac{1575 \pi}{256}+
    \lb \frac{20101 \pi}{128}-\frac{368}{3} \rb\frac{1}{v^2}
    -\lb \frac{3397}{3}-\frac{4921 \pi }{16}\rb\frac{1}{v^4}
    +\lb \frac{2655 \pi}{8}-922 \rb\frac{1}{v^6}\right.\nn\\
    &
    \left.-\lb \frac{1801}{6}-79 \pi \rb\frac{1}{v^8}+\lb 8 \pi-\frac{95}{6} \rb\frac{1}{v^{10}}\right]\hat{q}^2
    +\left[\frac{525 \pi }{256}
    +\lb \frac{1329\pi}{32}-28 \rb\frac{1}{v^2}
    -\lb 204-\frac{441 \pi}{8} \rb\frac{1}{v^4}\right.\nn\\
    &
    \left.+\lb \frac{105 \pi}{4}-\frac{267}{4} \rb\frac{1}{v^6}-\lb\frac{3 \pi}{2}- \frac{5}{4} \rb\frac{1}{v^8}\right]\hat{q}^4
    -\left(\frac{25 \pi }{256}+\frac{157 \pi }{128v^2}
    -\frac{5 \pi }{32 v^4}+\frac{\pi }{16v^6}\right)\hat{q}^6,\\
R_7=
    &
    S_7+\left[-\frac{198}{5}
    +\lb \frac{13381 \pi}{128}-\frac{20413}{15} \rb\frac{1}{v^2}
    +\lb \frac{96527 \pi}{64}-\frac{13091}{3} \rb\frac{1}{v^4}
    +\lb \frac{16249 \pi}{8}-\frac{40831}{6} \rb\frac{1}{v^6}\right.\nn\\
    &
    \left.+\lb \frac{4777 \pi}{4}-\frac{41507}{12} \rb\frac{1}{v^8}
    +\lb 208 \pi-\frac{91307}{120} \rb\frac{1}{v^{10}}
    +\lb 16 \pi-\frac{4049}{120} \rb\frac{1}{v^{12}}\right]\hat{q}^2\nn\\
    &
    +\left[18+\lb \frac{1565}{3}-\frac{4903 \pi}{128} \rb\frac{1}{v^2}
    +\lb \frac{14803}{12}-\frac{6811 \pi}{16} \rb\frac{1}{v^4}
    +\lb \frac{5147}{4}-\frac{3033 \pi}{8} \rb\frac{1}{v^6}\right.\nn\\
    &
    \left.+\lb \frac{6349}{24}-\frac{197 \pi}{2}\rb\frac{1}{v^8}
    +\lb 4 \pi-\frac{103}{24} \rb\frac{1}{v^{10}}\right]\hat{q}^4
    +\left[-2+\lb \frac{411 \pi}{128}-47 \rb\frac{1}{v^2}
    +\lb \frac{1143 \pi}{64}-\frac{205}{4} \rb\frac{1}{v^4}\right.\nn\\
    &
    \left.-\lb \frac{11}{8}+\frac{15 \pi}{16} \rb\frac{1}{v^6}
    +\lb \frac{3 \pi}{8}-\frac{1}{8} \rb\frac{1}{v^8}\right] \hat{q}^6,\\
R_8=
    &
    S_8+\left[-\frac{105105 \pi }{4096}
    +\lb \frac{14938}{15}-\frac{608893 \pi}{512} \rb\frac{1}{v^2}
    +\lb \frac{85417}{5}-\frac{1348113 \pi}{256} \rb\frac{1}{v^4}
    +\lb \frac{109123}{3}-\frac{766485 \pi}{64} \rb\frac{1}{v^6}\right.\nn\\
    &
    \left.+\lb \frac{189181}{6}-\frac{308185 \pi}{32} \rb\frac{1}{v^8}
    +\lb \frac{44805}{4}-3795 \pi \rb\frac{1}{v^{10}}
    +\lb \frac{221449}{120}-516 \pi \rb\frac{1}{v^{12}}
    +\lb \frac{2821}{40}-32 \pi \rb\frac{1}{v^{14}}\right]\hat{q}^2\nn\\
    &
    +\left[\frac{121275 \pi}{8192}
    +\lb \frac{301055 \pi}{512}-\frac{1396}{3} \rb\frac{1}{v^2}
    +\lb \frac{1081745 \pi}{512}-\frac{20473}{3} \rb\frac{1}{v^4}
    +\lb \frac{113345 \pi}{32}-\frac{128557}{12} \rb\frac{1}{v^6}\right.\nn\\
    &
    \left.+\lb \frac{117695 \pi}{64}-\frac{73183}{12} \rb\frac{1}{v^8}
    +\lb 320 \pi-\frac{21215}{24} \rb\frac{1}{v^{10}}
    +\lb \frac{301}{24}-10 \pi \rb\frac{1}{v^{12}}\right]\hat{q}^4\nn\\
    &
    +\left[-\frac{11025 \pi }{4096}
    +\lb 60-\frac{44751 \pi}{512} \rb\frac{1}{v^2}
    +\lb 710-\frac{28461 \pi}{128} \rb\frac{1}{v^4}
    +\lb \frac{2105}{4}-\frac{5757 \pi}{32} \rb\frac{1}{v^6}
    +\lb \frac{81}{8}+\frac{105 \pi}{32} \rb\frac{1}{v^8}\right.\nn\\
    &
    \left.+\lb \frac{7}{8}-\frac{3 \pi}{2} \rb\frac{1}{v^{10}}\right]\hat{q}^6
    +\left(\frac{1225 \pi }{16384}+\frac{803 \pi }{512 v^2}+\frac{39\pi }{512 v^4}+\frac{11 \pi }{64 v^6}-\frac{5 \pi }{128v^8}\right) \hat{q}^8,\\
R_9=
    &
    S_9+\left[-\frac{1144}{7}
    +\lb \frac{398377 \pi}{512}-\frac{993439}{105} \rb\frac{1}{v^2}
    +\lb \frac{1193677 \pi}{64}-\frac{884186}{15} \rb\frac{1}{v^4}
    +\lb \frac{1756347 \pi}{32}-\frac{1757143}{10} \rb\frac{1}{v^6}\right.\nn\\
    &
    +\lb \frac{277265 \pi}{4}-\frac{638909}{3} \rb\frac{1}{v^8}
    +\lb \frac{151885 \pi}{4}-\frac{2952011}{24} \rb\frac{1}{v^{10}}
    +\lb 11094 \pi-\frac{662131}{20} \rb\frac{1}{v^{12}}\nn\\
    &
    \left.+\lb 1232 \pi-\frac{2430129}{560} \rb\frac{1}{v^{14}}
    +\lb 64 \pi-\frac{81603}{560} \rb\frac{1}{v^{16}}\right]\hat{q}^2
    +\left[\frac{572}{5}
    +\lb \frac{29178}{5}-\frac{479403 \pi}{1024} \rb\frac{1}{v^2}\right.\nn\\
    &
    +\lb \frac{605559}{20}-\frac{610641 \pi}{64} \rb\frac{1}{v^4}
    +\lb \frac{290011}{4}-\frac{1447125 \pi}{64} \rb\frac{1}{v^6}
    +\lb \frac{513933}{8}-\frac{83985 \pi}{4} \rb\frac{1}{v^8}\nn\\
    &
    \left.+\lb \frac{963711}{40}-\frac{58815 \pi}{8} \rb\frac{1}{v^{10}}
    +\lb \frac{426581}{160}-948 \pi \rb\frac{1}{v^{12}}
    +\lb 24 \pi-\frac{5363}{160} \rb\frac{1}{v^{14}}\right]\hat{q}^4\nn\\
    &
    +\left[-\frac{88}{3}
    +\lb \frac{51385 \pi}{512}-\frac{3878}{3} \rb\frac{1}{v^2}
    +\lb \frac{101435 \pi}{64}-\frac{15251}{3} \rb\frac{1}{v^4}
    +\lb \frac{20495 \pi}{8}-\frac{66211}{8} \rb\frac{1}{v^6}\right.\nn\\
    &
    \left.+\lb \frac{9275 \pi}{8}-\frac{10331}{3} \rb\frac{1}{v^8}
    -\lb \frac{2321}{48}+\frac{35 \pi}{4} \rb\frac{1}{v^{10}}
    +\lb 5 \pi-\frac{187}{48} \rb\frac{1}{v^{12}}\right]\hat{q}^6\nn\\
    &
    +\left[2+\lb 74-\frac{11177 \pi}{2048} \rb\frac{1}{v^2}
    +\lb \frac{297}{2}-\frac{2837 \pi}{64} \rb\frac{1}{v^4}
    +\lb \frac{97}{4}-\frac{303 \pi}{64} \rb\frac{1}{v^6}
    +\lb \frac{11}{64}-\frac{13 \pi}{8} \rb\frac{1}{v^8}
    +\lb \frac{5 \pi}{16}-\frac{5}{64} \rb\frac{1}{v^{10}}\right] \hat{q}^8,\\
R_{10}=
    &
    S_{10}+\left[-\frac{6891885 \pi}{65536}
    +\lb \frac{140900}{21}-\frac{252273505 \pi}{32768} \rb\frac{1}{v^2}
    +\lb \frac{19342783}{105}-\frac{62200733 \pi}{1024} \rb\frac{1}{v^4}\right.\nn\\
    &
    +\lb \frac{11143912}{15}-\frac{122414497 \pi}{512} \rb\frac{1}{v^6}
    +\lb \frac{37511813}{30}-\frac{100535597 \pi}{256} \rb\frac{1}{v^8}
    +\lb \frac{6043237}{6}-\frac{41702245 \pi}{128} \rb\frac{1}{v^{10}}\nn\\
    &
    \left.+\lb \frac{10252363}{24}-\frac{530825 \pi}{4} \rb\frac{1}{v^{12}}
    +\lb \frac{459593}{5}-30578 \pi \rb\frac{1}{v^{14}}
    +\lb \frac{5586893}{560}-2864 \pi \rb\frac{1}{v^{16}}
    +\lb \frac{167371}{560}-128 \pi \rb\frac{1}{v^{18}}\right]\hat{q}^2\nn\\
    &
    +\left[\frac{2837835 \pi}{32768}
    +\lb \frac{11553985 \pi}{2048}-\frac{14306}{3} \rb\frac{1}{v^2}
    +\lb \frac{39435901 \pi}{1024}-\frac{1745708}{15} \rb\frac{1}{v^4}
    +\lb \frac{16127133 \pi}{128}-\frac{7822791}{20} \rb\frac{1}{v^6}\right.\nn\\
    &
    +\lb \frac{21081145 \pi}{128}-\frac{6302849}{12} \rb\frac{1}{v^8}
    +\lb \frac{1601545 \pi}{16}-\frac{7400429}{24} \rb\frac{1}{v^{10}}
    +\lb \frac{207435 \pi}{8}-\frac{674701}{8} \rb\frac{1}{v^{12}}\nn\\
    &
    \left.+\lb 2636 \pi-\frac{1201277}{160} \rb\frac{1}{v^{14}}
    +\lb \frac{13547}{160}-56 \pi \rb\frac{1}{v^{16}}\right]\hat{q}^4
    +\left[-\frac{945945 \pi }{32768}
    +\lb 1312-\frac{26840595 \pi}{16384} \rb\frac{1}{v^2}\right.\nn\\
    &
    +\lb 27705-\frac{18934095 \pi}{2048} \rb\frac{1}{v^4}
    +\lb \frac{140245}{2}-\frac{23203875 \pi}{1024} \rb\frac{1}{v^6}
    +\lb \frac{502879}{8}-\frac{5010525 \pi}{256} \rb\frac{1}{v^8}\nn\\
    &
    \left.+\lb \frac{69555}{4}-\frac{740715 \pi}{128} \rb\frac{1}{v^{10}}
    +\lb \frac{3007}{16}+\frac{75 \pi}{4} \rb\frac{1}{v^{12}}
    +\lb \frac{225}{16}-15 \pi \rb\frac{1}{v^{14}}\right]\hat{q}^6
    +\left[\frac{218295 \pi }{65536}
    +\lb \frac{2587225 \pi}{16384}-110 \rb\frac{1}{v^2}\right.\nn\\
    &
    +\lb \frac{2671225 \pi}{4096}-1910 \rb\frac{1}{v^4}
    +\lb \frac{425675 \pi}{512}-\frac{10105}{4} \rb\frac{1}{v^6}
    +\lb \frac{37975 \pi}{512}-\frac{613}{2} \rb\frac{1}{v^8}
    +\lb \frac{1195 \pi}{128}-\frac{131}{64} \rb\frac{1}{v^{10}}\nn\\
    &
    \left.+\lb \frac{45}{64}-\frac{25 \pi}{16} \rb\frac{1}{v^{12}}\right]\hat{q}^8
    -\left(\frac{3969 \pi }{65536}
    +\frac{62417 \pi }{32768v^2}
    +\frac{1507 \pi }{2048 v^4}+\frac{381 \pi }{1024 v^6}
    -\frac{77\pi }{512 v^8}+\frac{7 \pi }{256v^{10}}\right) \hat{q}^{10},\\
R_{11}=
    &
    S_{11}+\left[-\frac{41990}{63}
    +\lb \frac{163920193 \pi}{32768}-\frac{18362717}{315} \rb\frac{1}{v^2}
    +\lb \frac{2939299557 \pi}{16384}-\frac{20900769}{35} \rb\frac{1}{v^4}\right.\nn\\
    &
    +\lb \frac{471598515 \pi}{512}-\frac{122425265}{42} \rb\frac{1}{v^6}
    +\lb \frac{523008717 \pi}{256}-\frac{381772261}{60} \rb\frac{1}{v^8}
    +\lb \frac{282754857 \pi}{128}-\frac{280188473}{40} \rb\frac{1}{v^{10}}\nn\\
    &
    +\lb \frac{84979881 \pi}{64}-\frac{164824687}{40} \rb\frac{1}{v^{12}}
    +\lb 426027 \pi-\frac{2294376137}{1680} \rb\frac{1}{v^{14}}
    +\lb 80700-\frac{109249577}{448} \pi \rb\frac{1}{v^{16}}\nn\\
    &
    \left.+\lb 6528 \pi-\frac{909081107}{40320} \rb\frac{1}{v^{18}}
    +\lb 256 \pi-\frac{24593209}{40320} \rb\frac{1}{v^{20}}\right]\hat{q}^2
    +\left[\frac{4420}{7}
    +\lb \frac{748997}{15}-\frac{69100279 \pi}{16384} \rb\frac{1}{v^2}\right.\nn\\
    &
    +\lb \frac{188316941}{420}-\frac{137496109 \pi}{1024} \rb\frac{1}{v^4}
    +\lb \frac{22599985}{12}-\frac{76127315 \pi}{128} \rb\frac{1}{v^6}
    +\lb \frac{408829123}{120}-\frac{70065547 \pi}{64} \rb\frac{1}{v^8}\nn\\
    &
    +\lb \frac{35713057}{12}-\frac{59988655 \pi}{64} \rb\frac{1}{v^{10}}
    +\lb \frac{611485309}{480}-\frac{3294017 \pi}{8} \rb\frac{1}{v^{12}}
    +\lb \frac{303602051}{1120}-\frac{167543 \pi}{2} \rb\frac{1}{v^{14}}\nn\\
    &
    \left.+\lb \frac{1288199}{64}-7000 \pi \rb\frac{1}{v^{16}}
    +\lb -\frac{460919}{2240}+128 \pi\rb\frac{1}{v^{18}}\right]\hat{q}^4
    +\left[-260
    +\lb \frac{24935133 \pi}{16384}-\frac{91848}{5} \rb\frac{1}{v^2}\right.\nn\\
    &
    +\lb \frac{339175455 \pi}{8192}-\frac{556915}{4} \rb\frac{1}{v^4}
    +\lb \frac{153001887 \pi}{1024}-\frac{18959671}{40} \rb\frac{1}{v^6}
    +\lb \frac{105099615 \pi}{512}-\frac{5096833}{8} \rb\frac{1}{v^8}\nn\\
    &
    +\lb \frac{14695425 \pi}{128}-\frac{5865715}{16} \rb\frac{1}{v^{10}}
    +\lb \frac{1567767\pi}{64}-\frac{11871201}{160} \rb\frac{1}{v^{12}}
    -\lb \frac{41269}{64}+30 \pi \rb\frac{1}{v^{14}}\nn\\
    &
    \left.+\lb 42 \pi-\frac{14363}{320} \rb\frac{1}{v^{16}}\right]\hat{q}^6
    +\left[\frac{130}{3}
    +\lb \frac{8080}{3}-\frac{7128915 \pi}{32768} \rb\frac{1}{v^2}
    +\lb \frac{31615}{2}-\frac{37868925 \pi}{8192} \rb\frac{1}{v^4}\right.\nn\\
    &
    +\lb \frac{896225}{24}-\frac{23959375 \pi}{2048} \rb\frac{1}{v^6}
    +\lb \frac{4992479}{192}-\frac{2175535 \pi}{256} \rb\frac{1}{v^8}
    +\lb \frac{147215}{64}-\frac{156075 \pi}{256} \rb\frac{1}{v^{10}}\nn\\
    &
    \left.+\lb \frac{5305}{384}-\frac{2695 \pi}{64} \rb\frac{1}{v^{12}}
    +\lb \frac{25 \pi}{4}-\frac{1475}{384} \rb\frac{1}{v^{14}}\right]\hat{q}^8
    +\left[-2
    +\lb \frac{272293 \pi}{32768}-107 \rb\frac{1}{v^2}
    +\lb \frac{1524629 \pi}{16384}-\frac{685}{2} \rb\frac{1}{v^4}\right.\nn\\
    &
    \left.+\lb \frac{36479 \pi}{1024}-134 \rb\frac{1}{v^6}
    +\lb \frac{2897 \pi}{512}-\frac{191}{64} \rb\frac{1}{v^8}
    +\lb \frac{35}{128}-\frac{449 \pi}{256} \rb\frac{1}{v^{10}}
    +\lb \frac{35 \pi}{128}-\frac{7}{128} \rb\frac{1}{v^{12}}\right]\hat{q}^{10},\\
R_{12}=
    &
    S_{12}+\left[-\frac{224062839 \pi}{524288}
    +\lb \frac{12782582}{315}-\frac{5943708049 \pi}{131072} \rb\frac{1}{v^2}
    +\lb \frac{512938303}{315}-\frac{36689187229 \pi}{65536} \rb\frac{1}{v^4}\right.\nn\\
    &
    +\lb \frac{371978057}{35}-\frac{3484405131 \pi}{1024} \rb\frac{1}{v^6}
    +\lb \frac{6215189867}{210}-\frac{19175713287 \pi}{2048} \rb\frac{1}{v^8}
    +\lb \frac{2518076153}{60}-\frac{6878821887 \pi}{512} \rb\frac{1}{v^{10}}\nn\\
    &
    +\lb \frac{1329853851}{40}-\frac{2690974671 \pi}{256} \rb\frac{1}{v^{12}}
    +\lb \frac{1822427989}{120}-\frac{156211587 \pi}{32} \rb\frac{1}{v^{14}}
    +\lb \frac{6896200111}{1680}-\frac{2568507 \pi}{2} \rb\frac{1}{v^{16}}\nn\\
    &
    \left.+\lb \frac{1400016273}{2240}-206016 \pi \rb\frac{1}{v^{18}}
    +\lb \frac{2027115781}{40320}-14656 \pi \rb\frac{1}{v^{20}}
    +\lb \frac{50011907}{40320}-512 \pi \rb\frac{1}{v^{22}}\right]\hat{q}^2\nn\\
    &
    +\left[\frac{480134655 \pi}{1048576}
    +\lb \frac{5796051231 \pi}{131072}-\frac{1358414}{35} \rb\frac{1}{v^2}
    +\lb \frac{64014853131 \pi}{131072}-\frac{49590969}{35} \rb\frac{1}{v^4}\right.\nn\\
    &
    +\lb \frac{5323920879 \pi}{2048}-\frac{1136109421}{140} \rb\frac{1}{v^6}
    +\lb \frac{25106856243 \pi}{4096}-\frac{387671533}{20} \rb\frac{1}{v^8}
    +\lb \frac{3726840411 \pi}{512}-\frac{908981301}{40} \rb\frac{1}{v^{10}}\nn\\
    &
    +\lb \frac{2302073949 \pi}{512}-\frac{71179552}{5} \rb\frac{1}{v^{12}}
    +\lb \frac{6094473 \pi}{4}-\frac{756404989}{160} \rb\frac{1}{v^{14}}
    +\lb \frac{1014813 \pi}{4}-\frac{916040073}{1120} \rb\frac{1}{v^{16}}\nn\\
    &
    \left.+\lb 17952 \pi-\frac{116489451}{2240} \rb\frac{1}{v^{18}}
    +\lb \frac{1089209}{2240}-288 \pi \rb\frac{1}{v^{20}}\right]\hat{q}^4
    +\left[-\frac{58963905 \pi}{262144}
    +\lb \frac{249742}{15}-\frac{1279839941 \pi}{65536} \rb\frac{1}{v^2}\right.\nn\\
    &
    +\lb \frac{8167507}{15}-\frac{772894151 \pi}{4096} \rb\frac{1}{v^4}
    +\lb \frac{31483645}{12}-\frac{3446946385 \pi}{4096} \rb\frac{1}{v^6}
    +\lb \frac{604403483}{120}-\frac{1628991317 \pi}{1024} \rb\frac{1}{v^8}\nn\\
    &
    +\lb \frac{104767705}{24}-\frac{179329075 \pi}{128} \rb\frac{1}{v^{10}}
    +\lb \frac{429033197}{240}-\frac{4495667 \pi}{8} \rb\frac{1}{v^{12}}
    +\lb \frac{45035013}{160}-\frac{2957413 \pi}{32} \rb\frac{1}{v^{14}}\nn\\
    &
    \left.+\lb \frac{130107}{64}+\frac{35 \pi}{2} \rb\frac{1}{v^{16}}
    +\lb \frac{42273}{320}-112 \pi \rb\frac{1}{v^{18}}\right]\hat{q}^6
    +\left[\frac{52026975 \pi}{1048576}
    +\lb \frac{62279205 \pi}{16384}-\frac{9226}{3} \rb\frac{1}{v^2}\right.\nn\\
    &
    +\lb \frac{1009997265 \pi}{32768}-\frac{263849}{3} \rb\frac{1}{v^4}
    +\lb \frac{428673375 \pi}{4096}-\frac{1300655}{4} \rb\frac{1}{v^6}
    +\lb \frac{549174805 \pi}{4096}-\frac{10230721}{24} \rb\frac{1}{v^8}\nn\\
    &
    \left.+\lb \frac{996395 \pi}{16}-\frac{36862897}{192} \rb\frac{1}{v^{10}}
    +\lb \frac{927915 \pi}{256}-\frac{835753}{64} \rb\frac{1}{v^{12}}
    +\lb \frac{5245 \pi}{32}-\frac{27007}{384} \rb\frac{1}{v^{14}}
    +\lb \frac{6325}{384}-\frac{175 \pi}{8} \rb\frac{1}{v^{16}}\right]\hat{q}^8\nn\\
    &
    +\left[-\frac{2081079 \pi }{524288}
    +\lb 182-\frac{33843765 \pi}{131072} \rb\frac{1}{v^2}
    +\lb 4347-\frac{102982095 \pi}{65536} \rb\frac{1}{v^4}
    +\lb \frac{36015}{4}-\frac{12058815 \pi}{4096} \rb\frac{1}{v^6}\right.\nn\\
    &
    \left.+\lb \frac{20825}{8}-\frac{3112665 \pi}{4096} \rb\frac{1}{v^8}
    +\lb \frac{2955}{64}-\frac{51669 \pi}{1024} \rb\frac{1}{v^{10}}
    +\lb \frac{6153 \pi}{512}-\frac{453}{128} \rb\frac{1}{v^{12}}
    +\lb \frac{77}{128}-\frac{105 \pi}{64} \rb\frac{1}{v^{14}}\right]\hat{q}^{10}\nn\\
    &
    +\left(\frac{53361 \pi }{1048576}+\frac{293327 \pi }{131072v^2}
    +\frac{260087 \pi }{131072 v^4}+\frac{3591 \pi }{4096v^6}
    -\frac{3053 \pi }{8192 v^8}+\frac{137 \pi }{1024v^{10}}
    -\frac{21 \pi }{1024v^{12}}\right) \hat{q}^{12},\\
R_{13}=
    &
    S_{13}+\left[-\frac{29716}{11}
    +\lb \frac{3850927981\pi }{131072}-\frac{1151967139}{3465} \rb\frac{1}{v^2}
    +\lb \frac{48241418323\pi }{32768}-\frac{228157844}{45} \rb\frac{1}{v^4}\right.\nn\\
    &
    +\lb \frac{191009472785\pi }{16384}-\frac{4641305429}{126} \rb\frac{1}{v^6}
    +\lb \frac{165281509305\pi }{4096}-\frac{5299812025}{42} \rb\frac{1}{v^8}\nn\\
    &
    +\lb \frac{36786871665\pi }{512}-\frac{38062290169}{168} \rb\frac{1}{v^{10}}
    +\lb \frac{9406221615\pi }{128}-\frac{6899028911}{30} \rb\frac{1}{v^{12}}\nn\\
    &
    +\lb \frac{2832936855\pi }{64}-\frac{33546130901}{240} \rb\frac{1}{v^{14}}
    +\lb \frac{265688655\pi }{16}-\frac{34774883081}{672} \rb\frac{1}{v^{16}}
    +\lb 3688440 \pi-\frac{4517419595}{384} \rb\frac{1}{v^{18}}\nn\\
    &
    \left.+\lb 512320 \pi-\frac{3143099753}{2016} \rb\frac{1}{v^{20}}
    +\lb 32512 \pi-\frac{98378658109}{887040} \rb\frac{1}{v^{22}}
    +\lb 1024 \pi-\frac{2231583223}{887040} \rb\frac{1}{v^{24}}\right]\hat{q}^2\nn\\
    &
    +\left[3230
    +\lb \frac{23031478}{63}-\frac{8375639885 \pi}{262144} \rb\frac{1}{v^2}
    +\lb \frac{181003895}{36}-\frac{47726735105 \pi}{32768} \rb\frac{1}{v^4}\right.\nn\\
    &
    +\lb \frac{130318675}{4}-\frac{337724293305 \pi}{32768} \rb\frac{1}{v^6}
    +\lb \frac{16393662091}{168}-\frac{15980962035 \pi}{512} \rb\frac{1}{v^8}\nn\\
    &
    +\lb \frac{3594968549}{24}-\frac{48629763105 \pi}{1024} \rb\frac{1}{v^{10}}
    +\lb \frac{4015888405}{32}-\frac{5135347245 \pi}{128} \rb\frac{1}{v^{12}}
    +\lb \frac{5778014123}{96}-\frac{2437556415 \pi}{128} \rb\frac{1}{v^{14}}\nn\\
    &
    +\lb \frac{10879774471}{672}-\frac{10414935 \pi}{2} \rb\frac{1}{v^{16}}
    +\lb \frac{9402843}{4}-731220 \pi \rb\frac{1}{v^{18}}
    +\lb \frac{601597085}{4608}-44800 \pi\rb\frac{1}{v^{20}}\nn\\
    &
    \left.+\lb 640 \pi-\frac{36287861}{32256} \rb\frac{1}{v^{22}}\right]\hat{q}^4
    +\left[-\frac{12920}{7}
    +\lb \frac{1077663819 \pi }{65536}-\frac{6670747}{35} \rb\frac{1}{v^2}
    +\lb \frac{10948327239\pi }{16384}-\frac{162214931}{70} \rb\frac{1}{v^4}\right.\nn\\
    &
    +\lb \frac{16827455973\pi }{4096}-\frac{3639935497}{280} \rb\frac{1}{v^6}
    +\lb \frac{21287652225 \pi}{2048}-\frac{129901095}{4} \rb\frac{1}{v^8}
    +\lb \frac{1612432311 \pi}{128}-\frac{3181189483}{80} \rb\frac{1}{v^{10}}\nn\\
    &
    +\lb \frac{1006063863 \pi}{128}-\frac{1963895541}{80} \rb\frac{1}{v^{12}}
    +\lb \frac{77255493 \pi}{32}-\frac{17160260947}{2240} \rb\frac{1}{v^{14}}
    +\lb \frac{5124657 \pi}{16}-\frac{1096880473}{1120} \rb\frac{1}{v^{16}}\nn\\
    &
    \left.+\lb 120 \pi-\frac{5399973}{896} \rb\frac{1}{v^{18}}
    +\lb 288 \pi-\frac{1644563}{4480} \rb\frac{1}{v^{20}}\right]\hat{q}^6
    +\left[510
    +\lb \frac{713684}{15}-\frac{1059403293 \pi}{262144} \rb\frac{1}{v^2}\right.\nn\\
    &
    +\lb \frac{29624063}{60}-\frac{144481149 \pi}{1024} \rb\frac{1}{v^4}
    +\lb \frac{135966431}{60}-\frac{2926597377 \pi}{4096} \rb\frac{1}{v^6}
    +\lb \frac{820784273}{192}-\frac{2807125715 \pi}{2048} \rb\frac{1}{v^8}\nn\\
    &
    +\lb \frac{677092403}{192}-\frac{1140717145 \pi}{1024} \rb\frac{1}{v^{10}}
    +\lb \frac{2176275617}{1920}-\frac{23421157 \pi}{64} \rb\frac{1}{v^{12}}
    +\lb \frac{118899191}{1920}-\frac{35323 \pi}{2} \rb\frac{1}{v^{14}}\nn\\
    &
    \left.+\lb \frac{2312843}{7680}-\frac{9233 \pi}{16} \rb\frac{1}{v^{16}}
    +\lb 70 \pi-\frac{93689}{1536} \rb\frac{1}{v^{18}}\right]\hat{q}^8
    +\left[-60
    +\lb \frac{54476685 \pi}{131072}-\frac{14983}{3} \rb\frac{1}{v^2}\right.\nn\\
    &
    +\lb \frac{374307375 \pi}{32768}-\frac{122459}{3} \rb\frac{1}{v^4}
    +\lb \frac{680076975 \pi}{16384}-\frac{1060425}{8} \rb\frac{1}{v^6}
    +\lb \frac{183845005 \pi}{4096}-\frac{13331887}{96} \rb\frac{1}{v^8}\nn\\
    &
    \left.+\lb \frac{8831615 \pi}{1024}-\frac{10954055}{384} \rb\frac{1}{v^{10}}
    +\lb \frac{85593 \pi}{256}-\frac{6447}{16} \rb\frac{1}{v^{12}}
    +\lb \frac{20209}{768}-\frac{8069 \pi}{128} \rb\frac{1}{v^{14}}
    +\lb \frac{245 \pi}{32}-\frac{2989}{768} \rb\frac{1}{v^{16}}\right]\hat{q}^{10}\nn\\
    &
    +\left[2
    +\lb 146-\frac{3086115 \pi}{262144} \rb\frac{1}{v^2}
    +\lb \frac{2729}{4}-\frac{5704785 \pi}{32768} \rb\frac{1}{v^4}
    +\lb \frac{971}{2}-\frac{4412325 \pi}{32768} \rb\frac{1}{v^6}
    +\lb \frac{2725}{64}-\frac{25545 \pi}{1024} \rb\frac{1}{v^8}\right.\nn\\
    &
    \left.+\lb \frac{12375 \pi}{2048}-\frac{61}{64} \rb\frac{1}{v^{10}}
    +\lb \frac{139}{512}-\frac{471 \pi}{256} \rb\frac{1}{v^{12}}
    +\lb \frac{63 \pi}{256}-\frac{21}{512} \rb\frac{1}{v^{14}}\right]\hat{q}^{12},\\
R_{14}=
    &
    S_{14}+\left[-\frac{7250518275 \pi}{4194304}
    +\lb \frac{792361816}{3465}-\frac{525797225509 \pi}{2097152} \rb\frac{1}{v^2}
    +\lb \frac{4856676007}{385}-\frac{1169631625329\pi }{262144} \rb\frac{1}{v^4}\right.\nn\\
    &
    +\lb \frac{4243451234}{35}-\frac{5090627166819\pi}{131072} \rb\frac{1}{v^6}
    +\lb \frac{63786336433}{126}-\frac{5260989742955\pi }{32768} \rb\frac{1}{v^8}\nn\\
    &
    +\lb \frac{7796795759}{7}-\frac{5823433041435\pi}{16384} \rb\frac{1}{v^{10}}
    +\lb \frac{1191668102257}{840}-\frac{461317265205\pi }{1024} \rb\frac{1}{v^{12}}\nn\\
    &
    +\lb \frac{65906638889}{60}-\frac{179522771715\pi }{512} \rb\frac{1}{v^{14}}
    +\lb \frac{300070641453}{560}-\frac{10871742375\pi }{64} \rb\frac{1}{v^{16}}\nn\\
    &
    +\lb \frac{55741051801}{336}-\frac{212679975 \pi}{4} \rb\frac{1}{v^{18}}
    +\lb \frac{261783760691}{8064}-10195000 \pi \rb\frac{1}{v^{20}}
    +\lb \frac{8523820511}{2240}-1247328 \pi \rb\frac{1}{v^{22}}\nn\\
    &
    \left.+\lb \frac{23901371037}{98560}-71424 \pi \rb\frac{1}{v^{24}}
    +\lb \frac{4517099171}{887040}-2048 \pi \rb\frac{1}{v^{26}}\right]\hat{q}^2
    +\left[\frac{9570684123 \pi}{4194304}\right.\nn\\
    &
    +\lb \frac{160335967457 \pi}{524288}-\frac{86685352}{315} \rb\frac{1}{v^2}
    +\lb \frac{162874722689 \pi}{32768}-\frac{631164052}{45} \rb\frac{1}{v^4}\nn\\
    &
    +\lb \frac{2556201217555\pi }{65536}-\frac{30670759493}{252} \rb\frac{1}{v^6}
    +\lb \frac{1175361387975\pi }{8192}-\frac{38007593227}{84} \rb\frac{1}{v^8}\nn\\
    &
    +\lb \frac{568042018569\pi}{2048}-\frac{730006818661}{840} \rb\frac{1}{v^{10}}
    +\lb \frac{76757077845\pi}{256}-\frac{113323685947}{120} \rb\frac{1}{v^{12}}\nn\\
    &
    +\lb \frac{49279642545\pi }{256}-\frac{289364227151}{480} \rb\frac{1}{v^{14}}
    +\lb \frac{9394850625\pi }{128}-\frac{155704450201}{672} \rb\frac{1}{v^{16}}\nn\\
    &
    +\lb 16734270 \pi-\frac{10003627007}{192} \rb\frac{1}{v^{18}}
    +\lb 2026700 \pi-\frac{131064025129}{20160} \rb\frac{1}{v^{20}}
    +\lb 109376 \pi-\frac{51659391179}{161280} \rb\frac{1}{v^{22}}\nn\\
    &
    \left.+\lb \frac{412890517}{161280}-1408 \pi \rb\frac{1}{v^{24}}\right]\hat{q}^4
    +\left[-\frac{6241750515 \pi}{4194304}
    +\lb \frac{1130544}{7}-\frac{383387191875 \pi}{2097152} \rb\frac{1}{v^2}\right.\nn\\
    &
    +\lb \frac{52711116}{7}-\frac{1404145976715 \pi}{524288} \rb\frac{1}{v^4}
    +\lb \frac{403796016}{7}-\frac{4850329953645 \pi}{262144} \rb\frac{1}{v^6}\nn\\
    &
    +\lb \frac{10349663997}{56}-\frac{959824412775 \pi}{16384} \rb\frac{1}{v^8}
    +\lb \frac{589943295}{2}-\frac{771496521075 \pi}{8192} \rb\frac{1}{v^{10}}\nn\\
    &
    +\lb \frac{4072160885}{16}-\frac{165394088685 \pi}{2048} \rb\frac{1}{v^{12}}
    +\lb \frac{955615061}{8}-\frac{39106929015 \pi}{1024} \rb\frac{1}{v^{14}}
    +\lb \frac{13344454951}{448}-\frac{150445935 \pi}{16} \rb\frac{1}{v^{16}}\nn\\
    &
    \left.+\lb \frac{357227863}{112}-\frac{4160025 \pi}{4} \rb\frac{1}{v^{18}}
    +\lb \frac{15276333}{896}-720 \pi \rb\frac{1}{v^{20}}
    +\lb \frac{875667}{896}-720 \pi \rb\frac{1}{v^{22}}\right]\hat{q}^6\nn\\
    &
    +\left[\frac{2080583505 \pi}{4194304}
    +\lb \frac{57935094707 \pi}{1048576}-\frac{237308}{5} \rb\frac{1}{v^2}
    +\lb \frac{374702435457 \pi}{524288}-\frac{9981196}{5} \rb\frac{1}{v^4}\right.\nn\\
    &
    +\lb \frac{136647071953 \pi}{32768}-\frac{194672377}{15} \rb\frac{1}{v^6}
    +\lb \frac{174933658487 \pi}{16384}-\frac{505682821}{15} \rb\frac{1}{v^8}
    +\lb \frac{52882558365 \pi}{4096}-\frac{2585294483}{64} \rb\frac{1}{v^{10}}\nn\\
    &
    +\lb \frac{15116615899\pi }{2048}-\frac{22379022871}{960} \rb\frac{1}{v^{12}}
    +\lb \frac{470134931\pi }{256}-\frac{10952801069}{1920} \rb\frac{1}{v^{14}}
    +\lb \frac{4806921 \pi}{64}-\frac{165596657}{640} \rb\frac{1}{v^{16}}\nn\\
    &
    \left.+\lb \frac{7553 \pi}{4}-\frac{2933589}{2560} \rb\frac{1}{v^{18}}
    +\lb \frac{314197}{1536}-210 \pi \rb\frac{1}{v^{20}}\right]\hat{q}^8
    +\left[-\frac{328513185 \pi}{4194304}
    +\lb \frac{19040}{3}-\frac{16311312075 \pi}{2097152} \rb\frac{1}{v^2}\right.\nn\\
    &
    +\lb \frac{706718}{3}-\frac{5601268635 \pi}{65536} \rb\frac{1}{v^4}
    +\lb \frac{3581618}{3}-\frac{25247014965 \pi}{65536} \rb\frac{1}{v^6}
    +\lb \frac{12992801}{6}-\frac{5603612095 \pi}{8192} \rb\frac{1}{v^8}\nn\\
    &
    +\lb \frac{69497161}{48}-\frac{3809844805 \pi}{8192} \rb\frac{1}{v^{10}}
    +\lb \frac{86064347}{384}-\frac{35243623 \pi}{512} \rb\frac{1}{v^{12}}
    +\lb \frac{498469}{192}-\frac{931919 \pi}{512} \rb\frac{1}{v^{14}}\nn\\
    &
    \left.+\lb \frac{35833 \pi}{128}-\frac{113273}{768} \rb\frac{1}{v^{16}}
    +\lb \frac{14833}{768}-\frac{245 \pi}{8} \rb\frac{1}{v^{18}}\right]\hat{q}^{10}
    +\left[\frac{19324305 \pi }{4194304}
    +\lb \frac{6446055\pi }{16384}-280 \rb\frac{1}{v^2}\right.\nn\\
    &
    +\lb \frac{867597255 \pi}{262144}-8792 \rb\frac{1}{v^4}
    +\lb \frac{282541035 \pi}{32768}-\frac{52521}{2} \rb\frac{1}{v^6}
    +\lb \frac{137654265 \pi}{32768}-\frac{27475}{2} \rb\frac{1}{v^8}\nn\\
    &
    \left.+\lb \frac{422415 \pi}{1024}-\frac{15205}{16} \rb\frac{1}{v^{10}}
    +\lb \frac{535}{32}-\frac{113043 \pi}{2048} \rb\frac{1}{v^{12}}
    +\lb \frac{3717 \pi}{256}-\frac{2063}{512} \rb\frac{1}{v^{14}}
    +\lb \frac{273}{512}-\frac{441 \pi}{256} \rb\frac{1}{v^{16}}\right]\hat{q}^{12}\nn\\
    &
    -\left(\frac{184041 \pi }{4194304}+\frac{5386925 \pi}{2097152 v^2}
    +\frac{2087835 \pi }{524288 v^4}+\frac{570545 \pi}{262144 v^6}
    -\frac{23525 \pi }{32768 v^8}+\frac{6507 \pi }{16384v^{10}}
    -\frac{497 \pi }{4096 v^{12}}+\frac{33 \pi }{2048v^{14}}\right) \hat{q}^{14},\\
R_{15}=
    &
    S_{15}+\left[-\frac{1560090}{143}
    +\lb \frac{340293987205 \pi}{2097152}-\frac{1471598113}{819} \rb\frac{1}{v^2}
    +\lb \frac{11350546615939 \pi}{1048576}-\frac{132311745437}{3465} \rb\frac{1}{v^4}\right.\nn\\
    &
    +\lb \frac{16026424645869 \pi}{131072}-\frac{297995777173}{770} \rb\frac{1}{v^6}
    +\lb \frac{40025473857289 \pi}{65536}-\frac{2410345554823}{1260} \rb\frac{1}{v^8}\nn\\
    &
    +\lb \frac{26525655454535 \pi}{16384}-\frac{2567963506223}{504} \rb\frac{1}{v^{10}}
    +\lb \frac{20697278388165 \pi}{8192}-\frac{443828804171}{56} \rb\frac{1}{v^{12}}\nn\\
    &
    +\lb \frac{1243472226345 \pi}{512}-\frac{12838948917233}{1680} \rb\frac{1}{v^{14}}
    +\lb \frac{385595524305 \pi}{256}-\frac{31730749100761}{6720} \rb\frac{1}{v^{16}}\nn\\
    &
    +\lb 606156210 \pi-\frac{8559819989043}{4480} \rb\frac{1}{v^{18}}
    +\lb \frac{324415955 \pi}{2}-\frac{4084899351821}{8064} \rb\frac{1}{v^{20}}\nn\\
    &
    +\lb 27317120 \pi-\frac{15411846469111}{177408} \rb\frac{1}{v^{22}}
    +\lb 2984256 \pi-\frac{163480892411}{17920} \rb\frac{1}{v^{24}}
    +\lb 155648 \pi-\frac{24279041946077}{46126080} \rb\frac{1}{v^{26}}\nn\\
    &
    \left.+\lb 4096 \pi-\frac{474709559719}{46126080} \rb\frac{1}{v^{28}}\right]\hat{q}^2
    +\left[\frac{520030}{33}
    +\lb \frac{558094393}{231}-\frac{453796829565 \pi}{2097152} \rb\frac{1}{v^2}\right.\nn\\
    &
    +\lb \frac{10373089567}{220}-\frac{3497652641007 \pi}{262144} \rb\frac{1}{v^4}
    +\lb \frac{182975695523}{420}-\frac{18034226099139 \pi}{131072} \rb\frac{1}{v^6}\nn\\
    &
    +\lb \frac{325825175491}{168}-\frac{20291680546395 \pi}{32768}\rb\frac{1}{v^8}
    +\lb \frac{128507228313}{28}-\frac{23891530102155 \pi}{16384} \rb\frac{1}{v^{10}}\nn\\
    &
    +\lb \frac{1394216730049}{224}-\frac{2031978790035 \pi}{1024} \rb\frac{1}{v^{12}}
    +\lb \frac{818119180837}{160}-\frac{831854452275 \pi}{512} \rb\frac{1}{v^{14}}\nn\\
    &
    +\lb \frac{2914791387891}{1120}-\frac{106293678165 \pi}{128} \rb\frac{1}{v^{16}}
    +\lb \frac{318286394221}{384}-\frac{2102223375 \pi}{8} \rb\frac{1}{v^{18}}\nn\\
    &
    +\lb \frac{1715720714479}{10752}-51196380 \pi \rb\frac{1}{v^{20}}
    +\lb \frac{687075808093}{39424}-5442480 \pi \rb\frac{1}{v^{22}}
    +\lb \frac{912156014347}{1182720}-262272 \pi \rb\frac{1}{v^{24}}\nn\\
    &
    \left.+\lb 3072 \pi-\frac{6799588657}{1182720} \rb\frac{1}{v^{26}}\right]\hat{q}^4
    +\left[-\frac{104006}{9}
    +\lb \frac{305152568245 \pi}{2097152}-\frac{14758213}{9} \rb\frac{1}{v^2}\right.\nn\\
    &
    +\lb \frac{8587738068725 \pi}{1048576}-\frac{7316477575}{252} \rb\frac{1}{v^4}
    +\lb \frac{19882426496725 \pi}{262144}-\frac{121126993285}{504} \rb\frac{1}{v^6}\nn\\
    &
    +\lb \frac{39315663524235 \pi}{131072}-\frac{22541416171}{24} \rb\frac{1}{v^8}
    +\lb \frac{4961161009785 \pi}{8192}-\frac{640557093557}{336} \rb\frac{1}{v^{10}}\nn\\
    &
    +\lb \frac{2798348743365 \pi}{4096}-\frac{205667335109}{96} \rb\frac{1}{v^{12}}
    +\lb \frac{453138674835 \pi}{1024}-\frac{267516375655}{192} \rb\frac{1}{v^{14}}\nn\\
    &
    +\lb \frac{85142511825 \pi}{512}-\frac{12494746835}{24} \rb\frac{1}{v^{16}}
    +\lb \frac{135735015 \pi}{4}-\frac{432904194095}{4032} \rb\frac{1}{v^{18}}
    +\lb \frac{6414845 \pi}{2}-\frac{318024721277}{32256} \rb\frac{1}{v^{20}}\nn\\
    &
    \left.+\lb 2800 \pi-\frac{428420209}{9216} \rb\frac{1}{v^{22}}
    +\lb 1760 \pi-\frac{162383909}{64512} \rb\frac{1}{v^{24}}\right]\hat{q}^6
    +\left[4522
    +\lb \frac{4129017}{7}-\frac{108590638615 \pi}{2097152} \rb\frac{1}{v^2}\right.\nn\\
    &
    +\lb \frac{260396503}{28}-\frac{1367748775685 \pi}{524288} \rb\frac{1}{v^4}
    +\lb \frac{3754420781}{56}-\frac{5538552907005 \pi}{262144} \rb\frac{1}{v^6}\nn\\
    &
    +\lb \frac{295741006457}{1344}-\frac{1151898226415 \pi}{16384} \rb\frac{1}{v^8}
    +\lb \frac{68925037933}{192}-\frac{933776901575 \pi}{8192} \rb\frac{1}{v^{10}}\nn\\
    &
    +\lb \frac{38600073859}{128}-\frac{197097035835 \pi}{2048} \rb\frac{1}{v^{12}}
    +\lb \frac{50159186077}{384}-\frac{42409353455 \pi}{1024} \rb\frac{1}{v^{14}}\nn\\
    &
    +\lb \frac{273972066611}{10752}-\frac{1047962285 \pi}{128} \rb\frac{1}{v^{16}}
    +\lb \frac{3526526387}{3584}-\frac{2313375 \pi}{8} \rb\frac{1}{v^{18}}
    +\lb \frac{86040151}{21504}-\frac{11685 \pi}{2} \rb\frac{1}{v^{20}}\nn\\
    &
    \left.+\lb 600 \pi-\frac{13731205}{21504} \rb\frac{1}{v^{22}}\right]\hat{q}^8
    +\left[-\frac{4522}{5}
    +\lb \frac{19544855799 \pi}{2097152}-\frac{537803}{5} \rb\frac{1}{v^2}
    +\lb \frac{425205679437 \pi}{1048576}-\frac{29216037}{20} \rb\frac{1}{v^4}\right.\nn\\
    &
    +\lb \frac{179001734037 \pi}{65536}-\frac{347709929}{40} \rb\frac{1}{v^6}
    +\lb \frac{229155751461 \pi}{32768}-\frac{6996174717}{320} \rb\frac{1}{v^8}\nn\\
    &
    +\lb \frac{63860737185 \pi}{8192}-\frac{15735960657}{640} \rb\frac{1}{v^{10}}
    +\lb \frac{15145432575 \pi}{4096}-\frac{7388608919}{640} \rb\frac{1}{v^{12}}
    +\lb \frac{223610079 \pi}{512}-\frac{1804617143}{1280} \rb\frac{1}{v^{14}}\nn\\
    &
    +\lb \frac{2203299 \pi}{256}-\frac{35225211}{2560} \rb\frac{1}{v^{16}}
    \left.+\lb \frac{3543713}{5120}-\frac{4431 \pi}{4} \rb\frac{1}{v^{18}}
    +\lb \frac{441 \pi}{4}-\frac{416381}{5120} \rb\frac{1}{v^{20}}\right]\hat{q}^{10}\nn\\
    &
    +\left[\frac{238}{3}
    +\lb \frac{25543}{3}-\frac{1521029175 \pi}{2097152} \rb\frac{1}{v^2}
    +\lb \frac{1108009}{12}-\frac{819930945 \pi}{32768} \rb\frac{1}{v^4}
    +\lb \frac{9491881}{24}-\frac{8077077115 \pi}{65536} \rb\frac{1}{v^6}\right.\nn\\
    &
    +\lb \frac{111721631}{192}-\frac{382959035 \pi}{2048} \rb\frac{1}{v^8}
    +\lb \frac{81368179}{384}-\frac{541470785 \pi}{8192} \rb\frac{1}{v^{10}}
    +\lb \frac{5892029}{512}-\frac{1145015 \pi}{256} \rb\frac{1}{v^{12}}\nn\\
    &
    \left.+\lb \frac{191099 \pi}{512}-\frac{248939}{1536} \rb\frac{1}{v^{14}}
    +\lb \frac{103985}{3072}-\frac{1379 \pi}{16} \rb\frac{1}{v^{16}}
    +\lb \frac{147 \pi}{16}-\frac{4081}{1024} \rb\frac{1}{v^{18}}\right]\hat{q}^{12}\nn\\
    &
    +\left[-2
    +\lb \frac{33239375 \pi}{2097152}-191 \rb\frac{1}{v^2}
    +\lb \frac{313951835 \pi}{1048576}-\frac{4907}{4} \rb\frac{1}{v^4}
    +\lb \frac{100267485 \pi}{262144}-\frac{11039}{8} \rb\frac{1}{v^6}\right.\nn\\
    &
    +\lb \frac{14461015 \pi}{131072}-\frac{17665}{64} \rb\frac{1}{v^8}
    -\lb \frac{123}{128}+\frac{241235 \pi}{16384} \rb\frac{1}{v^{10}}
    +\lb \frac{59085 \pi}{8192}-\frac{469}{512} \rb\frac{1}{v^{12}}\nn\\
    &
    \left.+\lb \frac{261}{1024}-\frac{3927 \pi}{2048} \rb\frac{1}{v^{14}}
    +\lb \frac{231 \pi}{1024}-\frac{33}{1024} \rb\frac{1}{v^{16}}\right] \hat{q}^{14},
\end{align}
\end{subequations}

For lightlike rays, setting $v=1$ in the above equation produces
\begin{subequations}\label{rngammaexp7to15}
\begin{align}
R_{5,\gamma}=&S_{5,\gamma}-\left(469-\frac{945 \pi }{8}\right) \hat{q}^2+\left(\frac{79}{2}-\frac{105 \pi }{16}\right)\hat{q}^4,\\
R_{6,\gamma}=&S_{6,\gamma}-\left(\frac{227745 \pi }{256}-2493\right) \hat{q}^2+\left(\frac{31605 \pi}{256}-\frac{595}{2}\right) \hat{q}^4-\frac{315 \pi  \hat{q}^6}{256},\\
R_{7,\gamma}=&S_{7,\gamma}+ \left(\frac{647955 \pi }{128}-\frac{336457}{20}\right)
   \hat{q}^2+\left(\frac{13281}{4}-\frac{120015 \pi }{128}\right) \hat{q}^4+\left(\frac{2625 \pi
   }{128}-\frac{407}{4}\right) \hat{q}^6 ,\\
R_{8,\gamma}=&S_{8,\gamma}+
 \left(\frac{1982019}{20}-\frac{132837705 \pi }{4096}\right) \hat{q}^2+\left(\frac{68866875 \pi}{8192}-\frac{99891}{4}\right) \hat{q}^4\nn\\
 &+\left(\frac{5229}{4}-\frac{2009385 \pi }{4096}\right)
   \hat{q}^6+\frac{30345 \pi  \hat{q}^8}{16384},\\
R_{9,\gamma}=&S_{9,\gamma}+\left(\frac{99324225 \pi }{512}-\frac{34599415}{56}\right)
   \hat{q}^2+\left(\frac{15975817}{80}-\frac{63378315 \pi }{1024}\right) \hat{q}^4 \nn\\
   &+\left(\frac{2766225 \pi}{512}-\frac{436271}{24}\right) \hat{q}^6+\left(\frac{7963}{32}-\frac{114345 \pi
   }{2048}\right) \hat{q}^8 ,\\
R_{10,\gamma}=&S_{10,\gamma}+ \left(\frac{208366467}{56}-\frac{78147084015 \pi }{65536}\right)
   \hat{q}^2+\left(\frac{15189128955 \pi }{32768}-\frac{23002213}{16}\right)\hat{q}^4 \nn\\
 &+\left(\frac{1436721}{8}-\frac{1930944015 \pi }{32768}\right)
   \hat{q}^6+\left(\frac{113163435 \pi }{65536}-\frac{155331}{32}\right) \hat{q}^8-\frac{193347 \pi
   \hat{q}^{10}}{65536},\\
R_{11,\gamma}=&S_{11,\gamma}+ \left(\frac{235891610955 \pi }{32768}-\frac{30497954329}{1344}\right)
   \hat{q}^2+\left(\frac{2313962803}{224}-\frac{53538459975 \pi }{16384}\right)
   \hat{q}^4\nn\\
   &+\left(\frac{8797063275 \pi }{16384}-\frac{54734277}{32}\right)
   \hat{q}^6+\left(\frac{5388773}{64}-\frac{841575735 \pi }{32768}\right) \hat{q}^8+\left(\frac{4625775
   \pi }{32768}-\frac{37649}{64}\right) \hat{q}^{10} ,\\
R_{12,\gamma}=&S_{12,\gamma}+\left(\frac{131592603181}{960}-\frac{22914193617315 \pi }{524288}\right)
   \hat{q}^2+\left(\frac{23941565382735 \pi }{1048576}-\frac{2288375219}{32}\right)
   \hat{q}^4\nn\\
   &
   +\left(\frac{1407175289}{96}-\frac{1231101756885 \pi }{262144}\right)
   \hat{q}^6+\left(\frac{355936355775 \pi }{1048576}-\frac{67044835}{64}\right)
   \hat{q}^8\nn\\
   &+\left(\frac{1035463}{64}-\frac{2924276355 \pi }{524288}\right) \hat{q}^{10}+\frac{5127969
   \pi  \hat{q}^{12}}{1048576},\\
R_{13,\gamma}=&S_{13,\gamma}+\left(\frac{34596684547425 \pi }{131072}-\frac{17527835840519}{21120}\right)
   \hat{q}^2+\left(\frac{376091506903}{768}-\frac{40801130054685 \pi }{262144}\right)\hat{q}^4\nn\\
   &
   +\left(\frac{2515221764055 \pi }{65536}-\frac{7740478913}{64}\right)
   \hat{q}^6+\left(\frac{3022083897}{256}-\frac{977380328925 \pi }{262144}\right)
   \hat{q}^8\nn\\
   &
   +\left(\frac{14042373345 \pi }{131072}-\frac{44314579}{128}\right)
   \hat{q}^{10}+\left(\frac{347547}{256}-\frac{89396307 \pi }{262144}\right)
   \hat{q}^{12} ,\\
R_{14,\gamma}=&S_{14,\gamma}+\left(\frac{7053379738825}{1408}-\frac{6690887021944125 \pi }{4194304}\right)
   \hat{q}^2+\left(\frac{4402850339802915 \pi }{4194304}-\frac{12654212780671}{3840}\right)\hat{q}^4\nn\\
   &
   +\left(\frac{60935687417}{64}-\frac{1273048447896225 \pi }{4194304}\right)
   \hat{q}^6+\left(\frac{158640607530405 \pi }{4194304}-\frac{30311686613}{256}\right)
   \hat{q}^8\nn\\
   &
   +\left(\frac{675290915}{128}-\frac{7122235535415 \pi }{4194304}\right)
   \hat{q}^{10}+\left(\frac{70888442625 \pi }{4194304}-\frac{12801815}{256}\right)
   \hat{q}^{12}-\frac{35002539 \pi  \hat{q}^{14}}{4194304} ,\\
R_{15,\gamma}=&S_{15,\gamma}+\left(\frac{20181385376026875 \pi }{2097152}-\frac{201267377207613}{6656}\right)
   \hat{q}^2+\left(\frac{371049227959295}{16896}-\frac{14654459189895525 \pi }{2097152}\right)\hat{q}^4\nn\\
   &
   +\left(\frac{4863074421740535 \pi }{2097152}-\frac{11197800076079}{1536}\right)
   \hat{q}^6+\left(\frac{570686344201}{512}-\frac{742846835125395 \pi }{2097152}\right)\hat{q}^8\nn\\
   &
   +\left(\frac{46298654394993 \pi }{2097152}-\frac{178381330463}{2560}\right)
   \hat{q}^{10}+\left(\frac{1999224191}{1536}-\frac{851988352215 \pi }{2097152}\right)\hat{q}^{12}\nn\\
   &
   +\left(\frac{1675358685 \pi }{2097152}-\frac{1575575}{512}\right)\hat{q}^{14}.
\end{align}
\end{subequations}

When $x_0$ is expressed in terms of $b$ in Eq. \eqref{x0inbrn}, the high order terms are
\begin{subequations}\label{x0inbrn7to15}
\begin{align}
C_{\mathrm{RN},5}=
    &
C_{\mathrm{S},5}- 3\left(\frac{2}{v^2}+\frac{3}{v^4}+\frac{1}{4v^6}\right)\hat{q}^2 + \frac{1}{2}\left(\frac{1}{v^2}+\frac{3}{4 v^4}\right)\hat{q}^4,\\
C_{\mathrm{RN},6}=
    &C_{\mathrm{S},6}-16
   \left(\frac{1}{v^2}+\frac{3}{v^4}+\frac{1}{v^6}\right)\hat{q}^2+
   \left(\frac{3}{v^2}+\frac{6}{v^4}+\frac{1}{v^6}\right)\hat{q}^4,\\
C_{\mathrm{RN},7}=
    &
    C_{\mathrm{S},7}
   -\left(\frac{40}{v^2}+\frac{200}{v^4}+\frac{150}{v^6}+\frac{25}{2 v^8}-\frac{5}{16 v^{10}}\right)
   \hat{q}^2\nn\\
   &
   +\left(\frac{12}{v^2}+\frac{45}{v^4}+\frac{45}{2 v^6}+\frac{15}{16
   v^8}\right) \hat{q}^4
   -\left(+\frac{1}{2 v^2}+\frac{5}{4
   v^4}+\frac{5}{16 v^6}\right) \hat{q}^6,\\
C_{\mathrm{RN},8}=
    &C_{\mathrm{S},8}-48 \left(\frac{2}{v^2}+\frac{15}{v^{4}}+\frac{20} {v^{6}}+\frac{5}{
   v^8}\right) \hat{q}^2+40 \left(\frac{1}{v^{2}}+\frac{6 }{v^{4}}+\frac{6} {v^{6}}+\frac{1}{v^8}\right) \hat{q}^4
   -\left(\frac{4}{v^2}+\frac{18} {v^{4}}+\frac{12}
   {v^{6}}+\frac{1}{v^8}\right) \hat{q}^6,\\
C_{\mathrm{RN},9}=
    &C_{\mathrm{S},9}
   -\left(\frac{224}{v^2}+\frac{2352}{v^4}
   +\frac{4900}{v^6}+\frac{2450}{v^8}+\frac{735}{4 v^{10}}-\frac{49}{8 v^{12}}+\frac{7}{32
   v^{14}}\right)
   \hat{q}^2\nn\\
   &+\left(\frac{120}{v^2}+\frac{1050}{v^4}+\frac{1750}{v^6}+\frac{2625}{4
   v^8}+\frac{525}{16 v^{10}}-\frac{35}{64 v^{12}}\right)
   \hat{q}^4\nn\\
   &
   -\left(\frac{20}{v^2}+\frac{140}{v^4}+\frac{175}{v^6}+\frac{175}{4 v^8}+\frac{35}{32
   v^{10}}\right) \hat{q}^6
   +\left(\frac{1}{2 v^2}+\frac{21}{8
   v^4}+\frac{35}{16 v^6}+\frac{35}{128 v^8}\right)
   \hat{q}^8,\\
C_{\mathrm{RN},10}=
    &C_{\mathrm{S},10}-512 \left(\frac{1}{v^{2}}+\frac{14 }{v^{4}}+\frac{42 }{v^{6}}+\frac{35 }{v^{8}}+\frac{7
   }{v^{10}}\right) \hat{q}^2+336 \left(\frac{1}{v^{2}}+\frac{12 }{v^{4}}+\frac{30 }{v^{6}}+\frac{20 }{v^{8}}+\frac{3 }{v^{10}}\right) \hat{q}^4\nn\\
   &
   -80 \left(\frac{1}{v^{2}}+\frac{10
   }{v^{4}}+\frac{20 }{v^{6}}+\frac{10 }{v^{8}}+\frac{1}{v^{10}}\right) \hat{q}^6+\left(\frac{5}{v^2}+\frac{40}{v^{4}}+\frac{60 }{v^{6}}+\frac{20 }{v^{8}}+\frac{1}{v^{10}}\right)
   \hat{q}^8,\\
C_{\mathrm{RN},11}=
    &C_{\mathrm{S},11}
   -\left(\frac{1152}{v^2}+\frac{20736}{v^4}+
   \frac{84672}{v^6}+\frac{105840}{v^8}+\frac{39690}{v^{10}}+\frac{2646}{v^{12}}-\frac{189}
   {2 v^{14}}+\frac{81}{16 v^{16}}-\frac{45}{256
   v^{18}}\right)
   \hat{q}^2\nn\\
   &
   +\left(\frac{896}{v^2}+\frac{14112}{v^4}+\frac{49392}{v^6}+\frac{51450}{v^8}+\frac{15
   435}{v^{10}}+\frac{3087}{4 v^{12}}-\frac{147}{8 v^{14}}+\frac{63}{128
   v^{16}}\right)
   \hat{q}^4\nn\\
   &
   -\left(\frac{280}{v^2}+\frac{3780}{v^4}+\frac{11025}{v^6}+\frac{18375}{2
   v^8}+\frac{33075}{16 v^{10}}+\frac{2205}{32 v^{12}}-\frac{105}{128
   v^{14}}\right) \hat{q}^6\nn\\
   &
   +\left(\frac{30}{v^2}+\frac{675}{2
   v^4}+\frac{1575}{2 v^6}+\frac{7875}{16 v^8}+\frac{4725}{64 v^{10}}+\frac{315}{256
   v^{12}}\right) \hat{q}^8
   -\left(\frac{1}{2 v^2}+\frac{9}{2
   v^4}+\frac{63}{8 v^6}+\frac{105}{32 v^8}+\frac{63}{256
   v^{10}}\right) \hat{q}^{10},\\
C_{\mathrm{RN},12}=
    &C_{\mathrm{S},12}
    -1280\left(\frac{2}{v^2}+\frac{45
   }{v^{4}}+\frac{240 }{v^{6}}+\frac{420 }{v^{8}}+\frac{252 }{v^{10}}+\frac{42} {v^{12}}\right) \hat{q}^2\nn\\
   &
   +768\left(\frac{3}{v^2}+\frac{60}{v^{4}}+\frac{280
   }{v^{6}}+\frac{420 }{v^{8}}+\frac{210 }{v^{10}}+\frac{28} {v^{12}}\right) \hat{q}^4-448 \left(\frac{2}{v^2}+\frac{35 }{v^{4}}+\frac{140 }{v^{6}}+\frac{175 }{v^{8}}+\frac{70
   }{v^{10}}+\frac{7}{ v^{12}}\right) \hat{q}^6\nn\\
   &
   +140 \left(\frac{1}{v^{2}}+\frac{15 }{v^{4}}+\frac{50 }{v^{6}}+\frac{50 }{v^{8}}+\frac{15
   }{v^{10}}+\frac{1}{v^{12}}\right) \hat{q}^8-\left(\frac{6}{v^2}+\frac{75 }{v^{4}}+\frac{200 }{v^{6}}+\frac{150 }{v^{8}}+\frac{30 }{v^{10}}+\frac{1}{v^{12}}\right)
   \hat{q}^{10},\\
C_{\mathrm{RN},13}=
    &C_{\mathrm{S},13}
   -\left(\frac{5632}{v^2}+\frac{154880}{v^4}
   +\frac{1045440}{v^6}+\frac{2439360}{v^8}+\frac{2134440}{v^{10}}+\frac{640332}{v^{12}}
   +\frac{38115}{v^{14}}-\frac{5445}{4 v^{16}}\right.\nn\\
   &\left.+\frac{5445}{64 v^{18}}-\frac{605}{128
   v^{20}}+\frac{77}{512 v^{22}}\right)
   \hat{q}^2\nn\\
   &
   +\left(\frac{5760}{v^2}+\frac{142560}{v^4}+\frac{855360}{v^6}+\frac{1746360}{v^8}
   +\frac{1309770}{v^{10}}+\frac{654885}{2 v^{12}}+\frac{31185}{2 v^{14}}-\frac{13365}{32
   v^{16}}+\frac{4455}{256 v^{18}}-\frac{495}{1024
   v^{20}}\right)
   \hat{q}^4\nn\\
   &
   -\left(\frac{2688}{v^2}+\frac{59136}{v^4}+\frac{310464}{v^6}+\frac{543312}{v^8}
   +\frac{339570}{v^{10}}+\frac{67914}{v^{12}}+\frac{4851}{2 v^{14}}-\frac{693}{16
   v^{16}}+\frac{231}{256 v^{18}}\right)
   \hat{q}^6\nn\\
   &
   +\left(\frac{560}{v^2}+\frac{10780}{v^4}+\frac{48510}{v^6}+\frac{282975}{4
   v^8}+\frac{282975}{8 v^{10}}+\frac{169785}{32 v^{12}}+\frac{8085}{64
   v^{14}}-\frac{1155}{1024 v^{16}}\right)
   \hat{q}^8\nn\\
   &
   -\left(\frac{42}{v^2}+\frac{693}{v^4}+\frac{10395}{4 v^6}+\frac{24255}{8
   v^8}+\frac{72765}{64 v^{10}}+\frac{14553}{128 v^{12}}+\frac{693}{512
   v^{14}}\right) \hat{q}^{10}\nn\\
   &
   +\left(\frac{1}{2 v^2}+\frac{55}{8
   v^4}+\frac{165}{8 v^6}+\frac{1155}{64 v^8}+\frac{1155}{256 v^{10}}+\frac{231}{1024
   v^{12}}\right) \hat{q}^{12},\\
C_{\mathrm{RN},14}=
    &C_{\mathrm{S},14}-12288
   \left(\frac{1}{v^{2}}+\frac{33 }{v^{4}}+\frac{275 }{v^{6}}+\frac{825 }{v^{8}}+\frac{990 }{v^{10}}+\frac{462 }{v^{12}}+\frac{66 }{v^{14}}\right) \hat{q}^2\nn\\
   &
   +14080
   \left(\frac{1}{v^{2}}+\frac{30 }{v^{4}}+\frac{225 }{v^{6}}+\frac{600 }{v^{8}}+\frac{630 }{v^{10}}+\frac{252 }{v^{12}}+\frac{30 }{v^{14}}\right) \hat{q}^4-7680
   \left(\frac{1}{v^{2}}+\frac{27 }{v^{4}}+\frac{180 }{v^{6}}+\frac{420 }{v^{8}}+\frac{378 }{v^{10}}+\frac{126 }{v^{12}}+\frac{12 }{v^{14}}\right) \hat{q}^6\nn\\
   &
   +2016
   \left(\frac{1}{v^{2}}+\frac{24 }{v^{4}}+\frac{140 }{v^{6}}+\frac{280 }{v^{8}}+\frac{210 }{v^{10}}+\frac{56 }{v^{12}}+\frac{4 }{v^{14}}\right) \hat{q}^8-224
   \left(\frac{1}{v^{2}}+\frac{21 }{v^{4}}+\frac{105 }{v^{6}}+\frac{175 }{v^{8}}+\frac{105 }{v^{10}}+\frac{21 }{v^{12}}+\frac{1}{v^{14}}\right) \hat{q}^{10}\nn\\
   &
   +
   \left(\frac{7}{v^{2}}+\frac{126 }{v^{4}}+\frac{525 }{v^{6}}+\frac{700 }{v^{8}}+\frac{315 }{v^{10}}+\frac{42 }{v^{12}}+\frac{1}{v^{14}}\right) \hat{q}^{12},\\
C_{\mathrm{RN},15}=
    &C_{\mathrm{S},15}
   -\left(\frac{26624}{v^2}+\frac{1038336}{v^
   4}+\frac{10469888}{v^6}+\frac{39262080}{v^8}+\frac{61837776}{v^{10}}+\frac{41225184}{v^{
   12}}+\frac{10306296}{v^{14}}+\frac{552123}{v^{16}}\right.\nn\\
   &
   \left.-\frac{306735}{16
   v^{18}}+\frac{20449}{16 v^{20}}-\frac{5577}{64 v^{22}}+\frac{1183}{256
   v^{24}}-\frac{273}{2048 v^{26}}\right)\hat{q}^2
   +\left(\frac{33792}{v^2}+\frac{1208064}{v^4}+\frac{11073920}{v^6}+\frac{37374480}{v^8
   }\right.\nn\\
   &
   \left.+\frac{52324272}{v^{10}}+\frac{30522492}{v^{12}}+\frac{6540534}{v^{14}}+\frac{2335905}{
   8 v^{16}}-\frac{259545}{32 v^{18}}+\frac{51909}{128 v^{20}}-\frac{4719}{256
   v^{22}}+\frac{1001}{2048 v^{24}}\right)\hat{q}^4\nn\\
   &
   -\left(\frac{21120}{v^2}+\frac{686400}{v^4}+\frac{5662800}{v^6}+\frac{16988400}{v^8}
   +\frac{20810790}{v^{10}}+\frac{10405395}{v^{12}}+\frac{7432425}{4
   v^{14}}+\frac{1061775}{16 v^{16}}\right.\nn\\
   &
   \left.-\frac{353925}{256 v^{18}}+\frac{23595}{512
   v^{20}}-\frac{2145}{2048 v^{22}}\right)\hat{q}^6
   +\left(\frac{6720}{v^2}+\frac{196560}{v^4}+\frac{1441440}{v^6}+\frac{3783780}{v^8}+\frac{3972969}{v^{10}}\right.\nn\\
   &
   \left.+\frac{6621615}{4 v^{12}}+\frac{945945}{4 v^{14}}+\frac{405405}{64
   v^{16}}-\frac{45045}{512 v^{18}}+\frac{3003}{2048
   v^{20}}\right)
   \hat{q}^8\nn\\
   &+\left(-\frac{1008}{v^2}-\frac{26208}{v^4}-\frac{168168}{v^6}-\frac{378378}{v^8}-\frac{1324323}{4 v^{10}}-\frac{441441}{4 v^{12}}-\frac{189189}{16 v^{14}}-\frac{27027}{128
   v^{16}}+\frac{3003}{2048 v^{18}}\right)\hat{q}^{10}\nn\\
   &
   +\left(\frac{56}{v^2}+\frac{1274}{v^4}+\frac{7007}{v^6}+\frac{105105}{8
   v^8}+\frac{147147}{16 v^{10}}+\frac{147147}{64 v^{12}}+\frac{21021}{128
   v^{14}}+\frac{3003}{2048 v^{16}}\right)\hat{q}^{12}\nn\\
   &
   -\left(\frac{1}{2 v^2}+\frac{39}{4 v^4}+\frac{715}{16 v^6}+\frac{2145}{32
   v^8}+\frac{9009}{256 v^{10}}+\frac{3003}{512 v^{12}}+\frac{429}{2048
   v^{14}}\right) \hat{q}^{14}
    .\\
\end{align}
\end{subequations}

The high order terms of Eq. \eqref{angrninb} are
\begin{subequations}\label{angrninb7to15}
\begin{align}
R^\prime_5=
    &
    S^\prime_5- 4\left(\frac{7}{3}+\frac{140}{3 v^2}+\frac{70}{v^4}+\frac{28}{3 v^6}-\frac{1}{3 v^8}\right)\hat{q}^2 +
   2\left(1+\frac{15}{v^2}+\frac{15}{v^4}+\frac{1}{v^6}\right)\hat{q}^4 ,\\
R^\prime_6=
    &
    S^\prime_6-\frac{1575\pi}{256}\left(1+\frac{30}{v^2}+\frac{80}{v^4}+\frac{32}{v^6}\right)\hat{q}^2
    +\frac{105\pi}{256}\left(5+\frac{120}{v^2}+\frac{240}{v^4}+\frac{64}{v^6}\right)\hat{q}^4
    -\frac{5\pi}{256} \left(5+\frac{90}{v^2}+\frac{120}{v^4}+\frac{16}{v^6}\right) \hat{q}^6,\\
R^\prime_{7}=
    &
    S^\prime_{7}-\frac{6}{5} \left(33+\frac{1386}{v^2}+\frac{5775}{v^4}+\frac{4620}{v^6}+\frac{495}{v^8}
    -\frac{22}{v^{10}}+\frac{1}{v^{12}}\right)\hat{q}^2+2\left(9+\frac{315}{v^2}+\frac{1050}{v^4}
    +\frac{630}{v^6}+\frac{45}{v^8}-\frac{1}{v^{10}}\right)\hat{q}^4\nn\\
    &
    -2 \left(1+\frac{28}{v^2}+\frac{70}{v^4}+\frac{28}{v^6}+\frac{1}{v^8}\right) \hat{q}^6,\\
R^\prime_{8}=
    &
    S^\prime_{8}-\frac{105105 \pi }{4096} \left(1+\frac{56}{v^2}+\frac{336}{v^4}+\frac{448}{v^6}+\frac{128}{v^8}\right)\hat{q}^2
    +\frac{17325 \pi }{8192} \left(7+\frac{336}{v^2}+\frac{1680}{v^4}+\frac{1792}{v^6}+\frac{384}{v^8}\right)\hat{q}^4\nn\\
    &
    -\frac{1575\pi}{4096}\left(7+\frac{280}{v^2}+\frac{1120}{v^4}+\frac{896}{v^6}
    +\frac{128}{v^8}\right)\hat{q}^6
    +\frac{35 \pi}{16384}\left(35+\frac{1120}{v^2}+\frac{3360}{v^4}
    +\frac{1792}{v^6}+\frac{128}{v^8}\right)\hat{q}^8,\\
R^\prime_{9}=
    &
    S^\prime_{9}-\frac{8}{7} \left(143+\frac{10296}{v^2}+\frac{84084}{v^4}+\frac{168168}{v^6}+\frac{90090}{v^8}
    +\frac{8008}{v^{10}}-\frac{364}{v^{12}}+\frac{24}{v^{14}}-\frac{1}{v^{16}}\right) \hat{q}^2\nn\\
    &
    +\frac{4}{5} \left(143+\frac{9009}{v^2}+\frac{63063}{v^4}+\frac{105105}{v^6}+\frac{45045}{v^8}
    +\frac{3003}{v^{10}}-\frac{91}{v^{12}}+\frac{3}{v^{14}}\right)\hat{q}^4\nn\\
    &
    -\frac{8}{3} \left(11+\frac{594}{v^2}+\frac{3465}{v^4}+\frac{4620}{v^6}+\frac{1485}{v^8}
    +\frac{66}{v^{10}}-\frac{1}{v^{12}}\right)\hat{q}^6
    +2 \left(1+\frac{45}{v^2}+\frac{210}{v^4}+\frac{210}{v^6}+\frac{45}{v^8}
    +\frac{1}{v^{10}}\right) \hat{q}^8,\\
R^\prime_{10}=
    &
    S^\prime_{10}-\frac{6891885\pi}{65536}\left(1+\frac{90}{v^2}+\frac{960}{v^4}
    +\frac{2688}{v^6}+\frac{2304}{v^8}+\frac{512}{v^{10}}\right)\hat{q}^2\nn\\
    &
    +\frac{315315\pi}{32768}\left(9+\frac{720}{v^2}
    +\frac{6720}{v^4}+\frac{16128}{v^6}+\frac{11520}{v^8}
    +\frac{2048}{v^{10}}\right) \hat{q}^4
    -\frac{315315\pi}{32768}\left(3+\frac{210}{v^2}+\frac{1680}{v^4}+\frac{3360}{v^6}
    +\frac{1920}{v^8}+\frac{256}{v^{10}}\right)\hat{q}^6\nn\\
    &
    +\frac{10395 \pi}{65536}\left(21+\frac{1260}{v^2}+\frac{8400}{v^4}+\frac{13440}{v^6}
    +\frac{5760}{v^8}+\frac{512}{v^{10}}\right) \hat{q}^8\nn\\
    &
    -\frac{63\pi}{65536}\left(63+\frac{3150}{v^2}+\frac{16800}{v^4}+\frac{20160}{v^6}
    +\frac{5760}{v^8}+\frac{256}{v^{10}}\right)\hat{q}^{10},\\
R^\prime_{11}=
    &
    S^\prime_{11}-\frac{10}{63} \left(4199+\frac{461890}{v^2}+\frac{6235515}{v^4}+\frac{23279256}{v^6}
    +\frac{29099070}{v^8}+\frac{11639628}{v^{10}}+\frac{881790}{v^{12}}-\frac{38760}{v^{14}}
    \right.\nn\\
    &
    \left.+\frac{2907}{v^{16}}-\frac{190}{v^{18}}+\frac{7}{v^{20}}\right) \hat{q}^2\nn\\
    &
    +\frac{4}{7}\left(1105+\frac{109395}{v^2}+\frac{1312740}{v^4}+\frac{4288284}{v^6}
    +\frac{4594590}{v^8}+\frac{1531530}{v^{10}}+\frac{92820}{v^{12}}-\frac{3060}{v^{14}}
    +\frac{153}{v^{16}}-\frac{5}{v^{18}}\right)\hat{q}^4\nn\\
    &
    -4 \left(65+\frac{5720}{v^2}+\frac{60060}{v^4}+\frac{168168}{v^6}+\frac{150150}{v^8}
    +\frac{40040}{v^{10}}+\frac{1820}{v^{12}}-\frac{40}{v^{14}}+\frac{1}{v^{16}}\right)\hat{q}^6\nn\\
    &
    +\frac{10}{3}\left(13+\frac{1001}{v^2}+\frac{9009}{v^4}+\frac{21021}{v^6}
    +\frac{15015}{v^8}+\frac{3003}{v^{10}}+\frac{91}{v^{12}}-\frac{1}{v^{14}}\right)\hat{q}^8\nn\\
    &
    -2\left(1+\frac{66}{v^2}+\frac{495}{v^4}+\frac{924}{v^6}+\frac{495}{v^8}+\frac{66}{v^{10}}
    +\frac{1}{v^{12}}\right) \hat{q}^{10},\\
R^\prime_{12}=
    &
    S^\prime_{12}-\frac{32008977\pi}{524288}\left(7+\frac{924 }{v^{2}}+\frac{15400 }{v^{4}}+\frac{73920 }{v^{6}}+\frac{126720 }{v^{8}}+\frac{78848}{v^{10}}+\frac{14336 }{v^{12}}\right) \hat{q}^2\nn\\
    &
    +\frac{43648605 \pi}{1048576}\left(11+\frac{1320
   }{v^{2}}+\frac{19800 }{v^{4}}+\frac{84480 }{v^{6}}+\frac{126720 }{v^{8}}+\frac{67584 }{v^{10}}+\frac{10240 }{v^{12}}\right)
   \hat{q}^4\nn\\
    &-\frac{5360355\pi}{262144}\left(11+\frac{1188 }{v^{2}}+\frac{15840 }{v^{4}}+\frac{59136 }{v^{6}}+\frac{76032
   }{v^{8}}+\frac{33792 }{v^{10}}+\frac{4096 }{v^{12}}\right)\hat{q}^6\nn\\
   &+\frac{1576575 \pi}{1048576}\left(33+\frac{3168
   }{v^{2}}+\frac{36960 }{v^{4}}+\frac{118272 }{v^{6}}+\frac{126720 }{v^{8}}+\frac{45056 }{v^{10}}+\frac{4096 }{v^{12}}\right)
   \hat{q}^8\nn\\
   &-\frac{63063\pi}{524288}\left(33+\frac{2772 }{v^{2}}+\frac{27720 }{v^{4}}+\frac{73920 }{v^{6}}+\frac{63360
   }{v^{8}}+\frac{16896 }{v^{10}}+\frac{1024 }{v^{12}}\right)\hat{q}^{10}\nn\\
   &+\frac{231 \pi}{1048576}\left(231+\frac{16632 }{v^{2}}+\frac{138600 }{v^{4}}+\frac{295680 }{v^{6}}+\frac{190080 }{v^{8}}+\frac{33792 }{v^{10}}+\frac{1024 }{v^{12}}\right)
   \hat{q}^{12},\\
R^\prime_{13}=
    &
    S^\prime_{13}-\frac{4}{11} \left(7429+\frac{1158924}{v^2}+\frac{23371634}{ v^4}+\frac{140229804 }{v^6}+\frac{315517059}{ v^8}+\frac{280459608}{
    v^{10}}+\frac{89237148}{v^{12}}+\frac{5883768}{ v^{14}}\right.\nn\\
    &
    \left.-\frac{245157 }{v^{16}}+\frac{19228 }{v^{18}}-\frac{1518}{ v^{20}}+\frac{92 }{v^{22}}-\frac{3}{v^{24}}\right) \hat{q}^2
    +\frac{10}{3} \left(969+\frac{138567 }{v^2}+\frac{2540395}{ v^4}+\frac{13718133}{ v^6}+\frac{27436266}{v^8}\right.\nn\\
    &
    \left.+\frac{21339318}{ v^{10}}+\frac{5819814 }{v^{12}}+\frac{319770 }{v^{14}}-\frac{10659 }{v^{16}}+\frac{627}{ v^{18}}-\frac{33}{v^{20}}+\frac{1}{v^{22}}\right) \hat{q}^4\nn\\
    &
   -\frac{40}{7} \left(323+\frac{41990 }{v^2}+\frac{692835 }{v^4}+\frac{3325608}{
   v^6}+\frac{5819814}{ v^8}+\frac{3879876}{ v^{10}}+\frac{881790 }{v^{12}}+\frac{38760}{ v^{14}}-\frac{969 }{v^{16}}+\frac{38}{
   v^{18}}-\frac{1}{v^{20}}\right) \hat{q}^6\nn\\
   &+6 \left(85+\frac{9945}{ v^2}+\frac{145860 }{v^4}+\frac{612612 }{v^6}+\frac{918918}{ v^8}+\frac{510510}{
   v^{10}}+\frac{92820 }{v^{12}}+\frac{3060 }{v^{14}}-\frac{51 }{v^{16}}+\frac{1}{v^{18}}\right) \hat{q}^8\nn\\
   &-4 \left(15+\frac{1560}{ v^2}+\frac{20020}{
   v^4}+\frac{72072}{ v^6}+\frac{90090 }{v^8}+\frac{40040 }{v^{10}}+\frac{5460 }{v^{12}}+\frac{120 }{v^{14}}-\frac{1}{v^{16}}\right) \hat{q}^{10}\nn\\
   &+2\left(1+\frac{91 }{v^2}+\frac{1001 }{v^4}+\frac{3003 }{v^6}+\frac{3003 }{v^8}+\frac{1001 }{v^{10}}+\frac{91 }{v^{12}}+\frac{1}{v^{14}}\right)
   \hat{q}^{12},\\
R^\prime_{14}=
    &
    S^\prime_{14}-\frac{7250518275\pi}{4194304}\left(1+\frac{182 }{v^{2}}+\frac{4368 }{v^{4}}+\frac{32032 }{v^{6}}+\frac{91520 }{v^{8}}+\frac{109824
   }{v^{10}}+\frac{53248 }{v^{12}}+\frac{8192 }{v^{14}}\right) \hat{q}^2\nn\\
   &
   +\frac{736206471 \pi}{4194304}\left(13+\frac{2184 }{v^{2}}+\frac{48048 }{v^{4}}+\frac{320320 }{v^{6}}+\frac{823680 }{v^{8}}+\frac{878592 }{v^{10}}+\frac{372736 }{v^{12}}+\frac{49152
   }{v^{14}}\right) \hat{q}^4\nn\\
   &
   -\frac{480134655\pi}{4194304}\left(13+\frac{2002 }{v^{2}}+\frac{40040
   }{v^{4}}+\frac{240240 }{v^{6}}+\frac{549120 }{v^{8}}+\frac{512512 }{v^{10}}+\frac{186368 }{v^{12}}+\frac{20480 }{v^{14}}\right)
   \hat{q}^6\nn\\
   &
   +\frac{14549535 \pi}{4194304}\left(143+\frac{20020 }{v^{2}}+\frac{360360 }{v^{4}}+\frac{1921920 }{v^{6}}+\frac{3843840
   }{v^{8}}+\frac{3075072 }{v^{10}}+\frac{931840 }{v^{12}}+\frac{81920 }{v^{14}}\right) \hat{q}^8\nn\\
   &
   -\frac{2297295\pi}{4194304}\left(143+\frac{18018 }{v^{2}}+\frac{288288 }{v^{4}}+\frac{1345344 }{v^{6}}+\frac{2306304 }{v^{8}}+\frac{1537536 }{v^{10}}+\frac{372736
   }{v^{12}}+\frac{24576 }{v^{14}}\right)\hat{q}^{10}\nn\\
   &
   +\frac{45045 \pi}{4194304}\left(429+\frac{48048
   }{v^{2}}+\frac{672672 }{v^{4}}+\frac{2690688 }{v^{6}}+\frac{3843840 }{v^{8}}+\frac{2050048 }{v^{10}}+\frac{372736 }{v^{12}}+\frac{16384 }{v^{14}}\right)
   \hat{q}^{12}\nn\\
   &
   -\frac{429\pi}{4194304}\left(429+\frac{42042 }{v^{2}}+\frac{504504 }{v^{4}}+\frac{1681680 }{v^{6}}+\frac{1921920
   }{v^{8}}+\frac{768768 }{v^{10}}+\frac{93184 }{v^{12}}+\frac{2048 }{v^{14}}\right) \hat{q}^{14},\\
R^\prime_{15}=
    &
    S^\prime_{15}-\frac{14}{143} \left(111435+\frac{23401350 }{v^2}+\frac{659138025 }{v^4}+\frac{5800414620 }{v^6}+\frac{20508608835
   }{v^8}+\frac{31902280410 }{v^{10}}\right.\nn\\
   &
   \left.+\frac{21751554825 }{v^{12}}+\frac{5736673800 }{v^{14}}+\frac{334639305 }{v^{16}}-\frac{13123110
   }{v^{18}}+\frac{1036035 }{v^{20}}-\frac{89700 }{v^{22}}+\frac{6825 }{v^{24}}-\frac{378 }{v^{26}}+\frac{11 }{v^{28}}\right)
   \hat{q}^2\nn\\
   &
   +\frac{2}{33} \left(260015+\frac{50702925 }{v^2}
   +\frac{1318276050 }{v^4}+\frac{10634093470 }{v^6}+\frac{34181014725
   }{v^8}+\frac{47853420615 }{v^{10}}\right.\nn\\
   &\left.+\frac{29002073100 }{v^{12}}+\frac{6692786100 }{v^{14}}+\frac{334639305 }{v^{16}}-\frac{10935925
   }{v^{18}}+\frac{690690 }{v^{20}}-\frac{44850 }{v^{22}}+\frac{2275 }{v^{24}}-\frac{63 }{v^{26}}\right) \hat{q}^4\nn\\
   &-\frac{2}{9}
   \left(52003+\frac{9360540 }{v^2}+\frac{223092870 }{v^4}+\frac{1636014380 }{v^6}+\frac{4732755885 }{v^8}+\frac{5889651768
   }{v^{10}}+\frac{3123300180 }{v^{12}}\right.\nn\\
   &
   \left.+\frac{617795640 }{v^{14}}+\frac{25741485 }{v^{16}}-\frac{672980 }{v^{18}}+\frac{31878
   }{v^{20}}-\frac{1380 }{v^{22}}+\frac{35 }{v^{24}}\right) \hat{q}^6
   +2 \left(2261+\frac{373065 }{v^2}+\frac{8083075}{ v^4}\right.\nn\\
   &
   \left.+\frac{53348295
   }{v^6}+\frac{137181330 }{v^8}+\frac{149375226 }{v^{10}}+\frac{67897830 }{v^{12}}+\frac{11191950 }{v^{14}}+\frac{373065 }{v^{16}}-\frac{7315
   }{v^{18}}+\frac{231 }{v^{20}}-\frac{5 }{v^{22}}\right) \hat{q}^8\nn\\
   &
   -\frac{14}{5} \left(323+\frac{48450}{v^2}+\frac{944775
   }{v^4}+\frac{5542680 }{v^6}+\frac{12471030 }{v^8}+\frac{11639628 }{v^{10}}+\frac{4408950 }{v^{12}}+\frac{581400 }{v^{14}}+\frac{14535
   }{v^{16}}-\frac{190 }{v^{18}}+\frac{3 }{v^{20}}\right) \hat{q}^{10}\nn\\
   &
   +\frac{14}{3} \left(17+\frac{2295 }{v^2}+\frac{39780
   }{v^4}+\frac{204204 }{v^6}+\frac{393822 }{v^8}+\frac{306306 }{v^{10}}+\frac{92820 }{v^{12}}+\frac{9180 }{v^{14}}+\frac{153
   }{v^{16}}-\frac{1}{v^{18}}\right) \hat{q}^{12}\nn\\
   &
   -2 \left(1+\frac{120 }{v^2}+\frac{1820 }{v^4}+\frac{8008 }{v^6}+\frac{12870 }{v^8}+\frac{8008
   }{v^{10}}+\frac{1820 }{v^{12}}+\frac{120 }{v^{14}}+\frac{1}{v^{16}}\right) \hat{q}^{14}.\\
\end{align}
\end{subequations}

The full form of Eq. \eqref{angrnlightinb} is
\bea
I_{\mathrm{RN},\gamma}(b)&=&I_{\mathrm{S},\gamma}(b)
    -\frac{3 \pi}{4}\lb\frac{m}{b}\rb^{2}
    -16 \hat{q}^2\lb\frac{m}{b}\rb^{3}
    +\frac{105}{64} \pi \left(-18\hat{q}^2+\hat{q}^4\right)\lb\frac{m}{b}\rb^{4}
    +64\left(-8\hat{q}^2+\hat{q}^4\right)\lb\frac{m}{b}\rb^{5}\nn\\
    &&
    -\frac{1155}{256} \pi \left(195\hat{q}^2-39 \hat{q}^4+\hat{q}^6\right)\lb\frac{m}{b}\rb^{6}
    -\frac{256}{5}\left(288\hat{q}^2-80 \hat{q}^4+5 \hat{q}^6\right)\lb\frac{m}{b}\rb^{7}\nn\\
    &&
    +\frac{45045 \pi}{16384}\left(-9044\hat{q}^2+3230 \hat{q}^4-340 \hat{q}^6+5 \hat{q}^8\right)\lb\frac{m}{b}\rb^{8}
    +\frac{1024}{21} \left(-8448\hat{q}^2+3696 \hat{q}^4-560 \hat{q}^6+21 \hat{q}^8\right)\lb\frac{m}{b}\rb^{9}\nn\\
    &&
    -\frac{2909907 \pi}{65536}\left(15525\hat{q}^2-8050 \hat{q}^4+1610 \hat{q}^6-105 \hat{q}^8+\hat{q}^{10}\right)\lb\frac{m}{b}\rb^{10}\nn\\
   &&
    -\frac{4096}{3} \left(8320\hat{q}^2-4992 \hat{q}^4+1248 \hat{q}^6-120 \hat{q}^8+3 \hat{q}^{10}\right)\lb\frac{m}{b}\rb^{11}\nn\\
    &&
    +\frac{22309287 \pi}{1048576}\left(-890010\hat{q}^2+606825 \hat{q}^4-182700 \hat{q}^6+23625 \hat{q}^8-1050 \hat{q}^{10}+7
   \hat{q}^{12}\right)\lb\frac{m}{b}\rb^{12}\nn\\
   &&
    +\frac{16384}{33}\left(-626688\hat{q}^2+478720 \hat{q}^4-168960 \hat{q}^6+27720 \hat{q}^8-1848 \hat{q}^{10}+33
   \hat{q}^{12}\right)\lb\frac{m}{b}\rb^{13}\nn\\
   &&
    -\frac{717084225 \pi}{4194304}\left(3026933\hat{q}^2-2561251 \hat{q}^4+1038345 \hat{q}^6-207669 \hat{q}^8+18879 \hat{q}^{10}-609\hat{q}^{12}+3 \hat{q}^{14}\right)\lb\frac{m}{b}\rb^{14}\nn\\
    &&
    -\frac{65536}{2145}\left(277831680\hat{q}^2-257986560 \hat{q}^4+118243840 \hat{q}^6-28005120 \hat{q}^8+3267264
   \hat{q}^{10}-160160 \hat{q}^{12}\right.\nn\\
   &&
   \left.+2145 \hat{q}^{14}\right)\lb\frac{m}{b}\rb^{15}
+ \mathcal{O}\lb \frac{m}{b}\rb ^{16}, \label{angrnlightinbfull}\eea

\end{document}